\documentclass[11pt,english,expanded,copyright]{fsuthesis}

\usepackage{color}
\usepackage{float}
\usepackage{amsmath}
\usepackage{amssymb}
\usepackage{mathptmx}
\usepackage{graphicx}
\usepackage{esint}
\usepackage{caption}
\usepackage{subfig}
\usepackage[english]{babel}
\usepackage{lmodern}
\usepackage[T1]{fontenc}
\usepackage[babel=true]{microtype}

\usepackage{bm}
\usepackage{wasysym}

\usepackage{babel}
\usepackage[numbers]{natbib}
\addto\captionsenglish{
 \renewcommand{\contentsname}
   {Table of Contents}
}
\usepackage{afterpage}
\renewcommand{\[}{\begin{equation}}
\renewcommand{\]}{\end{equation}}
\newcommand{\lyxdot}{.}

\title{ \protect  IMPURITIES AND DEFECTS IN MOTT SYSTEMS}

\author{Shao Tang}

\college{College of Arts $\&$ Sciences}

\department{Department of Physics} 

\manuscripttype{Dissertation}              

\degree{Doctor of Philosophy}                 

\degreeyear{}

\defensedate{}

\begin{document}

\frontmatter

\maketitle



%

\tableofcontents
\listoffigures





\begin{abstract}

Disorder has intriguing consequences for correlated electronic materials,
which include several families of high-temperature superconductors
and resistive switching systems. In this dissertation, we study the
effects of impurities intertwined with correlations.

First, we study impurity healing effects in models of strongly correlated
superconductors. We show that in general both the range and the amplitude
of the spatial variations caused by nonmagnetic impurities are significantly
suppressed in the superconducting as well as in the normal states.
We explicitly quantify the weights of the local and the non-local
responses to inhomogeneities and show that the former are overwhelmingly
dominant over the latter. We find that the local response is characterized
by a well-defined healing length scale, which is restricted to only
a few lattice spacings over a significant range of dopings in the
vicinity of the Mott insulating state. We demonstrate that this healing
effect is ultimately due to the suppression of charge fluctuations
induced by Mottness. We also define and solve analytically a simplified
yet accurate model of healing, within which we obtain simple expressions
for quantities of direct experimental relevance.

Second, we address the question of why strongly correlated $d$-wave
superconductors, such as the cuprates, prove to be surprisingly robust
against the introduction of non-magnetic impurities. We show that,
very generally, both the pair-breaking and the normal state transport
scattering rates are significantly suppressed by strong correlations
effects arising in the proximity to a Mott insulating state. We also
show that the correlation-renormalized scattering amplitude is generically
enhanced in the forward direction, an effect which was previously
often ascribed to the specific scattering by charged impurities outside
the copper-oxide planes.

Finally, we provide the theoretical insights for resistive switching
systems and show how impurities and underlying correlations can play
significant roles in practical devices. We report the striking result of a connection between the resistive switching and {\em shock wave} formation, a classic topic of non-linear dynamics. We argue that the profile of oxygen vacancies that migrate during the commutation forms a shock wave that propagates through a highly resistive region of the device. We validate the scenario by means of model simulations and experiments in a manganese-oxide based memristor device and we extend our theory to the case of binary oxides. The shock wave scenario brings unprecedented physical insight and enables to rationalize the process of oxygen-vacancy-driven resistive change with direct implications for a key technological aspect -- the commutation speed.
\end{abstract}

\mainmatter

\chapter{Introduction}

Strongly correlated systems have become one of the central puzzles
in condensed matter physics after the discovery of unconventional
superconductivity in heavy fermion systems and copper oxide materials
\cite{kotliar1988superexchange,anderson1987resonating,anderson2004physics,anderson1997theory}.
The cuprates, which are widely considered as strongly correlated systems,
have many important features, such as the breakdown of conventional
Fermi-liquid state, quantum antiferromagnetism, pseudogap phase \cite{Daou2010},
spin-liquid phase, localization and the quantum critical point \cite{caprara2000charge,tallon2001doping},
etc. Strong electronic correlations are believed to be essential for
a complete understanding of many classes of unconventional superconductors
including the cuprates \cite{anderson1987resonating,anderson2004physics,lee2006doping,Dagotto2005},
heavy fermion superconductors \cite{varma85}, organic materials \cite{powellmckenzie06,powellmckenzie11}
and iron pnictides \cite{johnston10}. 

Among the many puzzling features of these systems is their behavior
in the presence of disorder\cite{Samietal2014,Rhodesetal}. In the case of the cuprates, experiments
have shown that these \textit{$d$}\textit{\emph{-wave superconductors
are quite robust against disorder as introduced by carrier doping
\cite{fujita2012spectroscopic,Dagotto2005,mcelroy2005atomic}. In
particular, there seems to be a \textquotedblleft quantum protection\textquotedblright{}
of the $d$-wave nodal points \cite{Anderson21042000}. Typically,
it has been confirmed that the low energy part of the density of state
at nodal points shows the clear $V$ shape and would not be smeared
in presence of doping disorder. Other anomalies were found in the
organics \cite{analytisetal06} and the pnictides \cite{lietal12}.
Although it is controversial whether conventional theory is able to
explain these features}} \cite{balatsky2006impurity}\textit{\emph{,}}
strong electronic interactions can give rise to these impurity screening
effects. Indeed, they have been captured numerically by\textit{\emph{
the Gutzwiller-projected wave function \cite{Fukushima20083046,PhysRevB.79.184510,Garg2008},
even though a deeper insight into the underlying mechanism is still
lacking. Similar impurity screening phenomena have been found as a
result of strong correlations in the metallic state of the Hubbard
model \cite{Andrade2010}.}}

In weakly interacting $d$-wave superconductors, Abrikosov-Gor'kov
(AG) theory predicts \textit{\emph{that a tiny amount of non-magnetic
impurities should bring the transition temperature $T_{c}$}} to zero
as shown in Table \ref{tab:Effects-of-potential}. In the case of
the cuprates, however, experiments have shown that these \textit{$d$}\textit{\emph{-wave
superconductors are very robust against disorder \cite{fujita2012spectroscopic,keimeretal15,mcelroy2005atomic,Dagotto2005}.}}
This feature was frequently ascribed to scattering by charged off-plane
impurities, which is mostly in the forward direction (see, e.g., \cite{PhysRevLett.113.057001}).
The puzzle was partially clarified, however, once strong electronic
interactions were shown to give rise to the impurity screening effects
seen in these experiments, as captured by\textit{\emph{ the Gutzwiller-projected
wave function \cite{Fukushima20083046,PhysRevB.79.184510,shaoetal15,Garg2008,shaoetal2016}}}.
\begin{table}
\caption{Effects of potential and magnetic scattering for different types of
superconductors are provided. ``+'' indicates the impurity is pair
breaking while ``-'' implies that it is not. At high enough concentration
of impurity, superconductivity would be suppressed in all cases.\label{tab:Effects-of-potential}}

\centering{}%
\begin{tabular}{|c|c|c|c|}
\hline 
 & $s$ & $p$ & $d$\tabularnewline
\hline 
Magnetic scattering & $+$ & $+$ & $+$\tabularnewline
\hline 
Non-magnetic scattering & $-$ & $+$ & $+$\tabularnewline
\hline 
\end{tabular}
\end{table}

\begin{figure}[H]
\centering{}\includegraphics[width=6in,height=2in]{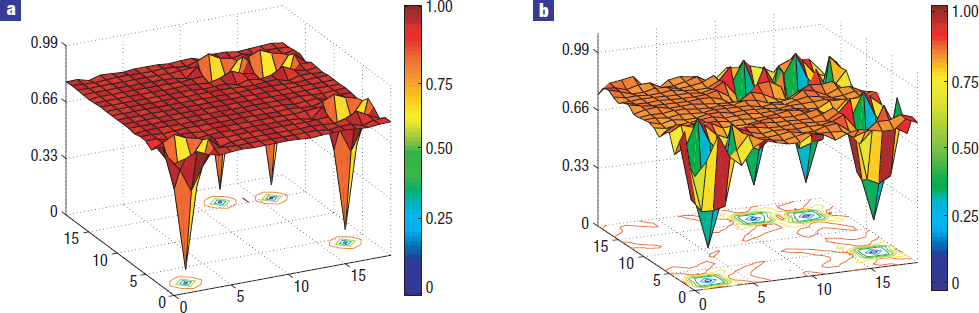}\caption{Spatial variations of the normalized local paring amplitude in cases
of with correlations (left panel) and without correlations (right
panel) (A. Garg et al. , 2008). \label{fig:Spatial-variations-of}}
\end{figure}

Numerically, exact diagnolization has been utilized to study the disorder
effect in the superconducting state, while the main drawback is the
inability to deal with a macroscopic system with a large amount of
lattice sites. Additionally, solving Bogoliubov\textendash de Gennes
and Andreev (BdG) equations self-consistently has also been applied
\cite{paramekanti2001projected,Paramekanti2004}. However, recent
findings showed a drastic contrast between the simple BdG method and
the BdG plus correlations within the Gutzwiller approximation \cite{Garg2008}.
In this work, A. Garg et al. generalized the conventional theory including
the correlation effects by projecting the ground state wave function
to the restricted subspace, where double occupancy of electrons is
not allowed on any lattice site. Remarkable results show that the
impurity effects are largely screened in the $t-t^{\prime}-J$ model
as shown in Fig.(\ref{fig:Spatial-variations-of}). Comparing (a)
with (b) of Fig.(\ref{fig:Spatial-variations-of}), they unambiguously
show how impurity effects are screened when correlation effects are
taken into account. 

\begin{figure}
\centering{}\includegraphics[width=4in,height=4in,keepaspectratio]{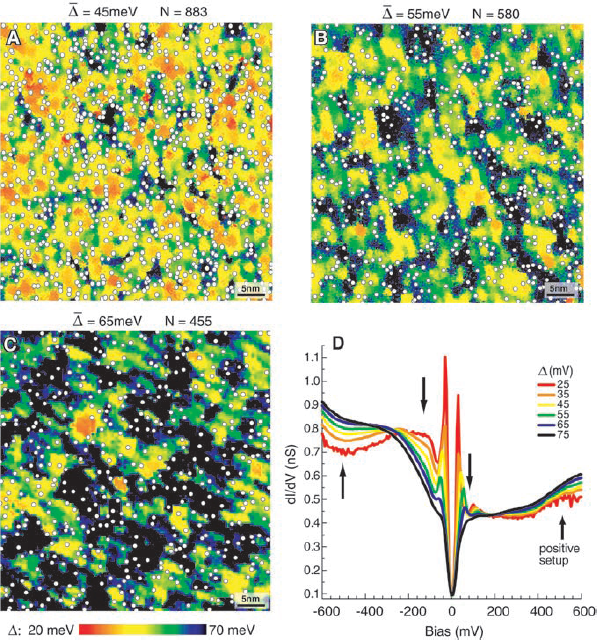}\caption{Three gap function maps are measured on samples with three different
oxygen doping levels as shown in plots A,B and C respectively. The
field of view size is identical and corresponds to 16,000 $\mathrm{CuO_{2}}$
plaquettes approximately. Average gap magnitudes $\bar{\Delta}$ are
shown at the top, together with values of $N$, which is the total
number of dopant atoms shown as white circles. Curves in plot D represent
the local density of states at different locations (K. McElroy et
al. , 2005).\label{fig:experiment gap}}
\end{figure}

Taking $\mathrm{Bi_{2}Sr_{2}CaCu_{2}O_{8+\delta}}$ as an example
\cite{mcelroy2005atomic}, the experimental results confirm that the
$d$ wave superconductivity is indeed robust against the doping disorder.
In (D) of Fig.(\ref{fig:experiment gap}), density of states curves
clearly indicate the significant spectral weight redistribution and
the strong coherence peak modification in high energy range. On the
contrary, weak scatterings between quasi-particles are dominant in
low energy scale seem to be protected from disorder effects. Similar
phenomena are common in all nonstoichiometric oxygen-doped high-Tc
cuprates and such energy-resolved inhomogeneity might be a robust
and general feature of disordered Mott systems.

\begin{figure}
\begin{centering}
\includegraphics[width=4in,height=4in,keepaspectratio]{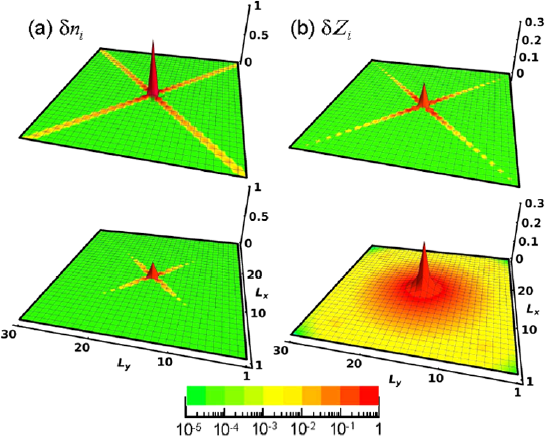}
\par\end{centering}

\caption{Spatial fluctuations of electron density (a) and quasiparticle weight
(b) vary as we approach the Mott transition critical point from top
to bottom (E. C. Andrade et al. , 2010). \label{fig:Spatial-fluctuations-of}}
\end{figure}

Analytically, it has been shown that, similar to the numerical results
obtained from the $t-t^{\prime}-J$ model,  the impurity screening
effect is also found in the Hubbard model. Following the Kotliar\textquoteright s
representation of four slave boson scheme \cite{kotliar1986new},
it has been shown that we can obtain analytical results regarding
the spatial fluctuations for different physical quantities at the
long wavelength limit. Considering the Hubbard model on a square lattice
at half-filling, there exists a metal-insulator transition when the
interaction strength $U$ is above the critical value $U_{c}$. From
Fig.(\ref{fig:Spatial-fluctuations-of}), as we approach the critical
point, it is clear that the density fluctuation is suppressed and
the correlation length for quasiparticle weight $Z_{i}$ diverges.
Starting from half-filling, the finite impurity potential tends to
push the site occupancy away from $1$ and the system becomes locally
more metallic by increasing the quasiparticle weight $Z_{i}$ \cite{Andrade2010}.

Despite this progress, it would be desirable to understand to what
extent this disorder screening is due only to the presence of strong
correlations or whether it is dependent on the details of the particular
model or system.\textit{\emph{ }}The main difference separating cuprates
from other materials is due to the comparable large $J$ term ($J/t=0.3$).
Thus, we shall start from the $t-J$ model \cite{RevModPhys.63.1}
which is then solved via a spatially inhomogeneous slave boson treatment
quantitatively \textit{\emph{\cite{coleman1984new,kotliar1986new,lee2006doping,kotliar1988superexchange,lee19982}. }}

At this stage, we shall briefly introduce the concept and main usage
of the slave boson method without technical details. This method is
a useful tool for analyzing strong correlation effects. In the large-$N$
limit in which there is a controlled treatment, the bosonic field
(the holon in this case) condenses at zero temperature. It is well
known that, within this approach, the effects of strong correlations
are determined by the condensed slave-boson and Lagrange multiplier
fields. They determine, through the constraint, the charge response
of the system. In the non-homogeneous situation we are considering,
the slave-boson and Lagrange multiplier fields show strong spatial
fluctuations and, as a result, the charge response is also non-homogeneous
and non-trivial. 

Furthermore, the slave-boson treatment provides information on the
superconducting gap and quasiparticle spectrum. As has been shown
in many contexts (the Kondo effect, heavy fermions, the Mott transition)
\cite{lee2006doping}, a condensed holon/charge field (together with
the Lagrange multiplier fields) signals the strong suppression of
charge fluctuations, perhaps the dominant feature in strongly correlated
systems. In the particular case of Mott systems, it is the single
most important force leading to the Mott insulating state. As the
pairing field and the single-particle Green\textquoteright s function
involve convolutions of the holon and the spinon fields, the fact
that the former is condensed simply means that it comes into the convolution
as a multiplicative factor. It is precisely this seemingly innocuous
multiplicative factor, ultimately coming from the suppression of charge
fluctuations, that is responsible for important effects in this theory,
including the healing we describe and the Mott transition itself.

\begin{figure}
\centering{}\includegraphics[width=0.95\textwidth]{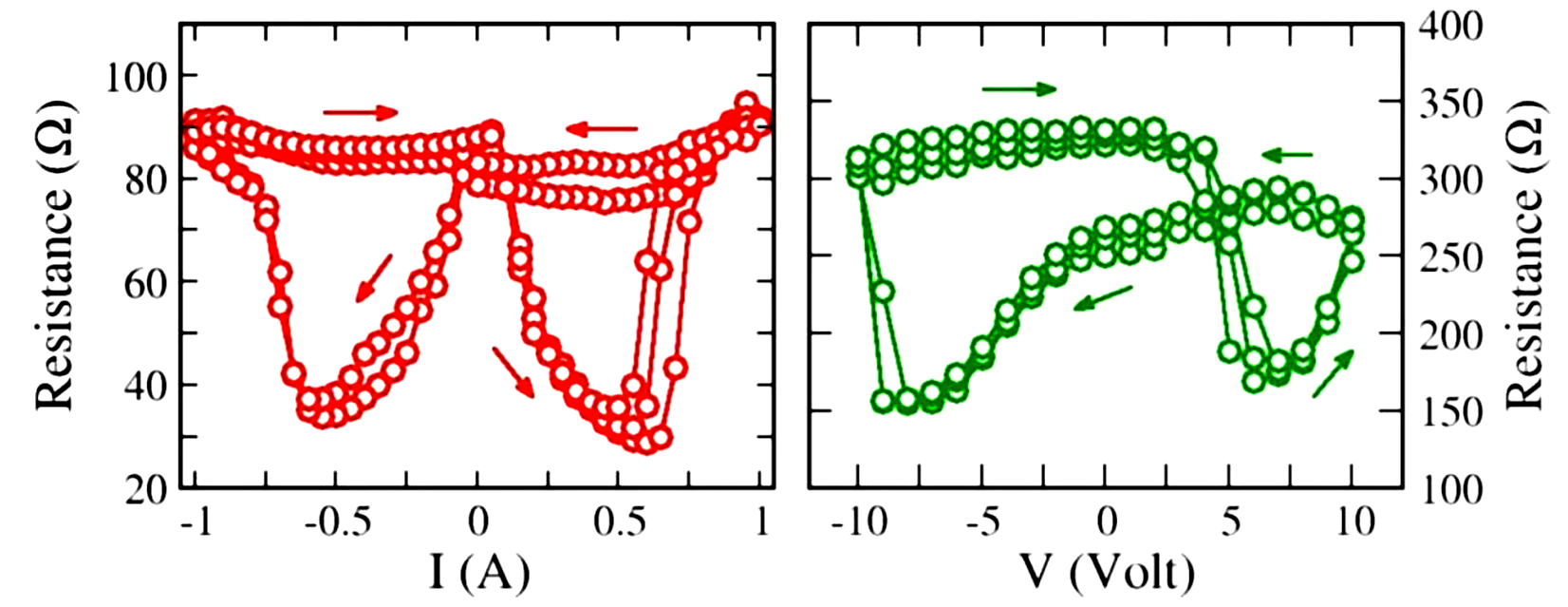}\caption{Experimental hysteresis loops measured in (left) manganite and (right)
cuprate devices (M. J. Rozenberg et al. , 2010). \label{fig:Experimental-hysteresis-loops} }
\end{figure}

Finally, besides a variety of theoretical interests, there are many
practical applications of correlated materials with impurities, which
includes the resistive switching devices. The resistance of certain
oxide films was surprisingly found to switch between low and high
values upon the application of external voltage pulses and displays
an interesting hysteresis loop as shown in Fig.(\ref{fig:Experimental-hysteresis-loops})
. It turns out that the impurity and correlation also play an important
role here. Typically, these materials are required to have a rather
high resistance in order to present a resistive switching phenomenon.
Mysteriously, the resistance is so high that it even exceeds the ``Mott-Ioffe-Regel
limit'' in many of these systems fabricated by the transition metal
oxides. Such metals is often considered as the ``bad metal'', which
have the resistance scales like $T$ over a wide temperature range
and thus indicates the importance of correlations in the underlying
electronic state.

The information age we live in is made possible by a physical underlayer of electronic hardware, which originates in condensed matter physics research. Despite the mighty   progress made in recent decades, the demand for faster and power efficient devices continues to grow. Progress of silicon based technology is nearing its physical limit, as minimum feature size of components is reaching a mere 10 nm. Thus, there is urgent need to identify novel materials and physical mechanisms for future electronic device applications.  The resistive switching behavior of transition metal oxides and the associated memristor device is emerging as a competitive technology for next generation electronics.
In this context, transition metal oxides (TMOs) are capturing a great deal of attention for non-volatile memory applications \cite{roadmap}. In particular, TMO are associated to the phenomenon of resistive  switching (RS) \cite{scholarpedia} and the memristor device \cite{HP} that is emerging  as a competitive technology for next generation electronics  \cite{roadmap,yang2008memristive,waser-aono,waser2009AdvMat,Sawa2008,inoue-sawa,MRS,baikalov2003field}. The RS effect is a large, rapid, non-volatile, and reversible change of the resistance, which may be used to encode logic information. In the simplest case one may associate  high and low resistance values to binary states, but multi-bit memory cells are also possible  \cite{IEEE,HP1}.
Typical systems where RS is observed are two-terminal capacitor-like devices,  where the dielectric might be a TMO and the electrodes are ordinary metals.  The phenomenon occurs in a strikingly large variety of systems. Ranging from simple binary compounds, such as NiO, TiO$_2$, ZnO, Ta$_2$O$_5$, HfO$_2$ and CuO, to more complex perovskite structures, such as superconducting cuprates and colossal magnetoresistive manganites   \cite{scholarpedia,waser-aono,Sawa2008,waser2009AdvMat,ghenzi2010hysteresis,tesler2017shock,Qiongetal}.
From a conceptual point of view, the main challenges for a non-volatile memory are:  (i) to change its resistance within nano seconds (required for modern electronics applications), (ii) to be able to retain the state for years (i.e. non-volatile), and  (iii) to reliably commute the state hundreds of thousands of times.
Through extensive experimental work in the past decade, a consensus has emerged around the notion that the change in  resistance is due to migration of ionic species, including oxygen vacancies, across different regions of the device,   affecting the local transport properties of the oxide  \cite{ nian2007evidence}.  In particular, the important role of highly resistive interfaces,  such as Schottky barriers, has also  been pointed out \cite{Sawa2008, inoue-sawa,chen}.
In contrast with the experimental efforts,  theoretical studies remain relatively scarce \cite{Larentis,Menzel,Bocquet,Hur,Huang,Noman,Strukov,Noh,Choi}. A few phenomenological models were proposed and numerically investigated, which captured different aspects  of the observed effects  \cite{prl2004,HP,Ielmini,Rozenberg2010}.  Significant progress has already been made in the past decade and devices are beginning to hit the market; however, it has been mainly the result of empirical trial and error. Hence, gaining theoretical insight is of essence.

\chapter{Mottness Induced Healing in Strongly Correlated Superconductors}

Despite the progress, it would be desirable to understand to what
extent this disorder screening is due only to the presence of strong
correlations or whether it is dependent on the details of the particular
model or system.\textit{\emph{ For example, are the effects of the
inter-site super-exchange, crucial to describe the cuprates, essential
for this phenomenon? To address these issues, it would be fruitful
to have an analytical treatment of the problem. We will describe in
this Letter how an expansion in the disorder potential is able to
provide important insights into these questions. In particular, we
show that the ``healing'' of the impurities is a sheer consequence
of the strong correlations and depend very little on the symmetry
of the superconducting (SC) state or the inclusion of inter-site magnetic
correlations.}}

We considered dilute nonmagnetic impurities in an otherwise homogenous,
strongly correlated electronic state. We avoided complications related
to the nucleation of possible different competing orders by the added
impurities, such as fluctuating or static charge- and spin-density-waves
\cite{fradkin2012high,Ghiringhelli17082012,PhysRevLett.96.017007,ubbens1992flux}
or the formation of local moments \cite{alloul2009defects}. Therefore,
we focused only on how a given strongly correlated state readjusts
itself in the presence of the impurities. We used a spatially inhomogeneous
slave boson treatment \textit{\emph{\cite{coleman1984new,kotliar1986new,lee2006doping,kotliar1988superexchange,lee19982},
which allowed us to perform a complete quantitative calculation. We
have allowed for either or both of $d$-wave SC and $s$-wave resonating
valence bond (RVB) orders.}}

Our analytical and numerical results demonstrate that (i) for sufficiently
weak correlations we recover the results of the conventional theory
\cite{balatsky2006impurity}, in which the variations of the different
fields induced by the impurities show oscillations with a long-ranged
power-law envelope; (ii) for strong interactions and in several different
broken symmetry states, the amplitude of the oscillations is strongly
suppressed by a common pre-factor $x$, the deviation from half-filling;
(iii) the spatial disturbances of the SC gap are healed over a precisely
defined length scale, which does not exceed a few lattice parameters
around the impurities; and (iv) this ``healing effect'' is intrinsically
tied to the proximity to the Mott insulating state, even though it
survives up to around 30\% doping.

\begin{figure*}[t]
\begin{centering}
\includegraphics[scale=0.7]{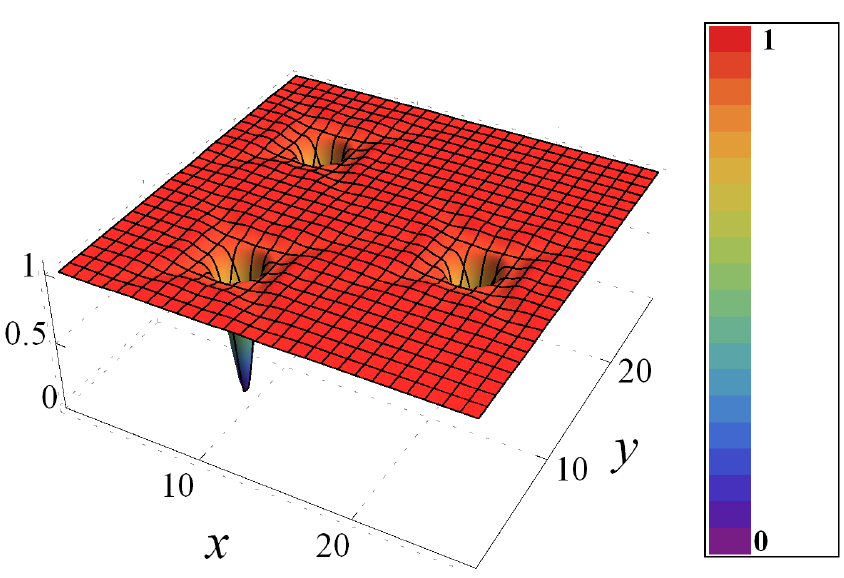}\includegraphics[scale=0.9]{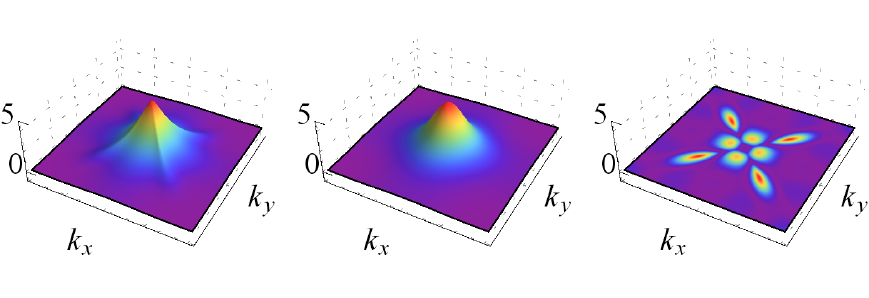}
\par\end{centering}

\begin{centering}
\includegraphics[scale=0.7]{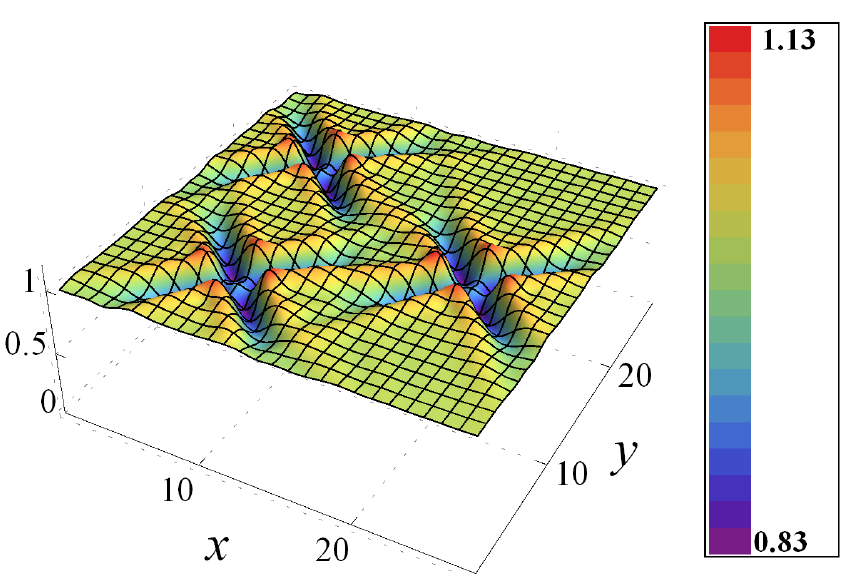}\includegraphics[scale=0.9]{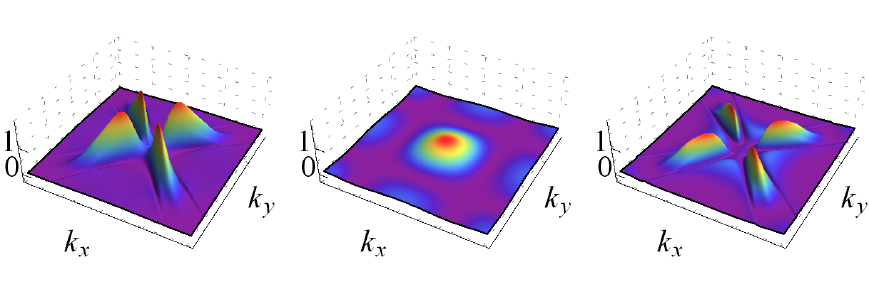}
\par\end{centering}

\caption{\label{fig:gapfluctuations}Spatial variations of normalized local
SC gap $\frac{\Delta_{i}}{\Delta_{0}}$ for three impurities (first
column) and the corresponding power spectra $S(\mathbf{k})$, $S(\mathbf{k})_{loc}$
and $S(\mathbf{k})_{nonloc}$ (second to fourth columns), in the presence
(top) and in the absence (bottom) of correlations for $x=0.2$. The
strong suppression of gap oscillations by correlations can be traced
to the dominance of the local, spherically symmetric power spectrum
{[}$S_{loc}\left(\mathbf{k}\right)${]} over the non-local anisotropic
part {[}$S_{nonloc}\left(\mathbf{k}\right)${]}.}
\end{figure*}

\section{Model and method}

We study the $t-t^{\prime}-J$ model on a cubic lattice in $d$ dimensions
with dilute nonmagnetic impurities

\begin{equation}
H=-\sum_{ij\sigma}t_{ij}c_{i\sigma}^{\dagger}c_{j\sigma}+J\sum_{ij}\mathbf{S}_{i}\cdot\mathbf{S}_{j}+\sum_{i}(\epsilon_{i}-\mu_{0})n_{i},
\end{equation}
where $t_{ij}$ are the hopping matrix elements between nearest-neighbor
($t$) and second-nearest-neighbor ($t^{\prime}$) sites, $c_{i\sigma}^{\dagger}\left(c_{i\sigma}\right)$
is the creation (annihilation) operator of an electron with spin projection
$\sigma$ at site $i$, $J$ is the super-exchange coupling constant
between nearest-neighbor sites, $n_{i}=\sum_{\sigma}c_{i\sigma}^{\dagger}c_{i\sigma}$
is the number operator, $\mu_{0}$ is the chemical potential and $\epsilon_{i}$
is the impurity potential. The no double occupancy constraint ($n_{i}\le1$)
is implied. We set the nearest-neighbor hopping $t$ as the energy
unit and choose $t^{\prime}=-0.25t$. To treat this model, we employ
the $U(1)$ slave boson theory \cite{coleman1984new,kotliar1988superexchange,lee2006doping,PhysRevB.36.857}.
Details can be found in \cite{lee2006doping} and we only describe
it very briefly here. It starts with the replacement $c_{i\sigma}^{\dagger}=f_{i\sigma}^{\dagger}b_{i}$,
where $f_{i\sigma}^{\dagger}$ and $b_{i}$ are auxiliary fermionic
(spinon) and bosonic fields, and the representation is faithful in
the subspace $n_{i}\le1$ if the constraint $\sum_{\sigma}f_{i\sigma}^{\dagger}f_{i\sigma}+b_{i}^{\dagger}b_{i}=1$
is enforced. This is implemented by a Lagrange multiplier $\lambda_{i}$
on each site. The $J$ term is then decoupled by Hubbard-Stratonovich
fields in the particle-particle ($\Delta_{ij}$) and particle-hole
($\chi_{ij}$) channels. The auxiliary bosonic fields are all treated
in the saddle-point approximation: $\left\langle b_{i}\right\rangle =r_{i}=\sqrt{Z_{i}}$
gives the quasiparticle residue, $\left\langle \lambda_{i}\right\rangle $
renormalizes the site energies and $\chi_{ij}=\sum_{\sigma}\left\langle f_{i\sigma}^{\dagger}f_{j\sigma}\right\rangle $
and $\Delta_{ij}=\left\langle f_{i\uparrow}f_{j\downarrow}-f_{i\downarrow}f_{j\uparrow}\right\rangle $
describe, respectively, the strength of a spinon singlet and the pairing
amplitude across the corresponding bonds. Note that we do not assume
these values are spatially uniform. This treatment is equivalent to
the Gutzwiller approximation \cite{anderson2004physics,Garg2008}.
In terms of Gorkov's spinor notation \cite{abrikosov1975methods}
with $\Psi_{i}(i\omega_{n})=\left[\begin{array}{cc}
f_{i\uparrow}^{\dagger}(i\omega_{n}) & f_{i\downarrow}(-i\omega_{n})\end{array}\right]^{\dagger}$, where $\omega_{n}$ is the fermionic Matsubara frequency, the spinon
Green's function is a $2\times2$ matrix: $\left[G_{ij}(i\omega_{n})\right]_{ab}=-\left\langle \Psi_{i}(i\omega_{n})\Psi_{j}^{\dagger}(i\omega_{n})\right\rangle _{ab}$.
Defining $h_{ij}\equiv-t_{ij}$, the saddle-point equations read as
follows

\begin{eqnarray}
\chi_{ij} & = & 2T\sum_{n}(G_{ij})_{11},\label{eq:chieq}\\
\Delta_{ij} & = & -2T\sum_{n}(G_{ij})_{12},\label{eq:order parameter}
\end{eqnarray}

\begin{eqnarray}
(r_{i}^{2}-1) & = & -2T\sum_{n}(G_{ii})_{11},\label{eq:constraint}\\
\lambda_{i}r_{i} & = & -2T\sum_{nl}h_{il}r_{l}(G_{il})_{11}=-\sum_{l}h_{il}r_{l}\chi_{il}.\label{eq:sum rule}
\end{eqnarray}

Note that we used Eq.~(\ref{eq:chieq}) in the second equality of
Eq.~(\ref{eq:sum rule}). At $T=0$ and in the clean limit $\epsilon_{i}=0$,
we have $Z=Z_{0}=x$. The Mott metal-insulator transition is signaled
by the vanishing of the quasi-particle weight $Z_{0}\rightarrow0$
at half-filling. It will be interesting to compare the results of
the above procedure with the ones obtained from solving only Eqs.~(\ref{eq:chieq}-\ref{eq:order parameter})
while setting $Z_{i}=1$ and $\lambda_{i}=0$. The two sets will be
called correlated and non-correlated, respectively. In order to be
able to compare them, we set $J=t/3$ in the correlated case and adjusted
$J$ in the non-correlated case in such a way that the two clean dimensionful
SC gaps coincide, as discussed in reference \cite{Garg2008}.

\section{Healing}

Although the detailed solutions of Eqs.~(\ref{eq:chieq}-\ref{eq:sum rule})
can be straightforwardly obtained numerically, we will focus on the
case of weak scattering by dilute impurities and expand those equations
up to first order in $\varepsilon_{i}$ around the homogeneous case.
It has been shown and we confirm that disorder induces long-ranged
oscillations in various physical quantities, specially near the nodal
directions in the $d$-wave SC state \cite{balatsky2006impurity}.
The linear approximation we employ is quite accurate for these extended
disturbances far from the impurities, since these are always small.
Besides, it provides more analytical insight into the results.

In general, we can expand the spatial variations of the various order
parameters in different symmetry channels through cubic harmonics:
$\delta\varphi_{ij}=\sum_{g}\delta\varphi_{i}\Gamma(g)_{ij}$ where
$\varphi_{ij}=\chi_{ij}$ or $\Delta_{ij}$ and $\Gamma(g)_{ij}$
are the basis functions for cubic harmonic $g$ of the square lattice
\footnote{$s$, $d_{x^{2}-y^{2}}$, $d_{xy}$, etc., with basis functions expressed
as: $\cos k_{x}+\cos k_{y}$, $\cos k_{x}-\cos k_{y}$ and $\sin k_{x}\sin k_{y}$,
etc.}. In the current discussion, we choose $\delta\chi_{ij}=\delta\chi_{i}\Gamma(s)_{ij}$
and $\delta\Delta_{ij}=\delta\Delta_{i}\Gamma(d_{x^{2}-y^{2}})_{ij}$,
as we are interested in oscillations with the same symmetry as the
ground state \cite{coleman1984new,kotliar1988superexchange,lee2006doping,PhysRevB.36.857}.
We also assume there is no phase difference between order parameters
on different bonds linked to same site. Then, we can define ``local''
spatial variations of the order parameters as $\delta\chi_{i}\equiv\frac{1}{2d}\sum_{j}\delta\chi_{ij}\Gamma(s)_{ij}$
and $\delta\Delta_{i}\equiv\frac{1}{2d}\sum_{j}\delta\Delta_{ij}\Gamma(d_{x^{2}-y^{2}})_{ij}$.
Details of the calculation can be found in Appendix \ref{sec:linearappox}.

We find that both $\delta\chi_{ij}$ and $\delta\Delta_{ij}$, as
well as the impurity-induced charge disturbance $\delta n_{i}$, are
proportional to $Z_{0}=x$, indicating the importance of strong correlations
for the healing effect. Indeed, we can trace back this behavior to
the readjustment of the $r_{i}$ and $\lambda_{i}$ fields, as encoded
in Eqs.~(\ref{eq:constraint}-\ref{eq:sum rule}). Besides, this
${\cal O}\left(x\right)$ suppression is a generic consequence of
the structure of the mean-field equations and holds for different
broken symmetry states, such as the flux phase state, $s$-wave superconductivity,
etc.

Let us focus in more detail on the spatial variations of the local
pairing field $\delta\Delta_{i}$. In the first column of Fig.~\ref{fig:gapfluctuations}
we show results for $\delta\Delta_{i}$ for three identical impurities.
The ``cross-like'' tails near the nodal directions \cite{PhysRevLett.76.2386}
are conspicuous in the absence of correlations (bottom) but are strongly
suppressed in their presence (top). While this suppression is further
enhanced as the Mott metal-insulator transition is approached ($x\to0$),
it is still quite significant even at optimal doping ($x=0.2$). This
is the ``healing'' effect previously reported \textit{\emph{\cite{Fukushima20083046,PhysRevB.79.184510,Garg2008}.}}
In order to gain insight into its underlying mechanism, we look at
the spatial correlation function of local gap fluctuations 
\begin{equation}
\left\langle \frac{\delta\Delta_{i}}{\Delta_{0}}\frac{\delta\Delta_{j}}{\Delta_{0}}\right\rangle _{disorder}=f\left(\mathbf{r}_{i}-\mathbf{r}_{j}\right),\label{eq:gapcorrfunc}
\end{equation}
where the brackets denote an average over disorder, after which lattice
translation invariance is recovered. The Fourier transform of $f\left(\mathbf{r}\right)$
can be written in the linear approximation as\vspace{-18pt}

\begin{equation}
f\left(\mathbf{k}\right)=\alpha W^{2}S\left(\mathbf{k}\right),\label{eq:gapcorrfuncink}
\end{equation}
where $W$ is the disorder strength, $\alpha$ depends on the detailed
bare disorder distribution, and the ``power spectrum'' (PS) $S\left(\mathbf{k}\right)$
is related to gap linear response function $M_{\Delta}\left(\mathbf{k}\right)$
by $S\left(\mathbf{k}\right)=M_{\Delta}^{2}\left(\mathbf{k}\right)$.
The latter is defined by Fourier transforming the kernel in $\delta\Delta_{i}=\Delta_{0}\sum_{j}M_{\Delta}\left(\mathbf{r}_{i}-\mathbf{r}_{j}\right)\varepsilon_{j}$,
which in turn can be easily obtained from the solution of the linearized
equations in Appendix \ref{sec:linearappox}. Inspired by the strongly
localized gap fluctuations at the top left of Fig.~\ref{fig:gapfluctuations},
we define the local component of the PS $S_{loc}\left(\mathbf{k}\right)\equiv M_{\Delta,loc}^{2}\left(\mathbf{k}\right)$,
where $M_{\Delta,loc}\left(\mathbf{k}\right)$ is obtained by \emph{restricting
the lattice sums up to the second nearest neighbor distance} ($\sqrt{2}a$)
in the linearized equations in Appendix \ref{sec:linearappox}. We
also define $S_{nonloc}\left(\mathbf{k}\right)=M_{\Delta,nonloc}^{2}\left(\mathbf{k}\right)\equiv\left[M_{\Delta}\left(\mathbf{k}\right)-M_{\Delta,loc}\left(\mathbf{k}\right)\right]^{2}$.
In the last three columns of Fig.~\ref{fig:gapfluctuations}, we
show, in this order, $S\left(\mathbf{k}\right)$, $S_{loc}\left(\mathbf{k}\right)$,
and $S_{nonloc}\left(\mathbf{k}\right)$ for the correlated (top)
and non-correlated (bottom) cases at $x=0.2$. Clearly, in the presence
of correlations the local PS is characterized by a smooth, spherically
symmetric bell-shaped function, whereas the non-local part is highly
anisotropic. Besides and more importantly, the non-local PS is negligibly
small in the correlated case. The full PS is thus \emph{overwhelmingly
dominated} by the local part, unlike in the non-correlated case. Below,
we extend the analysis to the underdoped and overdoped regimes, where
very similar behavior is found, even up to dopings of $x=0.3$. 

\begin{figure}
\begin{centering}
\includegraphics[scale=0.7]{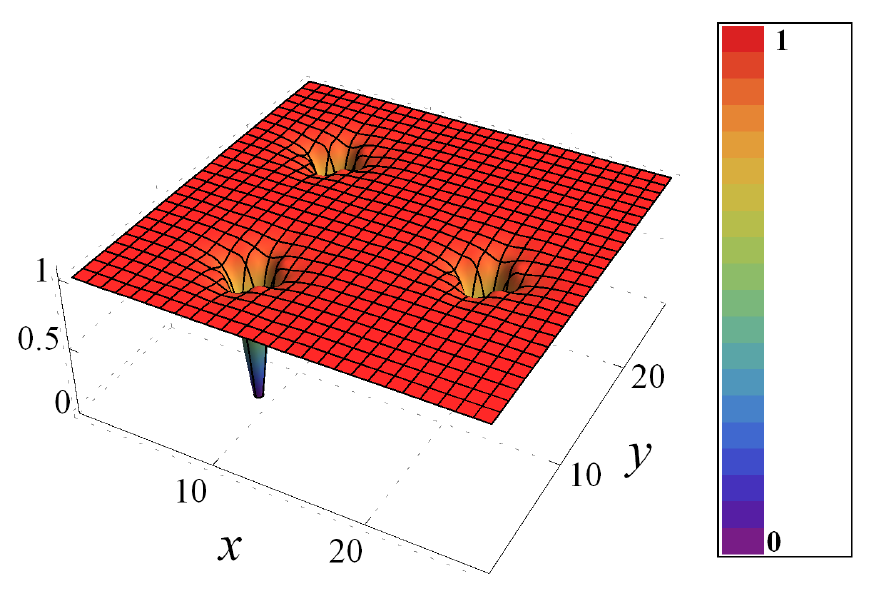}\includegraphics[scale=0.9]{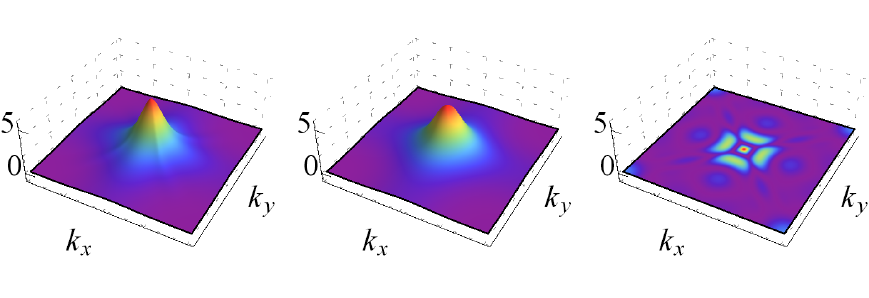}
\par\end{centering}

\begin{centering}
\includegraphics[scale=0.7]{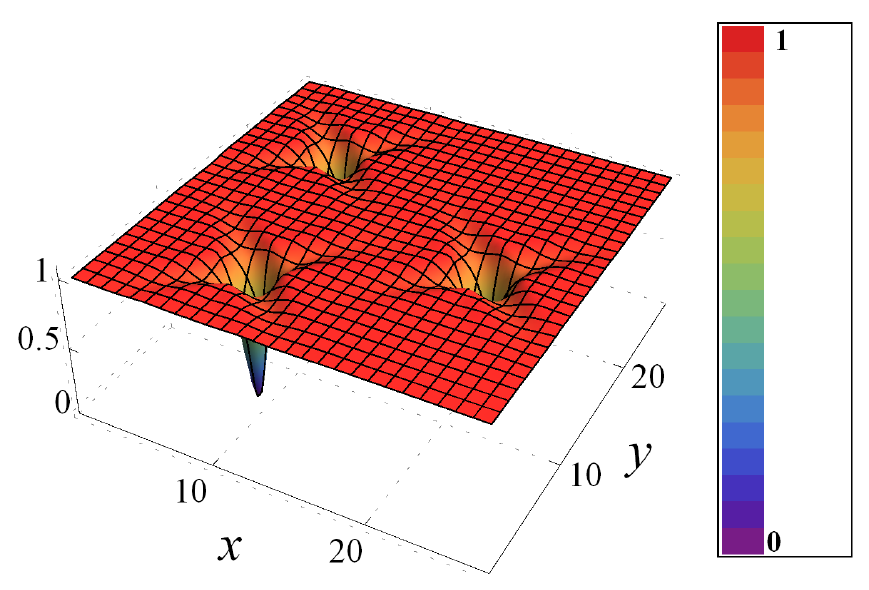}\includegraphics[scale=0.9]{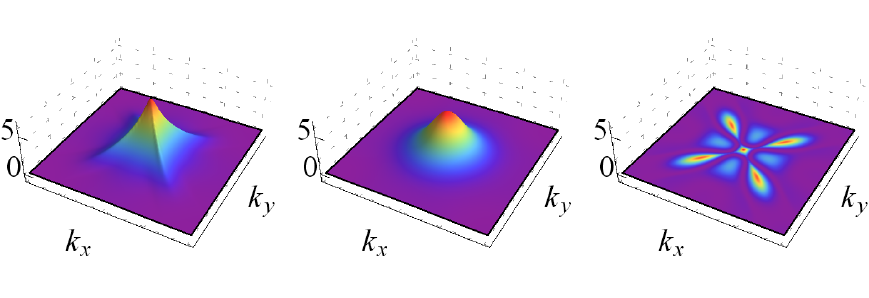}
\par\end{centering}

\begin{centering}
\includegraphics[scale=0.7]{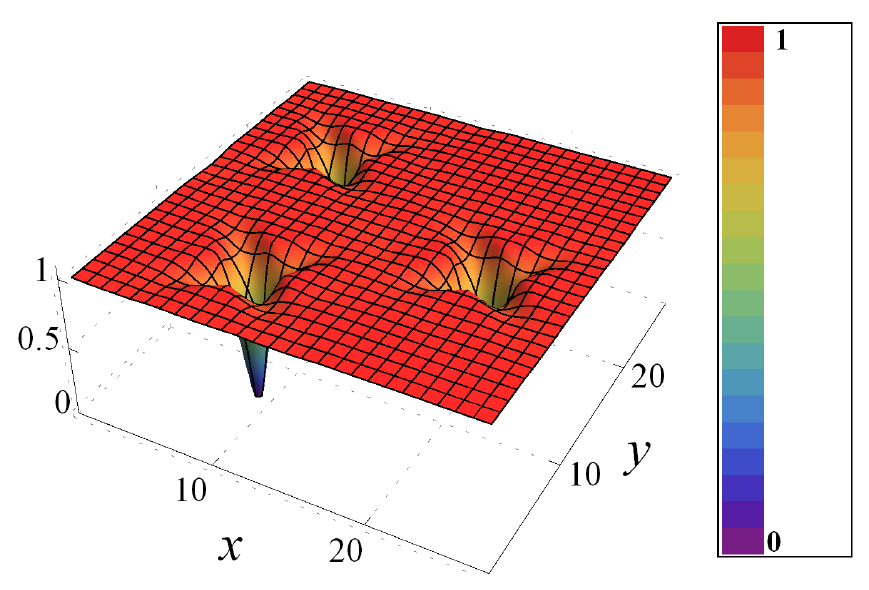}\includegraphics[scale=0.9]{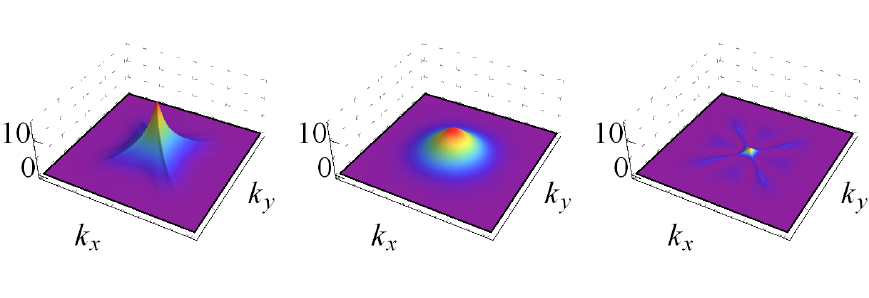}
\par\end{centering}

\caption{\label{fig:gapfluc=000026PSwithcorr}Spatial variations of normalized
local gap function $\frac{\delta\Delta_{i}}{\Delta_{0}}$ for three
impurities (first column) and the corresponding power spectra $S(\mathbf{k})$,
$S(\mathbf{k})_{loc}$ and $S(\mathbf{k})_{nonloc}$ (second to fourth
columns) for $x=0.15$ (first row), $x=0.25$ (second row), and $x=0.3$
(third row). The corresponding healing factors are (a) $h=0.23\%$,
(b) $h=1.77\%$, and (c) $h=2.74\%$.}
\end{figure}

The gap fluctuations $\delta\Delta_{i}$ for three impurities and
power spectra, for several dopings and in the presence of correlations,
are shown in Fig.~\ref{fig:gapfluc=000026PSwithcorr}. The strong
healing in the presence of correlations is conspicuous. It is important
to note that this suppression of gap fluctuations is not restricted
to small dopings and remains quite strong even at $x=0.3$, where
the healing factor does not exceed 3\%. As explained in the main text,
the healing effect originates in the dominance of the local spherically
symmetric part (third column in Fig.~\ref{fig:gapfluc=000026PSwithcorr})
over the anisotropic non-local response (fourth column in Fig.~\ref{fig:gapfluc=000026PSwithcorr}).

\begin{figure}
\begin{centering}
\includegraphics[width=0.8\textwidth]{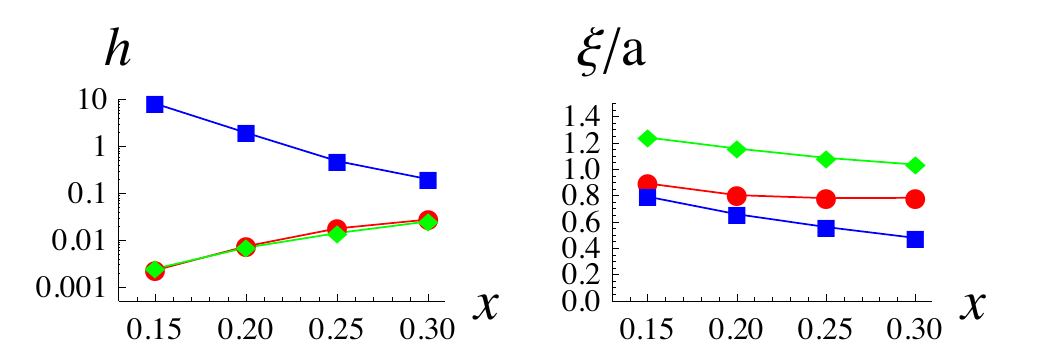}
\par\end{centering}

\caption{\label{fig:healcsiandh}Left panel: the healing factor $h$ as a function
of doping in the uncorrelated case (blue curve with squares), in the
correlated case (red curve with circles), and in the correlated case
without $\delta\chi_{i}$ fluctuations (green curve with diamonds).
Right panel: doping dependence of the SC ($\xi_{S}$, red curve with
circles) and normal state ($\xi_{N}$, blue curve with squares) healing
lengths. The green curve with diamonds gives $\xi_{S}$ calculated
within the minimal model (see text).}
\end{figure}

In order to quantify the localized nature of the healing effect, we
are led to a natural definition of a ``healing factor'' $h$ in
the $d$-wave SC state\vspace{-12pt}

\begin{equation}
h=\frac{\int S_{nonloc}\left(\mathbf{k}\right)d^{2}k}{\int S_{loc}\left(\mathbf{k}\right)d^{2}k},
\end{equation}
where the integration is over the first Brillouin zone. It measures
the relative weight of non-local and local parts of the gap PS. The
healing factor as a function of doping is shown on the left panel
of Fig.~\ref{fig:healcsiandh} for the non-correlated (blue) and
correlated (red) cases. The contrast is striking. When correlations
are present, $h$ is extremely small up to 30\% doping and the gap
disturbance is restricted to a small area around the impurities. In
contrast, without correlations significant pair fluctuations occur
over quite a large area for all dopings shown. We conclude that the
strong dominance of the local part over the highly anisotropic non-local
contribution caused by correlations is the \emph{key feature behind
the healing process}.

The shape of $S_{loc}\left(\mathbf{k}\right)$ shows that the gap
disturbance created by an impurity is healed over a well-defined distance,
the ``healing length'' $\xi_{S}$. This length scale can be obtained
by expanding the inverse of $M_{\Delta,loc}\left(\mathbf{k}\right)$
{[}or equivalently $M_{\Delta}\left(\mathbf{k}\right)${]} up to second
order in $k^{2}$, thus defining a Lorentzian in $\mathbf{k}$-space
\begin{equation}
M_{\Delta,loc}\left(\mathbf{k}\right)\approx\frac{1}{A+Bk^{2}}.\label{eq:lorentzian}
\end{equation}
The SC healing length is then given by $\xi_{S}=\sqrt{B/A}$. The
$x$ dependence of $\xi_{S}$ is shown in red on the right panel of
Fig.~\ref{fig:healcsiandh}. It is of the order of one lattice spacing
in the relevant range $0.15<x<0.3$. It should be noted that precisely
the same length scale also governs the healing of charge fluctuations
in the SC state, showing that this phenomenon is generic to the strongly
correlated state. A similar procedure can be carried out for the charge
fluctuations in the normal state, thus defining a normal state healing
length $\xi_{N}$. The blue curve of the right panel of Fig.~\ref{fig:healcsiandh}
shows the $x$ dependence of $\xi_{N}$, which is also of the order
of one lattice spacing.

\section{Mottness-induced healing}

The healing effect we have described comes almost exclusively from
the $\delta r_{i}$ and $\delta\lambda_{i}$ fluctuations: $h$ is
hardly affected by the $\delta\chi_{i}$ field. If we suppress the
$\delta\chi_{i}$ fluctuations completely, there is only a tiny change
in the results, as shown by the green curve of the left panel of Fig~\ref{fig:healcsiandh}.
The same is not true, however, if we turn off either $\delta r_{i}$
or $\delta\lambda_{i}$ or both. We conclude that the healing effect
in the $d$-wave SC state originates from the strong correlation effects
alone, rather than the spinon correlations.

\begin{figure}
\begin{centering}
\includegraphics[width=0.8\textwidth]{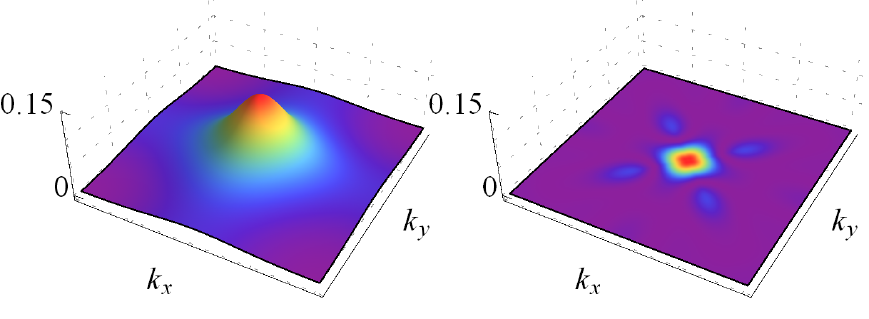}
\par\end{centering}

\caption{\label{fig:densityfluct}Local (left) and nonlocal (right) parts of
the charge-fluctuation power spectra $N(\mathbf{k})_{loc}$ and $N(\mathbf{k})_{nonloc}$
in the presence of strong correlations for $x=0.2$.}
\end{figure}

Within the linear approximation we are employing, all fluctuation
fields ($\delta\Delta$, $\delta r$, etc.) are proportional, in $\mathbf{k}$-space,
to the disorder potential $\varepsilon\left(\mathbf{k}\right)$. Therefore,
they are also proportional to each other. In particular, given the
centrality of the strong correlation fields, it is instructive to
write the gap fluctuations in terms of the slave boson fluctuations
\begin{equation}
\delta\Delta\left(\mathbf{k}\right)=-2\chi_{pc}\left(\mathbf{k}\right)r\delta r\left(\mathbf{k}\right)=\chi_{pc}\left(\mathbf{k}\right)\delta n\left(\mathbf{k}\right).\label{eq:MMgapfluc}
\end{equation}
In the last equality, we used $n_{i}=1-r_{i}^{2}$, which enables
us to relate two physically transparent quantities: the gap and the
charge fluctuations. Indeed, this will provide crucial physical insight
into the healing process. By focusing on the linear charge response
to the disorder potential $\delta n\left(\mathbf{k}\right)=n_{0}M_{n}\left(\mathbf{k}\right)\varepsilon\left(\mathbf{k}\right)$,
we can, in complete analogy with the gap fluctuations, define a PS
for the spatial charge fluctuations, $N\left(\mathbf{k}\right)=M_{n}^{2}\left(\mathbf{k}\right)$.
This PS can also be broken up into local {[}$N_{loc}\left(\mathbf{k}\right)=M_{n,loc}^{2}\left(\mathbf{k}\right)${]}
and non-local \{$N_{nonloc}\left(\mathbf{k}\right)=\left[M_{n}\left(\mathbf{k}\right)-M_{n,loc}\left(\mathbf{k}\right)\right]^{2}$\}
parts, as was done for the gap-fluctuation PS. These two contributions,
obtained from the solution of the full linearized equations, are shown
in Fig.~\ref{fig:densityfluct}. The charge PS in the correlated
$d$-wave SC state is also characterized by a smooth, almost spherically
symmetric local part and a negligibly small anisotropic non-local
contribution. Note also the strong similarity between the local PS
for gap (top row of Fig.~\ref{fig:gapfluctuations}) and charge fluctuations.
This shows a strong connection between the gap and charge responses.
Evidently, this is also reflected in real space, where the charge
disturbance is healed in the same strongly localized fashion as the
gap disturbance. In fact, the local part of the charge response function
$M_{n,loc}\left(\mathbf{k}\right)$ can be shown to be well approximated
by a Lorentzian and we can write for small $\mathbf{k}$
\begin{equation}
\delta\Delta_{loc}\left(\mathbf{k}\right)\approx-\chi_{pc}\left(\mathbf{k}=0\right)\frac{8r^{2}/\lambda}{k^{2}+\xi_{S}^{-2}}\varepsilon\left(\mathbf{k}\right),\label{eq:MMgapflucloc}
\end{equation}
where the SC healing length $\xi_{S}$ can be expressed in terms of
the Green's functions of the clean system in Appendix \ref{sec:linearappox}.
The relations implied by Eqs.~(\ref{eq:MMgapfluc}) and (\ref{eq:MMgapflucloc}),
as well as the doping dependence of the quantities in them, could
be tested in STM studies and would constitute an important test of
this theory. 

Eqs.~(\ref{eq:MMgapfluc}-\ref{eq:MMgapflucloc}) allow us to obtain
a clear physical picture of the healing mechanism. The spatial gap
fluctuations can be viewed as being ultimately determined by the charge
fluctuations. Furthermore, their ratio $\chi_{pc}\left(\mathbf{k}\right)$,
which is essentially a pair-charge correlation function, is a rather
smooth function \emph{of order unity},\emph{ only weakly renormalized
by interactions}. Therefore, it is the strong suppression of charge
fluctuations by ``Mottness'', as signaled by the $r^{2}$ factor in
Eq.~(\ref{eq:MMgapflucloc}), which is behind the healing of gap
fluctuations. This elucidates the physics of healing previously found
numerically \textit{\emph{\cite{Fukushima20083046,PhysRevB.79.184510,Garg2008}.
}}It also suggests that the healing phenomenon is generic to Mott
systems \cite{Andrade2010} and is not tied to the specifics of the
cuprates.

\section{A minimal model}

Interestingly, the crucial role played by the strong correlation fields
($r_{i}$ and $\lambda_{i}$) suggests a ``minimal model'' (MM) for
an accurate description of the healing process, which we define as
follows: (i) the spatially fluctuating strong correlation fields $r_{i}$
and $\lambda_{i}$ are first calculated for the self-consistently
determined, \emph{fixed, uniform} $\Delta$ and $\chi$, and then
(ii) the effects of their spatial readjustments are fed back into
the gap equation~(\ref{eq:order parameter}) in order to find $\delta\Delta_{i}$
as shown in Appendix \ref{sec:irrelevancespinon}. The accuracy of
this procedure can be ascertained by the behavior of the healing factor:
it is \emph{numerically indistinguishable} from the green curve of
the left panel of Fig.~\ref{fig:healcsiandh}. Furthermore, the value
of $\xi_{S}$ calculated within the MM differs from the one obtained
from the solution of the full linearized equations by at most 20\%
(red and green curves on the right panel of Fig.~\ref{fig:healcsiandh}).
Besides its accuracy, the advantage of this MM description lies in
the simplicity of the analytical expressions obtained. As shown in
Appendix \ref{sec:irrelevancespinon}, it provides simple expressions
for the important quantities $\chi_{pc}\left(\mathbf{k}\right)$ and
$\xi_{S}$.

\section{Conclusions}

In this work, we have found an inextricable link between the healing
of gap and charge disturbances in strongly correlated superconductors,
suggesting that this phenomenon is generic to any system close to
Mott localization. An important experimental test of this link would
be provided by STM studies of the organic superconductors \textit{\emph{\cite{analytisetal06}
and maybe the pnictides \cite{lietal12}. Whether it is also relevant
for heavy fermion systems \cite{FigginsMorr2011} is an open question
left for future study. }}

\chapter{Strong Correlations Generically Protect \MakeLowercase{$d$}-wave
Superconductivity against Disorder}

In this part, we provide a route to understand both qualitatively
and quantitatively whether disorder screening has any significant
influence on $T_{c}$ as well as on the normal state transport properties.
The transition temperature in the under-doped region of the hole-doped
cuprates is believed to be influenced by phase fluctuations, various
types of competing orders (such as charge- and spin-density waves),
stripe formation, etc. Consequently, impurities act as nucleations
centers, which complicates the analysis considerably. In the over-doped
region, however, $T_{c}$ is dominated by the superconducting gap
opening, thus offering a particularly favorable window into the interplay
between disorder and interactions.

In the presence of impurities, the strongly correlated state readjusts
itself and creates a renormalized disorder potential. In the dilute
limit, the AG theory can be extended to describe the effect of this
renormalized potential on $T_{c}$ degradation and transport properties.\textit{\emph{
We will describe in this part how electronic interactions lead to
a much slower decrease of $T_{c}$ as compared to the weak-coupling
theory. }}Our results demonstrate that (i) this effect is intrinsically
tied to the proximity to the Mott insulating state, although it is
significant even above optimal doping; (ii) the doping dependence
of normal state resistivity is different from that of the pair-breaking
scattering rate, which governs $T_{c}$; and (iii) the softening of
the disorder potential by interactions leads to a strong enhancement
of the forward scattering amplitude.

\section{Model and method}

We start with the $t-J$ model on a cubic lattice in $d$ dimensions
with dilute nonmagnetic impurities 

\begin{equation}
H=-t\sum_{\left\langle ij\right\rangle \sigma}c_{i\sigma}^{\dagger}c_{j\sigma}+J\sum_{\left\langle ij\right\rangle }\mathbf{S}_{i}\cdot\mathbf{S}_{j}+\sum_{i}(\epsilon_{i}-\mu_{0})n_{i},
\end{equation}
where $c_{i\sigma}^{\dagger}\left(c_{i\sigma}\right)$ is the creation
(annihilation) operator of an electron with spin projection $\sigma$
on site $i$, $t$ is the hopping matrix element between nearest neighbors,
$J$ is the super-exchange coupling constant between nearest-neighbor
sites, $n_{i}=\sum_{\sigma}c_{i\sigma}^{\dagger}c_{i\sigma}$ is the
number operator, $\mu_{0}$ is the chemical potential. The no-double-occupancy
constraint ($n_{i}\le1$) is implied. We work in units such that $\hbar=k_{B}=a=1$,
where $a$ is the lattice spacing and the total number of lattice
sites is $V$. For definiteness, we will set $J=t/3$. The impurities
are taken into account through a random on-site potential described
by $\epsilon_{i}$. We use a model of disorder in which we set the
potential $\epsilon_{i}=t$ and randomly place the impurities on lattice
sites with $n$ impurities per unit volume and no correlations between
their positions. Note that this model assumes random non-magnetic
scattering but does not describe the removal of magnetic ions. We
will focus on the two-dimensional case relevant to the cuprates, but
our results are easily generalizable to higher dimensions with few
modifications.

We proceed with $U\left(1\right)$ slave boson theory, details of
which can be found in \textit{\emph{\cite{coleman1984new,kotliar1986new,lee2006doping,kotliar1988superexchange,lee19982}}}.
Briefly, it starts with the replacement $c_{i\sigma}^{\dagger}\to f_{i\sigma}^{\dagger}b_{i}$,
where $f_{i\sigma}^{\dagger}$ and $b_{i}$ are auxiliary fermionic
(spinon) and bosonic (slave boson) fields. This substitution is faithful
if the constraint $n_{i}\le1$ is replaced by $\sum_{\sigma}f_{i\sigma}^{\dagger}f_{i\sigma}+b_{i}^{\dagger}b_{i}=1$.
The latter is enforced through Lagrange multiplier fields $\lambda_{i}$
on each site. The $J$ term is then decoupled through additional Hubbard-Stratonovitch
bosonic fields in the particle-particle ($\Delta_{ij}$) and particle-hole
($\chi_{ij}$) channels. The auxiliary bosonic fields are all treated
in the saddle-point approximation, which here is \emph{spatially inhomogeneous}
due to the presence of disorder: $\left\langle b_{i}\right\rangle =r_{i}$,
which governs the local quasiparticle residue $Z_{i}=r_{i}^{2}$,
$\left\langle \lambda_{i}\right\rangle $ (we will denote it simply
by $\lambda_{i}$) which renormalizes the site energies and $\chi_{ij}=\sum_{\sigma}\left\langle f_{i\sigma}^{\dagger}f_{j\sigma}\right\rangle $
and $\Delta_{ij}=\left\langle f_{i\uparrow}f_{j\downarrow}-f_{i\downarrow}f_{j\uparrow}\right\rangle $,
which describe, respectively, the strength of a spinon singlet and
the pairing amplitude across the corresponding bonds. We also made
the change $J\to\widetilde{J}=\frac{3}{8}J$. This choice is made
so that the saddle-point approximation of the above multi-channel
Hubbard-Stratonovitch transformation coincides with the mean-field
results \cite{lee2006doping}. We note, however, that the usual choice
$\widetilde{J}=\frac{1}{4}J$ would give rise to hardly noticeable
changes in the numerical results. We stress that the $f$-electrons
mentioned throughout the text are only auxiliary fermions, usually
called spinons, rather than the physical electrons. They are related
at the saddle point by $c_{i\sigma}^{\dagger}=r_{i}f_{i\sigma}^{\dagger}$.
Note that the non-trivial effects of this work come from the self-consistent
spatial readjustments of the condensed fields to the disorder potential. 

In the clean limit ($\epsilon_{i}=0$) and in the saddle-point approximation,
the bosonic fields are spatially uniform: $r_{i}=r_{0}$, $\lambda_{i}=\lambda_{0}$,
$\chi_{ij}=\chi\Gamma_{s}\left(i,j\right)$ and $\Delta_{ij}=\Delta_{0}\Gamma_{d}\left(i,j\right)$.
Here, $\Gamma_{s,d}\left(i,j\right)$ are the real space cubic harmonics
which, in $\mathbf{k}$-space, are given by $\Gamma_{s}\left(\mathbf{k}\right)=2\left(\cos k_{x}+\cos k_{y}\right)$
and $\Gamma_{d}\left(\mathbf{k}\right)=2\left(\cos k_{x}-\cos k_{y}\right)$.
As the doping level (measured with respect to half-filling) $x=1-\sum_{i}n_{i}/V=r_{0}^{2}$
is increased, the slave boson condensation temperature $T_{b}$ increases
monotonically from zero whereas the $\Delta$ field condenses at a
transition temperature $T_{\Delta}$ which decreases monotonically
from a finite value at $x=0$ to zero at an upper doping level $x_{max}$
\textit{\emph{\cite{kotliar1988superexchange,lee2006doping}}}. The
two curves meet at optimal doping $x_{opt}$. The dome below the two
curves is the superconducting dome. Our focus in this paper is on
the overdoped region $x>x_{opt}$, in which the superconducting transition
temperature $T_{c}=T_{\Delta}<T_{b}$.

Within this spatially inhomogeneous theory, we are able\textit{\emph{
to perform a complete quantitative calculation of the }}\textit{effective}\textit{\emph{
disorder potential. Details have been explained elsewhere \cite{shaoetal15}.}}
Here, we will focus on the effects of disorder on the superconducting
transition temperature $T_{c}$ and on transport properties in the
correlated normal state for $T\apprge T_{c}$ in the over-doped region.
For this purpose, we can set $\Delta_{ij}=0$. Moreover, in this range
of temperature and dopings the other bosons, $r_{i}$, $\lambda_{i}$
and $\chi_{ij}$, are thoroughly condensed and therefore fairly insensitive
to finite temperature effects. We are thus justified in approximating
them by their zero-temperature values.

We will focus on the case of weak scattering by dilute impurities
$n\ll1$, where a linear response theory is sufficient. In other words,
we calculate the spatial fluctuations of the various condensed fields
to first order in the perturbing potential $\epsilon_{i}$ \textit{\emph{\cite{shaoetal15}}}.
Extensive numerical calculations carried out both in the normal and
in the superconducting states have shown that the crucial spatial
fluctuations come from the $\lambda_{i}$ and $r_{i}$ fields whereas
fluctuations of $\chi_{ij}$ play only a negligible role \textit{\emph{\cite{shaoetal15}.
We will thus simply fix }}$\chi_{ij}$ at its clean limit value $\chi$
while allowing for the full self-consistent spatial adjustment of
the $\lambda_{i}$ and $r_{i}$ fields to the disordered situation.

Given this setup, the superconducting transition at $T_{c}$ corresponds
to the formation of the order parameter $\Delta_{ij}=\left\langle f_{i\uparrow}f_{j\downarrow}-f_{i\downarrow}f_{j\uparrow}\right\rangle $.
The condensing $f$-electrons, on the other hand, are governed, in
the clean limit by a dispersion relation renormalized by the slave
boson fields $r_{0}$ and $\chi$, $\widetilde{h}\left(\mathbf{k}\right)\equiv-\left(tx+\tilde{J}\chi\right)\Gamma_{s}\left(\mathbf{k}\right)$
and a renormalized chemical potential $\mu_{0}-\lambda_{0}\equiv-\nu_{0}$.
\emph{This theory, therefore, describes a BCS-type condensation of
the $f$-electrons}. In the presence of disorder, the various fields
will readjust themselves. The effect of dilute identical non-magnetic
impurities on $T_{c}$ will therefore be captured within the Abrikosov-Gor'kov
(AG) theory \cite{abrikosov1960contribution}. In that theory, the
only input needed is the scattering $T$-matrix due to a \emph{single}
impurity. For that purpose, we place a single impurity at the lattice
origin $\epsilon_{i}=t\delta_{i,0}$. Crucially, however, the $\lambda_{i}$
and $r_{i}$ fields will differ from their clean-limit value \emph{inside
an extended region around the impurity}, not only at the origin. The
effective $T$-matrix will thus reflect this non-trivial rearrangement.
As shown in reference \textit{\emph{\cite{shaoetal15}}}, the impurity
potential is ``healed'' within a length scale of a few lattice parameters,
the so-called healing length. Furthermore, it was shown that the healing
process/length is strongly influenced by electronic correlations and
`Mottness', even up to dopings $x\approx0.3$. Therefore, as will
be shown, the effective scattering will be strongly suppressed relative
to the non-correlated case.

We also look at the transport properties in the normal state around
$T_{c}$. Again, the AG analysis can be straightforwardly applied
in our case. The relevant input for the calculation of the resistivity
is the \emph{physical electron} scattering $T$-matrix for a single
impurity.

\section{Analytical results}

\subsection{The gap equation in AG theory}

\label{sub:GapEq}

Within the $U\left(1\right)$ slave boson theory for the $t-J$ model
of the cuprates, the superconducting transition \emph{in the over-doped
regime} is described by the usual BCS equation for a $d$-wave superconductor
in which it is the spinons (the auxiliary $f$-fermions) which pair
to form the condensate \cite{lee2006doping}. In other words, the
transition is signaled by a non-zero value of the $d$-wave pairing
order parameter $\Delta_{ij}=\left\langle f_{i\uparrow}f_{j\downarrow}-f_{i\downarrow}f_{j\uparrow}\right\rangle $.
We remind the reader that at the transition the other auxiliary fields
($r_{i}$, $\lambda_{i}$, $\chi_{ij}$) are well formed. The effects
of non-magnetic impurities on $T_{c}$ can then be treated by using
the AG theory \cite{abrikosov1960contribution}. 

Below, we discuss briefly the derivation of the linearized gap equation
following from the theory of superconducting alloys \cite{bennemann2008superconductivity}.
As the pairing field $\Delta_{ij}$ is assumed to be small near the
transition temperature $T_{c}$, we are justified to keep it up to
the linear order throughout the calculation. In superconducting state,
there are Green's function and two anomalous Green's functions involved.
Therefore, the diagrammatic approach is well suited in this case to
study the disorder effects. 

\begin{figure}
\begin{centering}
\subfloat[Diagrammatic representation of the Dyson equations for the green's
function and anomalous green's functions in superconducting state,
where impurity scattering events are indicated by the dashed line.]{\begin{centering}
\includegraphics[width=0.8\textwidth]{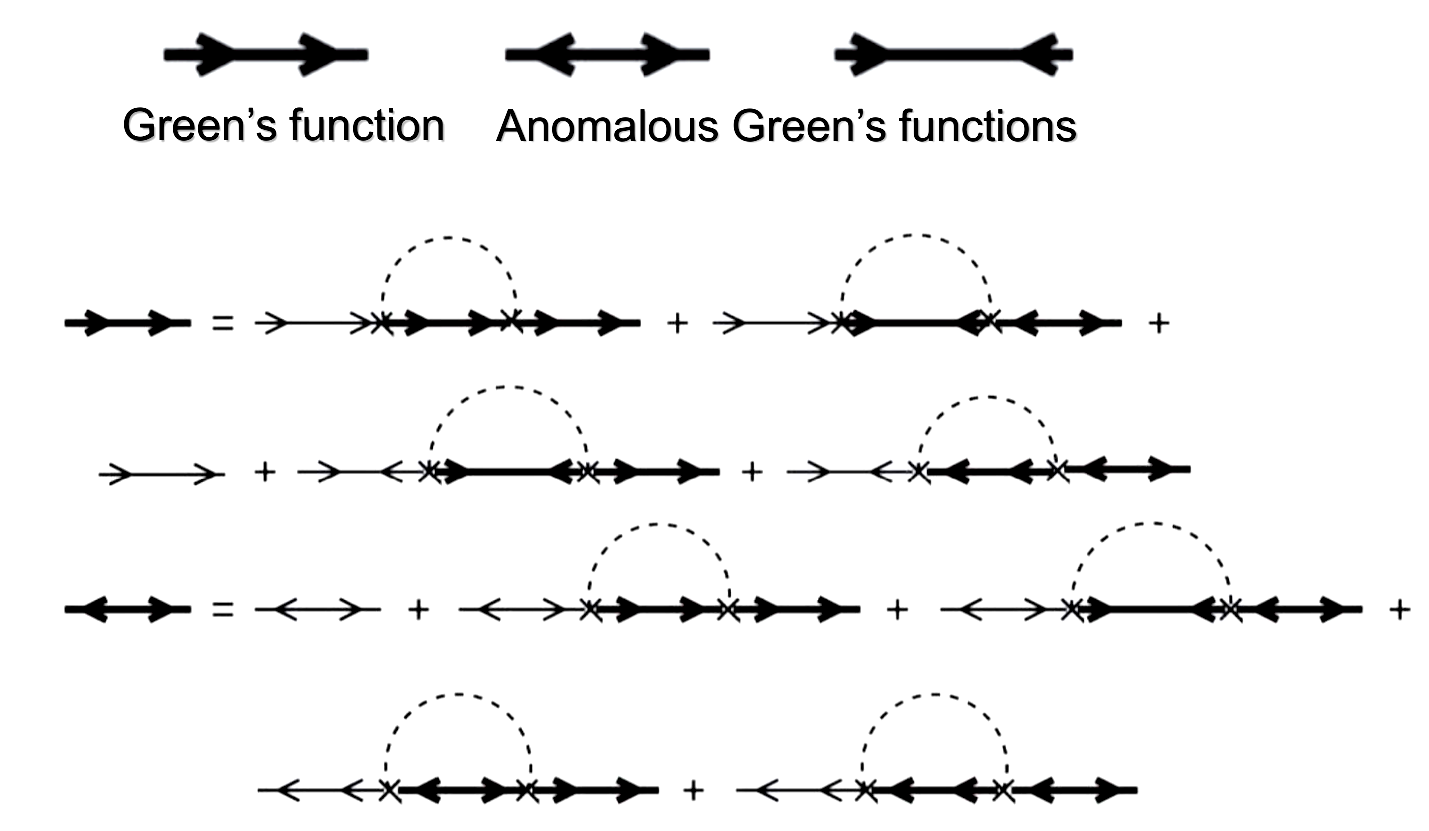}
\par\end{centering}

}
\par\end{centering}

\begin{centering}
\subfloat[Diagrammatic form of corrections to each of the two green's functions
(left) and ``vertex'' corrections (right)]{\begin{centering}
\includegraphics[width=0.8\textwidth]{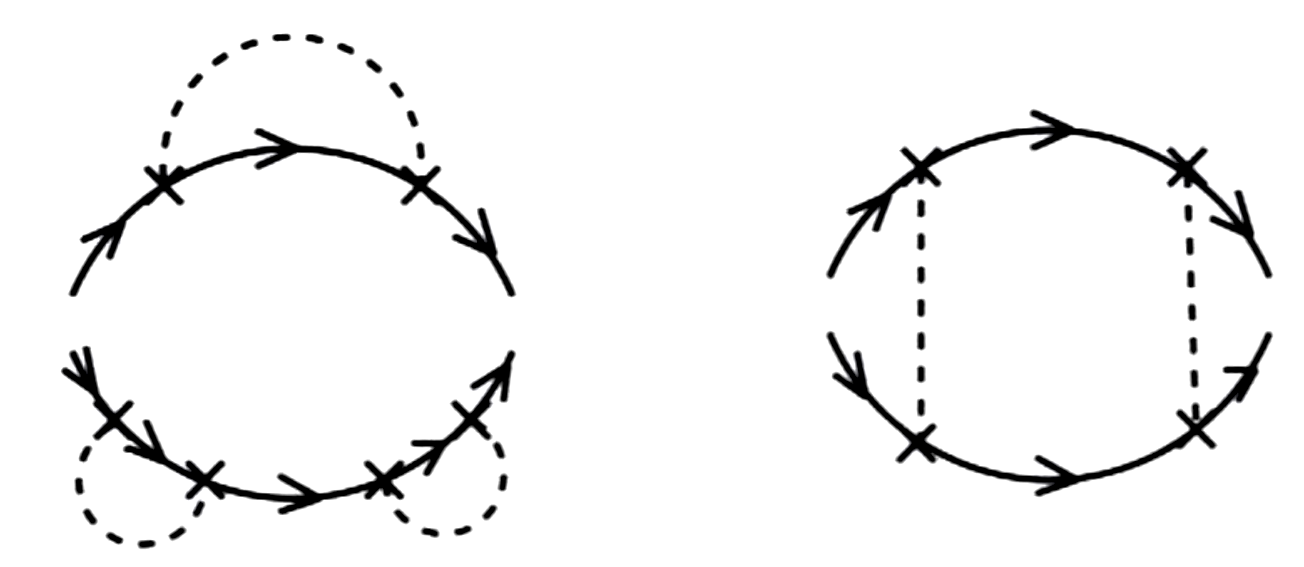}
\par\end{centering}

}
\par\end{centering}

\caption{Feynman diagrams in superconducting state with impurities. (L.P. Gor'kov
2008, Chapter 5 of Karl Heinz Bennemann et al. , 2008)\label{fig:Feynman-diagrams }}
\end{figure}
With the help of diagrammatic tricks, it is straightforward to keep
the leading order of the anomalous green's functions in the Dyson
equation according to (a) of Fig.(\ref{fig:Feynman-diagrams }). Furthermore,
the main physical processes are summarized in (b) of Fig.(\ref{fig:Feynman-diagrams }),
namely, (i) each of the green's functions for paring process are independently
renormalized by all possible impurity scattering diagrams which is
non-crossing, (ii) the impurity scattering events happens between
the two spinons which pair to form the condensate.

Therefore, within standard AG theory, the linearized gap equation
is written as \cite{bennemann2008superconductivity}

\begin{eqnarray}
\Delta_{0} & = & \frac{2\widetilde{J}kT_{c}}{d}\sum_{i\omega_{n}}\int\frac{d^{2}\mathbf{k}}{\left(2\pi\right){}^{2}}G^{f}\left(\mathbf{k},-i\omega_{n}\right)G^{f}\left(-\mathbf{k},i\omega_{n}\right)\Lambda\left(\mathbf{k},i\omega_{n}\right)\Gamma_{d}\left(\mathbf{k}\right),\label{eq:pairing}
\end{eqnarray}
where $\Delta_{0}$ is the superconducting gap amplitude, $\omega_{n}=\left(2n+1\right)\pi T_{c}$,
and $\Lambda\left(\mathbf{k},i\omega_{n}\right)$ is the vertex correction
function, which satisfies
\begin{equation}
\Lambda\left(\mathbf{k},i\omega_{n}\right)=\Delta_{0}\Gamma_{d}\left(\mathbf{k}\right)+n\int\frac{d^{2}\mathbf{k}^{\prime}}{\left(2\pi\right){}^{2}}\left|\left\langle \mathbf{k}\right|T^{f}\left|\mathbf{k}^{\prime}\right\rangle \right|^{2}G^{f}\left(\mathbf{k}^{\prime},-i\omega_{n}\right)G^{f}\left(-\mathbf{k}^{\prime},i\omega_{n}\right)\Lambda\left(\mathbf{\mathbf{k}^{\prime}},i\omega_{n}\right).\label{eq:vertex correction}
\end{equation}
In the last equation, $\left\langle \mathbf{k}\right|T^{f}\left|\mathbf{k}^{\prime}\right\rangle $
is the single-impurity scattering $T$-matrix for the spinons. The
disorder-averaged spinon Green's function in Eqs.~(\ref{eq:pairing})
and (\ref{eq:vertex correction}) is \cite{abrikosov1960contribution}
\begin{equation}
G^{f}\left(\mathbf{k},i\omega_{n}\right)=\frac{1}{i\omega_{n}\left(1+\frac{1}{2\tau_{\mathbf{k}}\left|\omega_{n}\right|}\right)-\widetilde{h}\left(\mathbf{k}\right)-\nu_{0}},\label{eq:spinongf}
\end{equation}
where $\widetilde{h}\left(\mathbf{k}\right)=-\left(tx+\widetilde{J}\chi\right)\Gamma_{s}\left(\mathbf{k}\right)$
is the renormalized dispersion, \textcolor{black}{$\nu_{0}=\lambda_{0}-\mu_{0}$}\textcolor{red}{{}
}\textcolor{black}{is the negative value of the renormalized chemical
potential, which controls the doping level, and} 
\begin{equation}
\frac{1}{\tau_{\mathbf{k}}}\equiv2\pi n\int\frac{d^{2}\mathbf{k}^{\prime}}{\left(2\pi\right)^{2}}\left|\left\langle \mathbf{k}\right|T^{f}\left|\mathbf{k}^{\prime}\right\rangle \right|^{2}\delta\left[\widetilde{h}\left(\mathbf{k}\right)-\widetilde{h}\left(\mathbf{k}^{\prime}\right)\right],\label{eq:gfscattime}
\end{equation}
is the quasiparticle scattering rate.

Following the arguments in \cite{abrikosov1960contribution,bennemann2008superconductivity},
Eq.~(\ref{eq:vertex correction}) can be solved to first order to
give

\begin{equation}
\Lambda\left(\mathbf{k},i\omega_{n}\right)=\Delta_{0}\left[\Gamma_{d}\left(\mathbf{k}\right)+\frac{1}{2\tau_{\mathbf{k}}^{d}\left|\omega_{n}\right|\left(1+\frac{1}{2\tau_{\mathbf{k}}\left|\omega_{n}\right|}\right)}\right],
\end{equation}
where 
\begin{eqnarray}
\frac{1}{\tau_{\mathbf{k}}^{d}} & \equiv & 2\pi n\int\frac{d^{2}\mathbf{k}^{\prime}}{\left(2\pi\right)^{2}}\left|\left\langle \mathbf{k}\right|T^{f}\left|\mathbf{k}^{\prime}\right\rangle \right|^{2}\delta\left(\widetilde{h}(\mathbf{k})-\widetilde{h}(\mathbf{k}^{\prime})\right)\Gamma_{d}(\mathbf{k}^{\prime}),\label{eq:gfdwavscattime}
\end{eqnarray}
is a different scattering rate. We will show later that {[}see Eq.~(\ref{eq:d wave channel scattering rate}){]},
when calculated at the approximately circular Fermi surface $\mathbf{k}=k_{F}\mathbf{\hat{\mathbf{k}}}$,
the quasiparticle scattering time is essentially isotropic $\tau_{\mathbf{k}}\approx\tau$,
whereas 
\begin{equation}
\frac{1}{\tau_{\mathbf{k}}^{d}}=\frac{1}{\tau^{d}}\Gamma_{d}\left(\mathbf{k}\right).
\end{equation}
We can thus write
\begin{equation}
\Lambda\left(\mathbf{k},i\omega_{n}\right)=\Delta_{0}\Gamma_{d}\left(\mathbf{k}\right)\left[1+\frac{1}{2\tau^{d}\left|\omega_{n}\right|\left(1+\frac{1}{2\tau\left|\omega_{n}\right|}\right)}\right].
\end{equation}
Eq.~(\ref{eq:vertex correction}) can now be solved to all orders
by noting that

\begin{eqnarray}
\Lambda\left(\mathbf{k},i\omega_{n}\right) & = & \Delta_{0}\Gamma_{d}\left(\mathbf{k}\right)\left\{ 1+\frac{1}{2\tau^{d}\left|\omega_{n}\right|\left(1+\frac{1}{2\tau\left|\omega_{n}\right|}\right)}+\left[\frac{1}{2\tau^{d}\left|\omega_{n}\right|\left(1+\frac{1}{2\tau\left|\omega_{n}\right|}\right)}\right]^{2}+\ldots\right\} \nonumber \\
 & = & \Delta_{0}\Gamma_{d}\left(\mathbf{k}\right)\left[1-\frac{1}{2\tau^{d}\left|\omega_{n}\right|\left(1+\frac{1}{2\tau\left|\omega_{n}\right|}\right)}\right]^{-1}\nonumber \\
 & = & \Delta_{0}\Gamma_{d}\left(\mathbf{k}\right)\frac{\left|\omega_{n}\right|+\frac{1}{2\tau}}{\left|\omega_{n}\right|+\frac{1}{2\tau}-\frac{1}{2\tau^{d}}}.
\end{eqnarray}
Plugging this result into Eq.~(\ref{eq:pairing}), we find that $T_{c}$
is determined by:

\begin{equation}
1=\frac{2\widetilde{J}kT_{c}}{d}\sum_{i\omega_{n}}\int\frac{d^{2}\mathbf{k}}{\left(2\pi\right){}^{2}}G^{f}\left(\mathbf{k},-i\omega_{n}\right)G^{f}\left(-\mathbf{k},i\omega_{n}\right)\frac{\left|\omega_{n}\right|+\frac{1}{2\tau}}{\left|\omega_{n}\right|+\frac{1}{2\tau}-\frac{1}{2\tau^{d}}}\Gamma_{d}^{2}\left(\mathbf{k}\right).\label{eq:pairing2}
\end{equation}
The momentum integral is, as usual, dominated by the region close
to the renormalized Fermi surface, which we assume to be approximately
circular in the over-doped region. We thus get, using polar coordinates
in the ($k_{x},k_{y}$)-plane,
\begin{equation}
1=\frac{\widetilde{J}kT_{c}}{d}\sum_{i\omega_{n}}\int\frac{d\theta}{2\pi}\frac{1}{\left|\omega_{n}\right|+\frac{1}{2\tau}-\frac{1}{2\tau^{d}}}\Gamma_{d}^{2}\left(k_{F}\mathbf{\hat{\mathbf{k}}}\right).\label{eq:pairing3}
\end{equation}
Using now $\Gamma_{d}\left(k_{F}\mathbf{\hat{\mathbf{k}}}\right)=2\left(\cos k_{x}-\cos k_{y}\right)\approx k_{x}^{2}-k_{y}^{2}\approx k_{F}^{2}\cos\left(2\theta\right)$
we get

\begin{equation}
1=\frac{\widetilde{J}kT_{c}m^{*}k_{F}^{2}}{d}\sum_{n\ge0}\frac{1}{\omega_{n}+\frac{1}{2\tau}-\frac{1}{2\tau^{d}}}.\label{eq:pairing4}
\end{equation}
As usual, the integral is formally divergent, but by comparing with
the equally divergent expression for the clean transition temperature
$T_{c0}$, we can get the ratio of clean ($T_{c0}$) to dirty ($T_{c}$)
transition temperatures \cite{abrikosov1960contribution}

\begin{equation}
\ln\frac{T_{c0}}{T_{c}}=\psi\left(\frac{1}{2}+\frac{\alpha}{2}\right)-\psi\left(\frac{1}{2}\right),
\end{equation}
where $\alpha=\frac{1}{2\pi T_{c}}\left(\frac{1}{\tau}-\frac{1}{\tau^{d}}\right)\equiv\frac{1}{2\pi T_{c}\tau_{pb}}$.
The leading behavior is

\begin{eqnarray}
T_{c} & = & T_{c0}-\frac{\pi}{8\tau_{pb}}.
\end{eqnarray}
The relevant scattering rates $\tau$ and $\tau_{d}$ will be calculated
in the next Section.

\subsection{$T$-matrix for the spinons and the pair-breaking scattering rate}

\label{sub:Tmatspinons}

As seen in Section~\ref{sub:GapEq}, the suppression of the superconducting
transition temperature $T_{c}$ by disorder requires the determination
of the scattering $T$-matrix of the $f$ fermions. We will find it
to first order in the disorder. In other words, the fields ($r_{i},\lambda_{i}$)
will be calculated to ${\cal O}\left(\epsilon_{i}\right)$. We define
the renormalized site energy for the $f$ electrons as $v_{i}\equiv\epsilon_{i}-\mu_{0}+\lambda_{i}$,
whose clean value limit is $\nu_{0}=\lambda_{0}-\mu_{0}$. Clean and
disordered $f$-fermion Green's functions are given by, respectively,

\begin{eqnarray}
\mathbf{G}_{0}^{f-1} & = & \left[i\omega_{n}\mathbf{1}+r_{0}^{2}t\bm{\Gamma}_{s}-v_{0}\mathbf{1}+\widetilde{J}\chi\bm{\Gamma}_{s}\right],\\
\mathbf{G}^{f-1} & = & \left[i\omega_{n}\mathbf{1}-\mathbf{v}+t\mathbf{r}\bm{\Gamma}_{s}\mathbf{r}+\widetilde{J}\chi\bm{\Gamma}_{s}\right],
\end{eqnarray}
where we have used boldface to denote matrices in the lattice site
basis, whose elements are $\mathbf{1}_{ij}=\delta_{ij}$, $\mathbf{r}_{ij}=r_{i}\delta_{ij}$,
$\mathbf{v}_{ij}=v_{i}\delta_{ij}$, and $\left(\bm{\Gamma}_{s}\right)_{ij}$
is equal to 1 if sites $i$ and $j$ are nearest neighbors and zero
otherwise. We remind the reader that we are neglecting spatial fluctuations
of the $\chi_{ij}$ field. The spinon $T$-matrix is defined through
\begin{equation}
\mathbf{G}^{f}=\mathbf{G}_{0}^{f}+\mathbf{G}_{0}^{f}\mathbf{T}^{f}\mathbf{G}_{0}^{f}=\mathbf{G}_{0}^{f}\left(\mathbf{1}+\mathbf{T}^{f}\mathbf{G}_{0}^{f}\right),
\end{equation}
from which we obtain
\begin{equation}
\mathbf{G}^{f-1}=\left(\mathbf{1}+\mathbf{T}^{f}\mathbf{G}_{0}^{f}\right)^{-1}\mathbf{G}_{0}^{f-1},
\end{equation}
and, to first order in the disorder,
\begin{equation}
\mathbf{G}^{f-1}\approx\left(\mathbf{1}-\mathbf{T}^{f}\mathbf{G}_{0}^{f}\right)\mathbf{G}_{0}^{f-1}=\mathbf{G}_{0}^{f-1}-\mathbf{T}^{f}.
\end{equation}
Thus, again to first order,

\begin{eqnarray}
\mathbf{T}^{f} & = & \mathbf{G}_{0}^{f-1}-\mathbf{G}^{f-1}\nonumber \\
 & = & \left(\mathbf{v}-v_{0}\mathbf{1}\right)-t\mathbf{r}\bm{\Gamma}_{s}\mathbf{r}+r_{0}^{2}t\bm{\Gamma}_{s}\nonumber \\
 & = & \delta\mathbf{v}-tr_{0}\left(\delta\mathbf{r}\bm{\Gamma}_{s}+\bm{\Gamma}_{s}\delta\mathbf{r}\right),
\end{eqnarray}
where $\delta\mathbf{r}\equiv\left(\mathbf{r}-r_{0}\mathbf{1}\right)$
and $\delta\mathbf{v}\equiv\left(\mathbf{v}-v_{0}\mathbf{1}\right)$.
Defining $\delta v_{i}\equiv\epsilon_{i}+\lambda_{i}-\lambda_{0}$
and $\delta r_{i}=r_{i}-r_{0}$, we have, in components,

\begin{equation}
\mathbf{T}_{ij}^{f}=\delta v_{i}\delta_{ij}-r_{0}\left(\delta r_{i}+\delta r_{j}\right)t_{ij}.
\end{equation}
All we need now is to find $\delta r_{i}$ and $\delta v_{i}$ to
first order in $\epsilon_{i}$. This was already obtained in reference
\cite{shaoetal15} (see, in particular, the Supplemental Material).
After setting in those equations the gap and its fluctuations to zero
$\bm{\Delta}=\delta\bm{\Delta}=0$ (normal state) and $\delta\bm{\chi}=0$
(as being negligible), we obtain in $\mathbf{k}$-space
\begin{eqnarray}
\Pi^{a}\left(\mathbf{k}\right)\delta v\left(\mathbf{k}\right)+r_{0}\left[1+\Pi^{b}\left(\mathbf{k}\right)\right]\delta r\left(\mathbf{k}\right) & = & 0,\label{eq:first}
\end{eqnarray}

\begin{eqnarray}
\left[\lambda_{0}-\frac{\lambda_{0}}{2d}\Gamma_{s}\left(\mathbf{k}\right)\right]\delta r\left(\mathbf{k}\right)+r_{0}\delta v\left(\mathbf{k}\right) & = & r_{0}\epsilon\left(\mathbf{k}\right),\label{eq:second}
\end{eqnarray}
where
\begin{eqnarray}
\boldsymbol{\Pi^{a}}\left(\mathbf{k}\right) & = & \frac{1}{V}\sum_{\mathbf{q}}\frac{f\left[\widetilde{h}\left(\mathbf{q}+\mathbf{k}\right)\right]-f\left[\widetilde{h}\left(\mathbf{q}\right)\right]}{\widetilde{h}\left(\mathbf{q}+\mathbf{k}\right)-\widetilde{h}\left(\mathbf{q}\right)},\\
\boldsymbol{\Pi^{b}}\left(\mathbf{k}\right) & = & \frac{1}{V}\sum_{\mathbf{q}}\frac{f\left[\widetilde{h}\left(\mathbf{q}+\mathbf{k}\right)\right]-f\left[\widetilde{h}\left(\mathbf{q}\right)\right]}{\widetilde{h}\left(\mathbf{q}+\mathbf{k}\right)-\widetilde{h}\left(\mathbf{q}\right)}\left[h\left(\mathbf{q}+\mathbf{k}\right)+h\left(\mathbf{q}\right)\right],
\end{eqnarray}
where $d$ is the lattice dimension ($d=2$, for our purposes), $f\left(x\right)$
is the Fermi-Dirac function, $\Gamma_{s}\left(\mathbf{k}\right)=2\left(\cos k_{x}+\cos k_{y}\right)$,
and $h\left(\mathbf{k}\right)=-t\Gamma_{s}\left(\mathbf{k}\right)$
is the bare energy dispersion. Solving Eqs.~(\ref{eq:first})-(\ref{eq:second})
for $\delta r\left(\mathbf{k}\right)$ and $\delta v\left(\mathbf{k}\right)$

\begin{eqnarray}
\delta v\left(\mathbf{k}\right) & = & -x\frac{\Pi\left(\mathbf{k}\right)\epsilon\left(\mathbf{k}\right)}{\lambda_{0}-\frac{\lambda_{0}}{2d}\Gamma_{s}\left(\mathbf{k}\right)-x\Pi\left(\mathbf{k}\right)},\\
\delta r\left(\mathbf{k}\right) & = & r_{0}\frac{\epsilon\left(\mathbf{k}\right)}{\lambda_{0}-\frac{\lambda_{0}}{2d}\Gamma_{s}\left(\mathbf{k}\right)-x\Pi\left(\mathbf{k}\right)},
\end{eqnarray}
where we used $x=r_{0}^{2}$ and defined 
\begin{equation}
\Pi\left(\mathbf{k}\right)\equiv\frac{1+\Pi^{b}\left(\mathbf{k}\right)}{\Pi^{a}\left(\mathbf{k}\right)}.
\end{equation}
Therefore

\begin{eqnarray}
\left\langle \mathbf{k}\right|T^{f}\left|\mathbf{k}^{\prime}\right\rangle  & = & \delta v\left(\mathbf{k}^{\prime}-\mathbf{k}\right)+r_{0}\left[h\left(\mathbf{k}\right)+h\left(\mathbf{k}^{\prime}\right)\right]\delta r\left(\mathbf{k}^{\prime}-\mathbf{k}\right)\\
 & = & x\left[\frac{h\left(\mathbf{k}\right)+h\left(\mathbf{k}^{\prime}\right)-\Pi\left(\mathbf{k}^{\prime}-\mathbf{k}\right)}{\lambda_{0}-\frac{\lambda_{0}}{2d}\Gamma_{s}\left(\mathbf{k}^{\prime}-\mathbf{k}\right)-x\Pi\left(\mathbf{k}^{\prime}-\mathbf{k}\right)}\right]\epsilon\left(\mathbf{k}^{\prime}-\mathbf{k}\right).\label{eq:T-matrix}
\end{eqnarray}

The result in Eq.~(\ref{eq:T-matrix}) is general. For the $T_{c}$
calculation within the AG theory, we only need it for a single impurity
{[}see Eq.~(\ref{eq:vertex correction}){]}. We therefore set $\epsilon_{i}=t\delta_{i,0}$
or $\epsilon\left(\mathbf{k}\right)=t$. We must now plug Eq.~(\ref{eq:T-matrix})
into Eqs.~(\ref{eq:gfscattime}) and (\ref{eq:gfdwavscattime}).
Since the superconducting pairing is mostly affected by the scattering
near Fermi surface, we can set $\mathbf{k}=k_{F}\mathbf{\hat{k}}$
in Eq.~(\ref{eq:T-matrix}). For computations, $k_{F}$ is taken
as the magnitude of the Fermi momentum averaged over the approximately
circular Fermi surface. We can thus make the following simplifications:
$h\left(\mathbf{k}\right)+h\left(\mathbf{k}^{\prime}\right)=2E_{F}$,
$\Pi^{b}\left(\mathbf{k}\right)=2E_{F}\Pi^{a}\left(\mathbf{k}\right)$,
$\Pi\left(\mathbf{k}\right)=\left[\Pi^{a}\left(\mathbf{k}\right)\right]{}^{-1}+2E_{F}$,
and $\Pi^{a}\left(\mathbf{k}-\mathbf{k}^{\prime}\right)=-\rho^{\ast}g_{L}\left(y\right)$,
where $y=\frac{\left|\mathbf{k}-\mathbf{k}^{\prime}\right|}{2k_{F}}=\left|\sin\left(\frac{\varphi}{2}\right)\right|$
, $\varphi=\theta-\theta^{\prime}$ is the angle between $\mathbf{k}$
and $\mathbf{k}^{\prime}$, $E_{F}$ is the bare Fermi energy (obtained
by solving the mean-field equations with the constraint of an electron
filling of $1-x$), the function $g_{L}\left(y\right)$ is defined
as \cite{RevModPhys.54.437} 
\begin{equation}
g\left(y\right)\equiv\begin{cases}
1 & y\leq1,\\
1-\sqrt{1-y^{-2}} & y>1,
\end{cases}
\end{equation}
and $\rho^{*}=\frac{m^{*}}{2\pi}$ is the renormalized (spinon) density
of states. Finally, defining

\begin{equation}
g\left(y\right)\equiv\frac{t^{2}}{\left\{ \rho^{\ast}\lambda_{0}k_{F}^{2}y^{2}g_{L}\left(y\right)+x\left[1-2\rho^{\ast}E_{F}g_{L}\left(y\right)\right]\right\} ^{2}},
\end{equation}
we obtain

\begin{eqnarray}
\frac{1}{\tau^{d}\left(\theta\right)} & = & x^{2}\frac{nm^{*}}{2\pi}\int_{0}^{2\pi}d\theta^{\prime}g\left[\left|\sin\left(\frac{\theta-\theta^{\prime}}{2}\right)\right|\right]\cos\left(2\theta^{\prime}\right)\nonumber \\
 & = & x^{2}\frac{nm^{*}}{2\pi}\int_{0}^{2\pi}dug\left[\left|\sin\left(\frac{u}{2}\right)\right|\right]\cos\left[2\left(\theta-u\right)\right]\nonumber \\
 & = & \cos2\theta\left[x^{2}\frac{nm^{*}}{2\pi}\int_{0}^{2\pi}dug\left[\left|\sin\left(\frac{u}{2}\right)\right|\right]\cos2u\right]\label{eq:d wave channel scattering rate}\\
 & \equiv & \cos2\theta\frac{1}{\tau^{d}}=\Gamma_{d}\left(\mathbf{k}\right)\frac{1}{\tau_{d}}.\nonumber 
\end{eqnarray}
In the last step we dropped the term in $\sin2\theta$ since this
term vanishes after integration in Eq.(\ref{eq:pairing}). Analogously,
\begin{equation}
\frac{1}{\tau}=x^{2}\frac{nm^{*}}{2\pi}\int_{0}^{2\pi}d\theta g\left[\left|\sin\left(\frac{\theta}{2}\right)\right|\right],\label{eq:quasipartscattime}
\end{equation}
and 
\begin{equation}
\frac{1}{\tau_{pb}}=x^{2}\frac{nm^{*}}{2\pi}\int_{0}^{2\pi}d\theta g\left[\left|\sin\left(\frac{\theta}{2}\right)\right|\right]\left(1-\cos2\theta\right).\label{eq:pairbreakscatttime}
\end{equation}

\subsection{$T$-matrix for the physical electrons and the normal state transport
scattering rate}

\label{sub:Tmatnormal}

In order to describe transport in the normal state, we must analyze
the physical electron scattering $T$-matrix. The calculation is analogous
to the one in Section~\ref{sub:Tmatspinons}. The bare and renormalized
Green's functions for the physical electrons are given by
\begin{eqnarray}
\mathbf{G}_{0}^{e-1} & = & r_{0}^{-2}\left[i\omega_{n}\mathbf{1}+r_{0}^{2}t\bm{\Gamma}_{s}-v_{0}\mathbf{1}+\widetilde{J}\chi\bm{\Gamma}_{s}\right],\\
\mathbf{G}^{e-1} & = & \mathbf{r}^{-1}\left[i\omega_{n}\mathbf{1}-\mathbf{v}+t\mathbf{r}\bm{\Gamma}_{s}\mathbf{r}+\widetilde{J}\chi\bm{\Gamma}_{s}\right]\mathbf{r}^{-1}.
\end{eqnarray}
Proceeding to first order in the disorder as before yields
\begin{eqnarray}
\mathbf{T}^{e} & = & \mathbf{G}_{0}^{e-1}-\mathbf{G}^{e-1}\nonumber \\
 & = & r_{0}^{-2}\delta\mathbf{v}-2v_{0}r_{0}^{-3}\delta\mathbf{r}+\widetilde{J}\chi\left(r_{0}^{-2}\bm{\Gamma}_{s}-\mathbf{r}^{-1}\bm{\Gamma}_{s}\mathbf{r}^{-1}\right)\nonumber \\
 & = & x^{-1}\left[\delta\mathbf{v}-2v_{0}r_{0}^{-1}\delta\mathbf{r}+\widetilde{J}\chi r_{0}^{-1}\left(\delta\mathbf{r}\bm{\Gamma}_{s}+\bm{\Gamma}_{s}\delta\mathbf{r}\right)\right].
\end{eqnarray}

\begin{eqnarray}
\left\langle \mathbf{k}\right|T^{e}\left|\mathbf{k}^{\prime}\right\rangle  & = & -\left\{ \frac{\Pi\left(\mathbf{k}^{\prime}-\mathbf{k}\right)+\frac{2v_{0}}{x}+\frac{\widetilde{J}\chi}{tx}\left[h\left(\mathbf{k}\right)+h\left(\mathbf{k}^{\prime}\right)\right]}{\lambda_{0}-\frac{\lambda_{0}}{2d}\Gamma_{s}\left(\mathbf{k}^{\prime}-\mathbf{k}\right)-x\Pi\left(\mathbf{k}^{\prime}-\mathbf{k}\right)}\right\} \epsilon\left(\mathbf{k}^{\prime}-\mathbf{k}\right).\label{eq:T-matrix-e}
\end{eqnarray}

This scattering $T$-matrix can now be used to calculate the transport
scattering rate that enters the expression for the conductivity in
the normal state
\begin{equation}
\frac{1}{\tau_{\mathbf{k}}^{tr}}\equiv2\pi xn\int\frac{d^{2}\mathbf{k}^{\prime}}{\left(2\pi\right)^{2}}\left|\left\langle \mathbf{k}\right|T^{e}\left|\mathbf{k}^{\prime}\right\rangle \right|^{2}\delta\left[\widetilde{h}\left(\mathbf{k}\right)-\widetilde{h}\left(\mathbf{k}^{\prime}\right)\right]\left(1-\cos\varphi\right),\label{eq:condscattrate}
\end{equation}
where, as before, $\varphi$ is the angle between $\mathbf{k}$ and
$\mathbf{k}^{\prime}$ and the $x$ factor in front of the integral
comes from the quasiparticle weight of the physical electron Green's
function. We now focus again on wave vectors close to the approximately
circular Fermi surface and implement the same approximations used
in the previous Section. We first note that the numerators of the
fractions is Eqs.~(\ref{eq:T-matrix}) and (\ref{eq:T-matrix-e}),
though seemingly different, are actually the same, since
\[
\frac{2v_{0}}{x}+\frac{\widetilde{J}\chi}{tx}\left[h\left(\mathbf{k}\right)+h\left(\mathbf{k}^{\prime}\right)\right]=-2\frac{tx+\widetilde{J}\chi}{tx}E_{F}+2\frac{\widetilde{J}\chi}{tx}E_{F}=-2E_{F},
\]
where the negative of the renormalized chemical potential $v_{0}=-\frac{m}{m^{*}}E_{F}=-\frac{tx+\widetilde{J}\chi}{t}E_{F}$.
Like the quasiparticle scattering rate, the transport scattering rate
does not depend on the wave vector direction within the assumed approximations
and we get 
\begin{equation}
\frac{1}{\tau^{tr}}=\frac{xnm^{*}}{2\pi}\int_{0}^{2\pi}d\theta g\left[\left|\sin\left(\frac{\theta}{2}\right)\right|\right]\left(1-\cos\theta\right).\label{eq:transscatttime}
\end{equation}

A straightforward calculation up to first order in the impurity potential
gives the $T$-matrix in momentum space for $f$ fermions and physical
($e$) electrons, respectively, as shown previously in Subsection
\ref{sub:Tmatspinons} and \ref{sub:Tmatnormal}

\begin{equation}
\left\langle \mathbf{k}\right|T^{f}\left|\mathbf{k}^{\prime}\right\rangle =xt\left[\frac{h\left(\mathbf{k}\right)+h\left(\mathbf{k}^{\prime}\right)-\Pi\left(\mathbf{k}^{\prime}-\mathbf{k}\right)}{\lambda_{0}-\frac{\lambda_{0}}{2d}\Gamma_{s}\left(\mathbf{k}^{\prime}-\mathbf{k}\right)-x\Pi\left(\mathbf{k}^{\prime}-\mathbf{k}\right)}\right],\label{eq: Tf}
\end{equation}

\begin{equation}
\left\langle \mathbf{k}\right|T^{e}\left|\mathbf{k}^{\prime}\right\rangle =-t\left\{ \frac{\Pi\left(\mathbf{k}^{\prime}-\mathbf{k}\right)+\frac{2v_{0}}{x}+\frac{\widetilde{J}\chi}{tx}\left[h\left(\mathbf{k}\right)+h\left(\mathbf{k}^{\prime}\right)\right]}{\lambda_{0}-\frac{\lambda_{0}}{2d}\Gamma_{s}\left(\mathbf{k}^{\prime}-\mathbf{k}\right)-x\Pi\left(\mathbf{k}^{\prime}-\mathbf{k}\right)}\right\} ,\label{eq:T matrix conductivity}
\end{equation}
where $h\left(\mathbf{k}\right)=-t\Gamma_{s}\left(\mathbf{k}\right)$
is the bare energy dispersion,
\begin{equation}
\Pi\left(\mathbf{k}\right)\equiv\frac{1+\Pi^{b}\left(\mathbf{k}\right)}{\Pi^{a}\left(\mathbf{k}\right)},\label{eq:pi}
\end{equation}
with 
\begin{eqnarray*}
\Pi^{a}\left(\mathbf{k}\right) & = & \frac{1}{V}\sum_{\mathbf{q}}\frac{f\left[\widetilde{h}\left(\mathbf{q}+\mathbf{k}\right)\right]-f\left[\widetilde{h}\left(\mathbf{q}\right)\right]}{\widetilde{h}\left(\mathbf{q}+\mathbf{k}\right)-\widetilde{h}\left(\mathbf{q}\right)},\\
\Pi^{b}\left(\mathbf{k}\right) & = & \frac{1}{V}\sum_{\mathbf{q}}\frac{f\left[\widetilde{h}\left(\mathbf{q}+\mathbf{k}\right)\right]-f\left[\widetilde{h}\left(\mathbf{q}\right)\right]}{\widetilde{h}\left(\mathbf{q}+\mathbf{k}\right)-\widetilde{h}\left(\mathbf{q}\right)}\left[h\left(\mathbf{q}+\mathbf{k}\right)+h\left(\mathbf{q}\right)\right],
\end{eqnarray*}
and $f\left(x\right)$ is the Fermi-Dirac function at $T=0$.

In order to assess the role of electronic correlations we will compare
our full results as described above with a corresponding non-correlated
system in which $J=0$. In the latter case, the $T$-matrix is given
simply by the lattice Fourier transform of the bare disorder potential,
$\left\langle \mathbf{k}\right|T_{0}\left|\mathbf{k}^{\prime}\right\rangle =\epsilon\left(\mathbf{k}^{\prime}-\mathbf{k}\right)=t$,
and there is no distinction between auxiliary and physical fermions.
The two sets of results will be called correlated and non-correlated,
respectively. Even at this point, the renormalizations due to strong
correlations are clear: the $\mathbf{k}$-dependent factors in Eqs.~(\ref{eq: Tf})
and (\ref{eq:T matrix conductivity}), which reflect the spatial readjustments
of the $r_{i}$ and $\lambda_{i}$ fields, make the bare potential
``softer'' and more non-local. Note also the extra $x$ factor in
Eq.~(\ref{eq: Tf}) as compared to Eq.~(\ref{eq:T matrix conductivity}).

At low temperatures, only scattering very close to the Fermi level
is relevant. We will thus calculate the $T$-matrices at the Fermi
surface. Furthermore, we are interested in the over-doped region,
where the Fermi surface anisotropy becomes increasingly less pronounced
as the doping increases. Therefore, we will simplify the actual lattice
dispersion in favor of an isotropic one corresponding to the continuum
limit, $h\left(\mathbf{k}\right)\approx-4t+tk^{2}$. This is equivalent
to a bare effective mass $m=1/2t$ and a renormalized one $m^{*}\equiv1/\left(2tx+2\tilde{J}\chi\right)$.
Finally, we call $E_{F}$ and $k_{F}$ the Fermi energy and momentum
for the bare dispersion $h\left(\mathbf{k}\right)$, respectively,
while $\widetilde{E}_{F}=\frac{m^{\ast}}{m}E_{F}$ is the Fermi energy
for the renormalized dispersion of the $f$-fermions.

\section{Discussions}

\subsection{Pair breaking parameter}

Once the scattering matrix has been determined, it is a trivial matter
to write down the predictions of the Abrikosov-Gor'kov (AG) theory
for the suppression of the superconducting transition temperature
$T_{c}$ as shown in Subsection \ref{sub:GapEq}

\begin{equation}
\ln\frac{T_{c0}}{T_{c}}=\psi\left(\frac{1}{2}+\frac{\alpha}{2}\right)-\psi\left(\frac{1}{2}\right),
\end{equation}
where $T_{c0}$ is the transition temperature in clean limit, $\alpha\equiv1/\left(2\pi T_{c}\tau_{pb}\right)$,
and $\tau_{pb}$ is the pair breaking scattering time. The latter
is given in the continuum limit by
\begin{equation}
\frac{1}{\tau_{pb}}=\frac{x^{2}nm^{*}}{2\pi}\int_{0}^{2\pi}d\theta g\left[\left|\sin\left(\frac{\theta}{2}\right)\right|\right]\left(1-\cos2\theta\right),\label{eq:pairbreakingrate}
\end{equation}
where 
\begin{equation}
g\left(y\right)\equiv\frac{t^{2}}{\left\{ \rho^{\ast}\lambda_{0}k_{F}^{2}y^{2}g_{L}\left(y\right)+x\left[1-2\rho^{\ast}E_{F}g_{L}\left(y\right)\right]\right\} ^{2}},\label{eq:gfunction}
\end{equation}
where $\rho^{*}=\frac{m^{*}}{2\pi}$ is the renormalized density of
states and
\begin{equation}
g_{L}\left(y\right)\equiv\begin{cases}
1 & y\leq1,\\
1-\sqrt{1-y^{-2}} & y>1.
\end{cases}
\end{equation}
The factor of $1-\cos2\theta$ comes from the vertex corrections for
$d$-wave pairing and can be generalized to other pairing symmetries
by changing $\cos2\theta$ to the corresponding lattice harmonic.
The leading behavior for low impurity concentrations is

\begin{equation}
T_{c}=T_{c0}-\pi/8\tau_{pb}.\label{eq:leading order}
\end{equation}

\begin{figure}
\centering{}\includegraphics[width=0.8\textwidth]{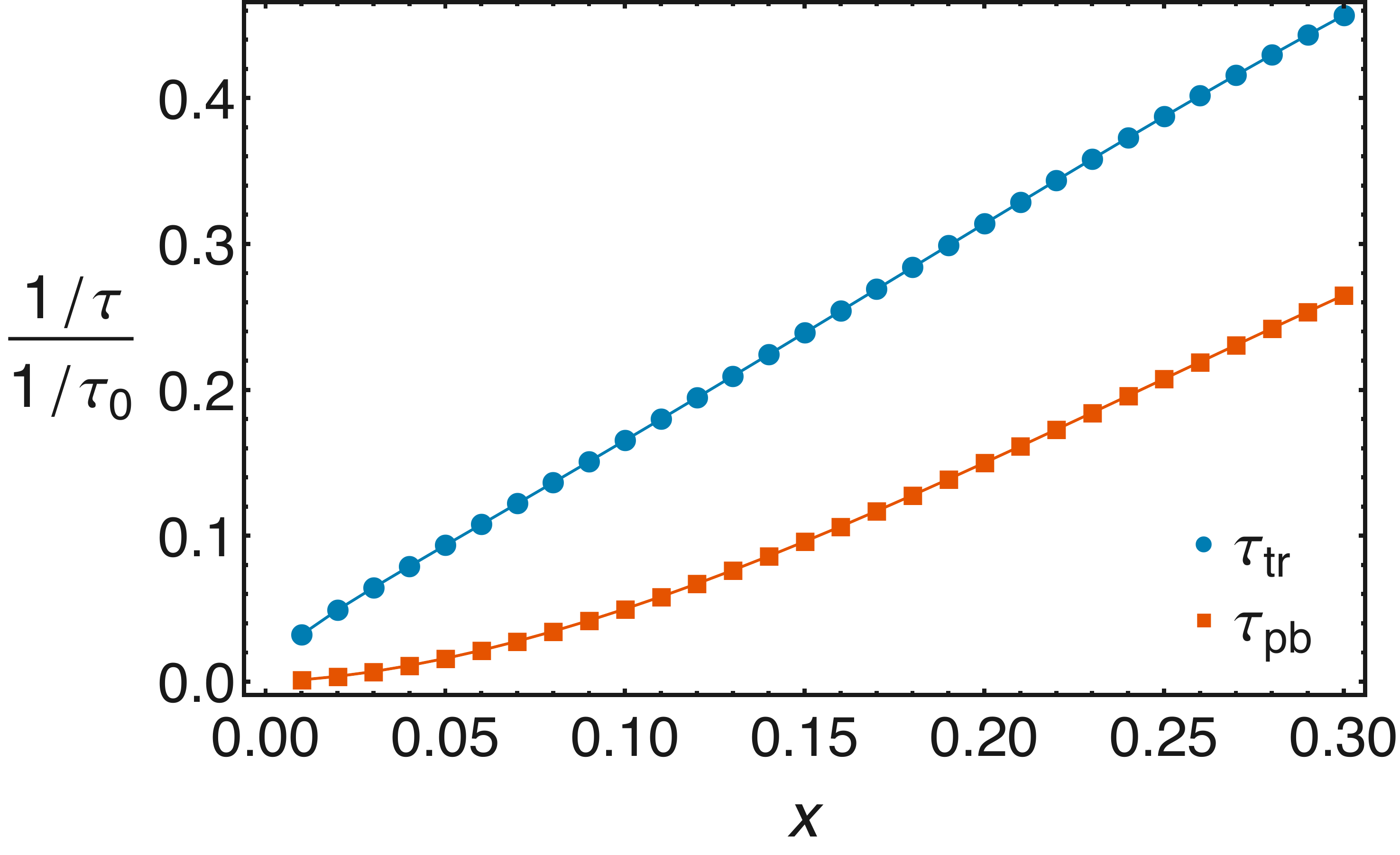}\caption{\label{fig:scattering rates}The pair-breaking and transport scattering
rates normalized by the non-correlated value $1/\tau_{0}$ as a function
of the doping level.}
\end{figure}

Fig.~\ref{fig:scattering rates} shows the ratio of the pair-breaking
scattering rate $1/\tau_{pb}$ in the correlated case to the non-correlated
one. Note that, for the non-correlated case, 
\begin{equation}
\frac{1}{\tau_{0}}=\frac{nm}{2\pi}\int_{0}^{2\pi}d\theta t^{2}\left(1-\cos2\theta\right)=nmt^{2}.\label{eq:noncorr-rate}
\end{equation}
Clearly, pair breaking is strongly suppressed by electronic correlations.
While this suppression is enhanced as the density-driven Mott transition
is approached ($x\to0$), it is still quite significant up to dopings
of $x\approx0.3$. As a result, the $T_{c}$ degradation is expected
to be considerably slower in that case and we expect the $d$-wave
superconductivity to be more robust than predicted by the weak coupling
theory. Equivalently, the critical impurity concentration $n_{c}$
at which $T_{c}$ vanishes is enhanced when compared to the non-correlated
case, $5-10$ times in the range of dopings from 0.15 to 0.3. We note
that this suppression of pair-breaking by the impurities is completely
dominated by the $x^{2}$ dependence of Eq.~(\ref{eq:pairbreakingrate}).
Indeed, in the whole range of dopings from $\sim0.01$ to $\sim0.3$,
the product of the effective mass $m^{\ast}$ and the angular integral
in Eq.~(\ref{eq:pairbreakingrate}) varies very little (roughly from
5 to 3). Thus, in a manner very reminiscent of the strong healing
of gap fluctuations found in reference \textit{\emph{\cite{shaoetal15},
here the robustness of $T_{c}$ can also be attributed to}} `Mottness'.

\subsection{Transport property in correlated normal state}

The normal state resistivity is governed by the impurity induced transport
scattering rate, which can be evaluated straightforwardly via Eq.~(\ref{eq:T matrix conductivity})
to give:

\begin{equation}
\frac{1}{\tau_{tr}}=\frac{xnm^{*}}{2\pi}\int_{0}^{2\pi}d\theta g\left[\left|\sin\left(\frac{\theta}{2}\right)\right|\right]\left(1-\cos\theta\right).\label{eq:transportrate}
\end{equation}
The non-correlated transport scattering rate defined as 
\begin{equation}
\frac{1}{\tau_{0}^{tr}}=\frac{nm}{2\pi}\int_{0}^{2\pi}d\theta t^{2}\left(1-\cos\theta\right),
\end{equation}
which coincides with the above $1/\tau_{0}$ for the bare isotropic
scattering impurity potential we used. As shown in Fig.~\ref{fig:scattering rates},
the transport rate is also suppressed by electronic correlations and
`Mottness'. In contrast to Eq.~(\ref{eq:pairbreakingrate}), however,
the dependence is almost linear in $x$. This is because, as before,
the product of $m^{\ast}$ and the angular integral in Eq.~(\ref{eq:transportrate})
is almost doping independent. As a result, as seen in Fig.~\ref{fig:scattering rates},
for a wide range of doping levels the suppression of the pair-breaking
scattering rate is much more significant than the transport one.

\begin{figure}
\centering{}\includegraphics[width=0.8\textwidth]{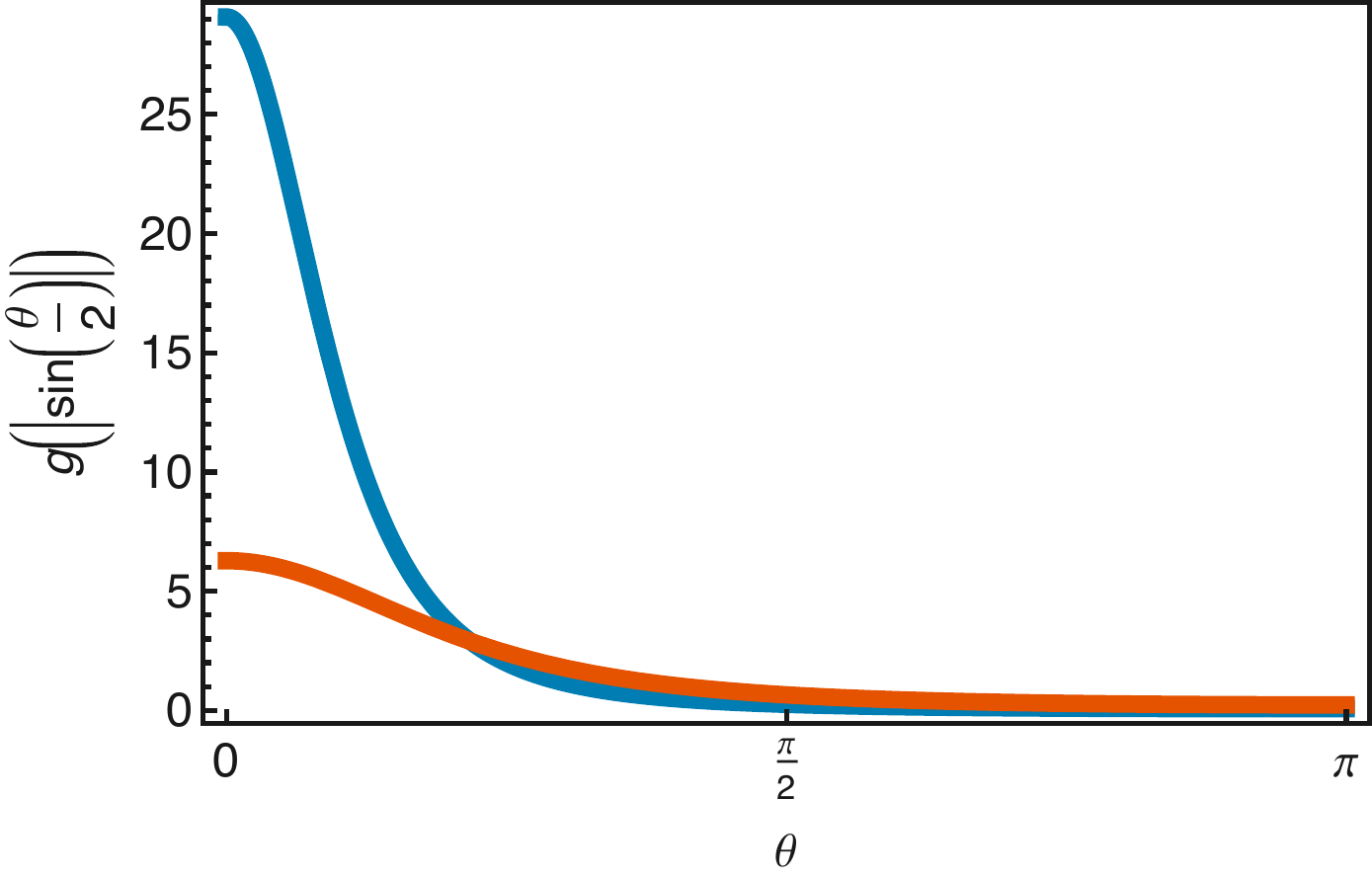}\caption{\label{fig:efftive T matrices}The angular dependence of the renormalized
$T$-matrices at $x=0.15$ (blue) and $x=0.3$ (red). }
\end{figure}

\subsection{Enhanced forward scattering}

The doping dependence of the scattering rates illustrated in Fig.~\ref{fig:scattering rates}
makes it clear that the dominant effect comes from the explicit $x$
dependence in Eqs.~(\ref{eq:pairbreakingrate}) ($\sim x^{2}$) and
(\ref{eq:transportrate}) ($\sim x$). The $x$-dependence coming
from $m^{\ast}$ times the angular integrals over the scattering matrices
is very weak. However, this does not mean that the angular dependence
of the $T$-matrices is not affected by strong correlations, as we
will now show.

In Fig.~\ref{fig:efftive T matrices} we show, for two doping levels,
the angular dependence of the function $g\left[\left|\sin\left(\frac{\theta}{2}\right)\right|\right]$
{[}defined in Eq.~(\ref{eq:gfunction}){]}, which is integrated over
in Eqs.~(\ref{eq:pairbreakingrate}) and (\ref{eq:transportrate}).
This should be compared to the bare impurity result, which is $\sim t^{2}$
and thus $\theta$-independent. Clearly, there is a large enhancement
of forward scattering, indicating a ``softening'' of the impurity
scattering by correlations, even for point-like impurities in the
plane.

This function is weighted by $1-\cos2\theta$ and $1-\cos\theta$
in the integrations in Eqs.~(\ref{eq:pairbreakingrate}) and (\ref{eq:transportrate}),
respectively. These weight functions amplify the contributions from
the regions $\theta\approx\pi/2$ and $\theta\approx\pi$, respectively,
which are, however, hardly affected by correlations. As a result,
even with the softening of the impurity scattering, the angular integrals
are not renormalized significantly in the range from $\sim0.15$ to
$\sim0.3$, when compared to the non-correlated bare impurity result:
$\sim0.8-1$ in Eq.~(\ref{eq:pairbreakingrate}) and $\sim0.3-0.5$
in Eq.~(\ref{eq:transportrate}). The conclusion, then, is that strong
correlations enhance significantly the forward scattering region even
for point-like in-plane impurities, but this is \emph{not} the reason
for the robustness of $T_{c}$ or the resilience of the normal state
conductivity.

\section{Conclusions}

\textit{\emph{We have shown how the weak-coupling AG theory of $T_{c}$
suppression and normal state resistivity by dilute non-magnetic impurities
is modified in a strongly correlated metal. Even though the renormalized
scattering amplitude is strongly enhanced in the forward direction,
the most significant effect comes from the suppression of the electron
fluid compressibility by `Mottness', which is effective even relatively
far from the Mott insulating state.}} Given its simplicity, we suggest
that this phenomenon is generic to other systems close to Mott localization.

\chapter{Shock Waves and Commutation Speed of Memristors}

\begin{figure}
\centering{}\includegraphics[width=0.6\textwidth]{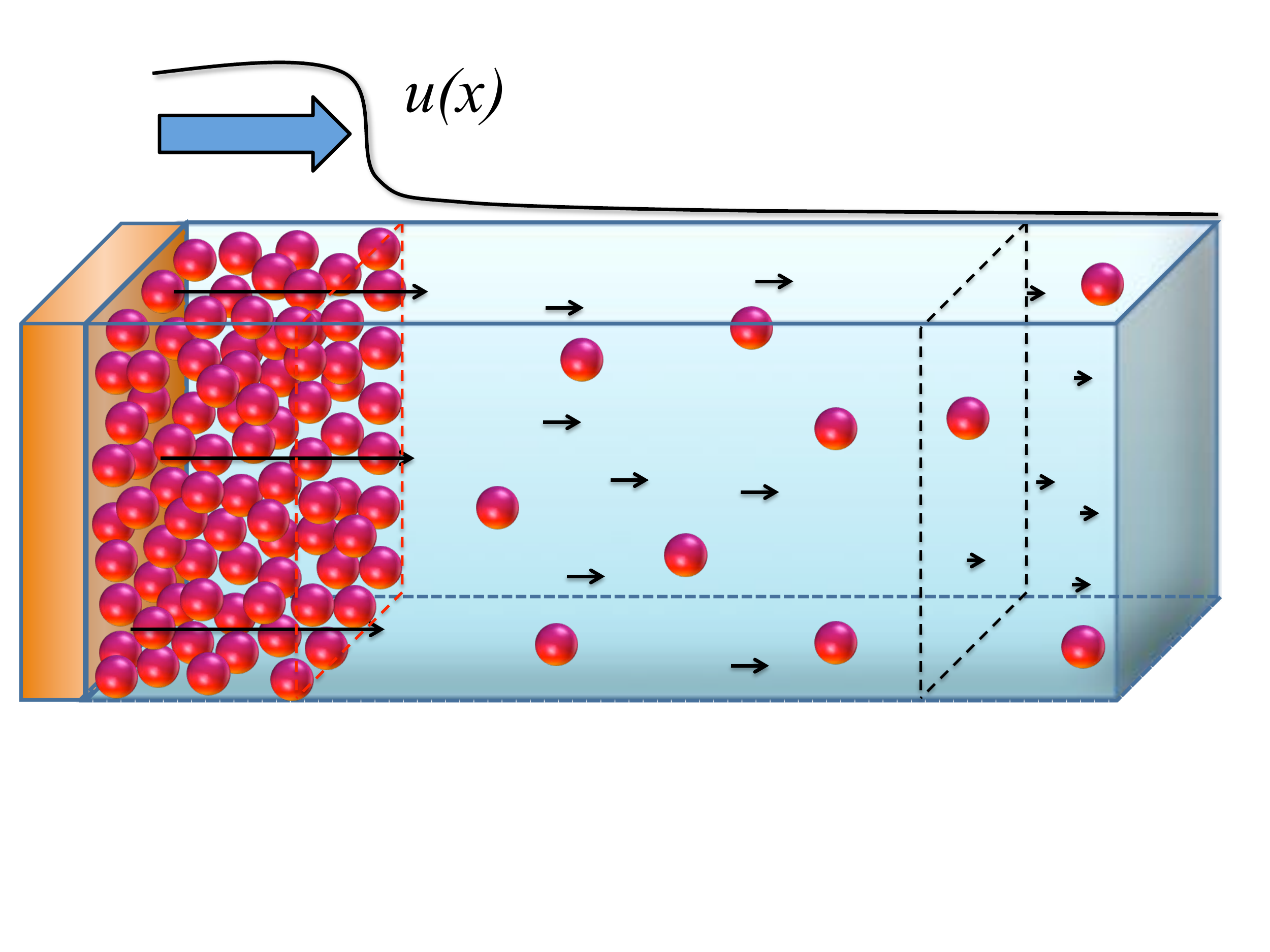}  
\caption{\label{fig:shock} Schematic representation of the shock wave evolution.
The orange region indicates the metallic electrode and the blue indicates the TMO dielectric. Small spheres
denote the ionic defects (oxygen vacancies) whose density profile form a shock wave. It evolves
through a highly resistive (Schottky) interface and eventually leaks over the more conductive bulk, producing
the resistive change. Black arrows depict the strength of the local electric fields.}
\end{figure}

In this context gaining theoretical insight is of essence. Thus, in the present work we shall address one of the
key aspects of the RS phenomenon, namely, the issue of the commutation speed of the resistance change. 
Our first striking result is a connection between the RS phenomenon and {\em shock wave} formation,
a classic topic of non-linear dynamics \cite{debnath2011nonlinear}. In fact, we shall argue that the profile of oxygen vacancies that
migrate during the resistive change forms a shock wave that propagates through the Schottky barrier and
leaks onto the bulk of the device, which we schematically illustrate in Fig.\ref{fig:shock}.
We further validate the scenario by means of numerical simulations on a successful model of 
RS and by novel experiments done on a manganese based memristor device. Both model calculations
and experiments reveal a striking scaling behaviour as predicted by the shock wave scenario.

\section{Generalized Burgers' equation}
When ions migrate through a {\em conducting} medium under the influence
of strong applied voltage, they are likely to undergo a nonlinear
diffusion process, as we explain in the following. The total ionic current
$\mathbf{j}(t,\mathbf{x})=\mathbf{j}_{diffusion}+\mathbf{j}_{drift}$
can be expressed as the sum of a diffusion current $\mathbf{j}_{diffusion}=-D\nabla u$
and a drift current $\mathbf{j}_{drift}$, which is induced by the
local electric field $\mathbf{E}$ and the local concentration
$u$.  Together with the continuity equation $\partial_{t}u+\nabla\cdot\mathbf{j}(t,\mathbf{x})=0$, this
immediately gives us a generalized diffusion equation of the Nernst-Planck
type. This would represent a familiar drift-diffusion equation if the
local electric field $\mathbf{E}$ was held constant, i.e. independent of the local ion 
concentration $u$. In contrast, in (poorly) conducting media and under voltage pulses, the local electric field may strongly 
depend on the local ion concentration; this effect is the key source of nonlinearity causing 
the formation of shock waves and very sudden resistance switching \cite{inoue-sawa,MRS}. 

Since electrons move much faster than the ions, we can view the ions as
static when considering the electronic current $\mathbf{I}$,
which obeys a steady-state condition $\nabla\cdot\mathbf{I}=0$.
The local electric field is then simply determined, through Ohm\textquoteright{}s
law, by the local resistivity $\mathbf{E}=\rho(u)\mathbf{I}$,
which may be a strong function of the local ion concentration $u$.
In particular, in bad metals such as the transition metal oxides,
the migrating ions (e.g. oxygen vacancies) act as scattering centers
for the conduction electrons. In such situations, we expect $\rho(u)$
to be a \textit{monotonically increasing} function of the local ion vacancy
density $u(t,\mathbf{x})$. Therefore, the redistribution
of the local ion density results in the change of local resistivities
and, consequently, of the local electric fields, which further promotes
the non-linear effect in the drift. 

Under the experimentally-relevant case where the transverse currents may be neglected,
the problem simplifies to a one-dimensional
non-linear diffusion equation,
\begin{equation}
\partial_{t}u+f\left(u\right)\partial_{x}u=D\partial_{xx}u,\label{burgers}
\end{equation}
where $f\left(u\right)\equiv\partial_{u}j_{drift}\left(u,I\right)$, and
$I(t)$ is the magnitude of the electronic current.
Equation (\ref{burgers}) can be considered a generalization of the
famous \emph{Burgers' }\textit{\emph{equation}}\emph{, }which corresponds
to the special case $f\left(u\right)\propto u$. Its most significant
feature is the presence of a density-dependent drift term, {\em which physically
means that the ``crest of the wave'' experiences a stronger external
force than the ``trough''}. This generally leads to the formation
of a sharply defined \emph{shock-wave }front in the 
$u(t,x)$ profile, which assumes a universal
form at long times, completely independent of the - quickly ``forgotten''
- initial conditions.  Although the process is driven
by the drift term, the stability
of the shock wave form is provided by the existence of the diffusion term
which prevents the shock wave from self-breaking \cite{taylor2011partial,debnath2011nonlinear} \footnote{ While the drift current contains the nonlinear term that leads to a steep wave front, according to Fick's law, the diffusion current flows in such a way as to reduce the steepness of the density profile. Therefore, the diffusion
current stabilizes the propagation of the shock wave and it protects the solution from entering the self-breaking region (multi-valued
solution). In the scenario of the standard Burgers' equation, where $f(u)\sim u$ in Eq(1), the equation can be explicitly solved via Cole-Hopf transformation (see references \cite
  {taylor2011partial,debnath2011nonlinear}. In that case we can check the effects of the diffusion current against nonlinear drift exactly.} . 
Remarkably, the formation
of shock waves proves to be robust in a much more general family
of models with the nonlinear drift term specified by the function
$f(u)$, any \emph{monotonically increasing} function of
$u$. The qualitative behaviour can be established by using the well-known
method of characteristics \cite{debnath2011nonlinear,courant1962methods}, that
we illustrate with an example in Fig.~\ref{fig:shock-sketch} (see Appendix~\ref{sec:charact} for a more detailed explanation).

\begin{figure}
\centering{}\includegraphics[width=0.6\textwidth]{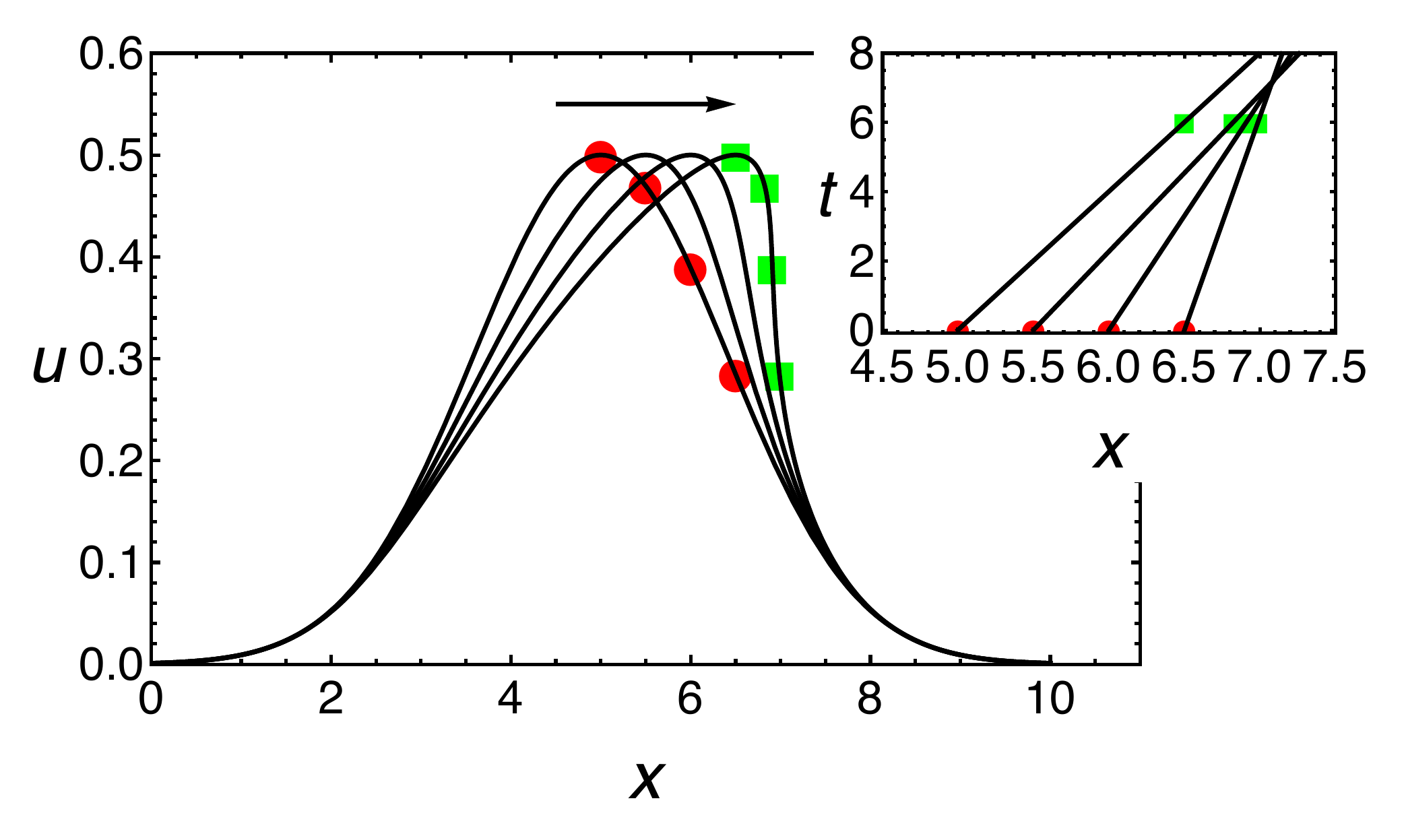}
\caption{\label{fig:shock-sketch} Time evolution of a generic density profile according to the 
generalized Burgers' equation (\ref{burgers}). Curves correspond to profiles at $t=$ 0, 2, 4 and 6. 
The (arbitrary) initial profile is taken Gaussian and we adopt $f(u) \propto u^2$.
The color dots track constant-$u$ points and show the formation of the universal steep
shock wave profile.  Inset: $t-x$ coordinates of the constant-$u$ tracking points (characteristics). 
The velocity of each point only depends on its $u$ value. The crossing of the characteristics 
signals the formation of the multivalued steep shock-wave front.}
\end{figure}

The drift current is generally given by the expression $j_{drift}=ug(E)$. The
form of the function $g(E)$ is material-dependent, and here we envision
two limiting situations. In homogeneous conductors, we should have
simple ``Ohmic''  behavior as $g(E)\sim E$
while in granular materials, we expect exponential dependence due to
activated transport, corresponding to: $g(E)\sim\sinh\left(E/E_{0}\right)$, 
where $E_{0}$ is a parameter describing the activation process.

\begin{figure}[h!]
\centering{}\includegraphics[width=0.6\textwidth]{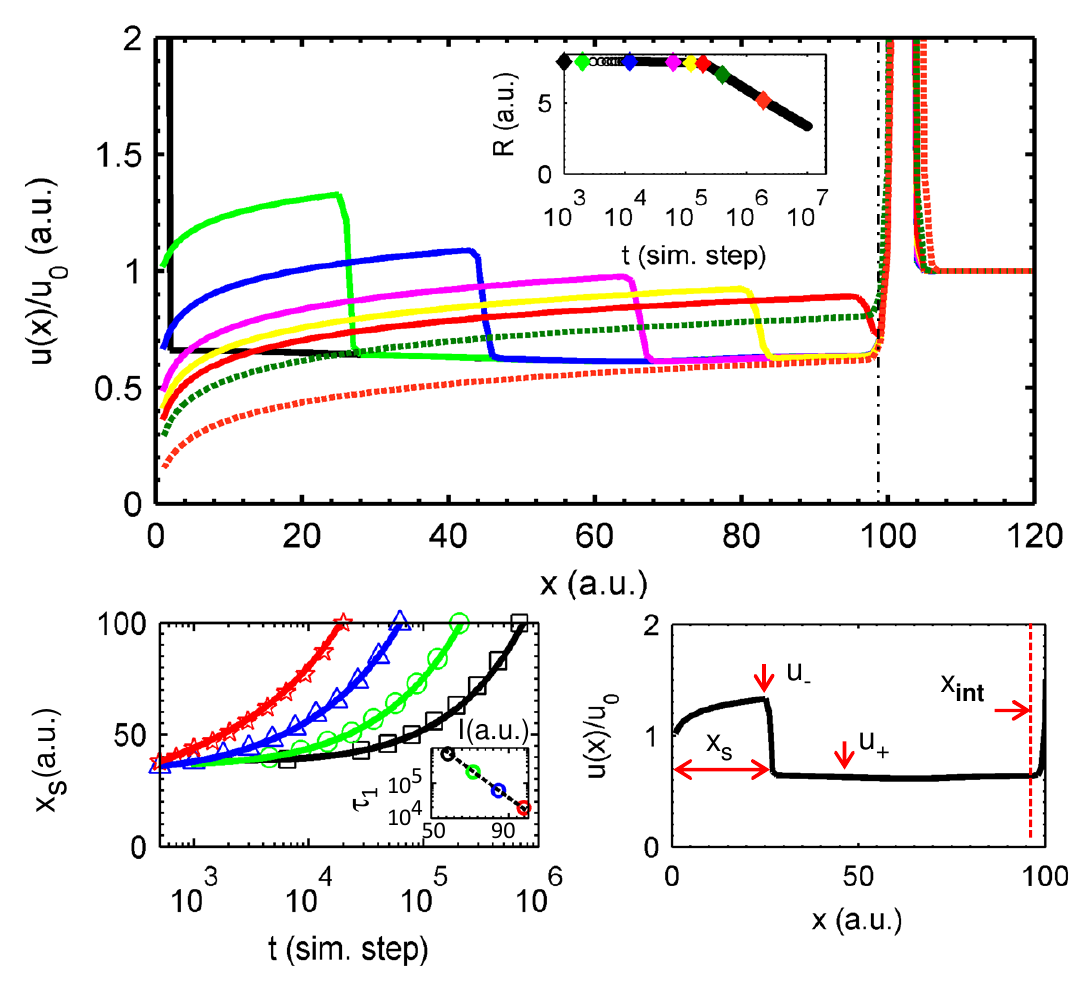}  

\caption{\label{fig:shock wave dynamics} Top panel: Snap shots of the time evolution of the density profile $u(t,x)$
 within the active interfacial region in a simulation of the VEOVM model (see Appendix~\ref{sec:VEOVMmod} and ~\ref{sec:simdet} for details). 
The current used is $I[a.u]$=71.5, with $A_{SB}$=1000 and $A_{B}$=1.   
The time steps of the successive profiles can be read-off from the corresponding color dots
in the inset. The initial state $u(0,x)$ (black line) exhibits a vacancy pile-up next to the electrode at $x$=1. 
The SB-bulk interface is denoted with a vertical dash-dot line at $x_{int}$=100.
The large accumulation of vacancies on the right of $x_{int}$ (bulk side) results from 
the initial ``forming'' cycles performed on an originally uniform profile of density $u_0$.
The result of the forming cycles is an approximately fixed background on top of which
the density profile further evolves (see details in \cite{Rozenberg2010}). 
Inset: Resistance of the device as a function of time. Color dots indicate the value of $R(t)$ 
at the corresponding snapshots of the main panel. 
Bottom-left panel: evolution of the shock wave front position $x_s(t)$ for different 
currents ($I[a.u]$=58.5, 71.5, 84.5 and 97.5). Dots are from numerical simulations and
the solid lines are analytic fits from integration of Eq.~(\ref{eq:shock wave velocity}).  
Inset: characteristic impact-time $\tau_{1}$ as a function of applied currents from the numerical simulation (circles) 
and analytic fit (dotted-line) in semi-log scale. Bottom-right panel: Shock wave parameters.  }
\end{figure}

Remarkably, these general ideas find an explicit realization 
in the context of RS in transition metal oxide memristors,
such as manganites \cite{chen,lee}.
In fact, their transport properties are very sensitively dependent on the oxygen stoichiometry,
i.e. on the concentration of oxygen vacancies [$V_O$].
Thus, it is now widely accepted that the mechanism of the bipolar (i.e. polarity dependent) RS in those
systems is due to the induced changes in the spatial distribution of 
$[V_O({\bf x})] \equiv u({\bf x})$ by means of externally applied strong electric stress \cite{inoue-sawa,MRS}. 
In particular, the accumulation of vacancies within highly resistive regions between the oxide and 
the metallic electrode, such as Schottky barrier (SB) interfaces,
greatly increases the (two-terminal) resistance across the device \cite{Rozenberg2010}. 
This accumulation can be achieved by applying strong voltage pulses across the device, 
leading to the high resistance state $R_{HI}$. Abrupt resistance switching from such  
high-resistance state to a significantly lower resistance state can be accomplished by 
reversing the voltage applied, which removes a significant fraction of vacancies 
from the SB region. The precise characterization of this resistance 
switching process is the main subject of this paper. 

We should mention that an important assumption is 
that the nonlinear drift term plays the dominant role as compared to the normal diffusion, i.e. we shall not
be concerned with the resistive changes involving thermal effects \cite{inoue-sawa,MRS}.
This restriction enables us to apply our analytical tools in a simple manner, allowing us to obtain a simplified mathematical description 
of the migration process, as we show in the following. 

\section{Model system}
For concreteness, we adopt the \emph{voltage-enhanced oxygen-vacancy migration model} \cite{Rozenberg2010} (VEOVM), which 
corresponds to granular materials with activated transport process and has been
previously used for manganite devices \cite{Rozenberg2010}.
Within the framework of this model, we shall perform numerical simulations to validate our shock-wave scenario. 
The VEOVM simply assumes that the local resistance of the cell at (discretized) position $x$ along the conductive path of the device
is simply given as a linear function of the local vacancy concentration, namely,
\begin{equation}
r(x)=A_{\alpha}u(x)
\label{resistance}
\end{equation}
with $\alpha={SB,B}$, where $SB$ denotes the highly resistive (Schottky barrier) region and $B$ the 
more conductive bulk \cite{chen}.  
The values of these constants are taken $A_{SB}\gg A_{B}=1$, which allows us to neglect the bulk resistance \cite{Rozenberg2010}.  
The discretized conducting path assumes the metal-electrode at $x$=0,
and $x=x_{int}$ denotes the point within the dielectric where the SB meets the bulk region.  
Under the action of the external stress (electric current $I$), the local fields at each cell position $x$ are computed at every discrete time step $t$. 
The field-driven migration of vacancies is simulated computing the local ionic migration rates from cell $x$ to $x$+$\Delta x$ as \cite{Rozenberg2010}
\begin{equation}
P(x,x+\Delta x) = u(x)[1- u(x+\Delta x)]\exp{\left(\frac{-V_0 + qIr(x)}{k_B T}\right)},
\label{model}
\end{equation}
where, for simplicity, we take the ionic charge $q$=1 and $k_B T$=1. The
parameter $V_0$ denotes the activation energy for ionic diffusion. The new profile $u(t,x)$ is updated from
the migration rates, and from (\ref{resistance}) 
we get the new total (two point) $R(t)$ as the discrete $x$-integral of the local cell's resistance $r(t,x)$. 
Here, for simplicity, we focus on a single active SB-bulk interface, while the more general situation with two barriers may be 
analyzed following a similar line of argument \cite{Rozenberg2010}. 
The applied external electric stress that we adopt is a constant current, in both, simulations and experiments (see below). 

As described in Ref.\cite{Rozenberg2010}, the initial vacancy concentration profile is assumed to be constant $[u(x)]=[u^0]$. 
The ``forming'' or initialization of the memory is done by first applying a few 
current loops of alternate polarity, $\pm I^0$, until the migration of vacancies evolves towards a {\em limit cycle},
with a well defined profile $u_0(x)$.
After this, the system begins to repetitively switch between two values: $R_{HI}$ and $R_{LO}$.
In the first, most of the vacancies reside within the high-resistance region SB, and in the second vacancies accumulate in the more conductive bulk. 
The $R_{HI}$ state with the vacancies piled up in the first cell, at $x$=1, defines the initial state for 
the shock wave propagation (see Fig.\ref{fig:shock wave dynamics}).

\section{Shock wave formation: the ``propagation phase''} 
We apply an external field with polarity pointing from the SB to the bulk and observe the evolution of the vacancy profile as a function of
the (simulation) time. As can be observed in Fig.\ref{fig:shock wave dynamics} there is a rapid evolution of the profile 
into a shock wave form with a sharply defined front. 
We also notice that the total resistance remains approximately constant during an initial phase,
and suddenly starts to decrease after the front hits the internal SB-bulk interface at $x_{int}$ (inset of 
top panel of Fig.\ref{fig:shock wave dynamics}).  
We shall analyze these key features in the following.

First, we focus on the propagation of the shock wave front position $x_s(t)$, as shown in Fig.\ref{fig:shock wave dynamics} (bottom-left panel) 
for different values of the electronic current $I$.  We observe that the characteristic time $\tau_1$  for the shock wave to travel through the
Schottky barrier and reach the SB-bulk point $x_{int}$ decreases exponentially with the magnitude of $I$. 
To obtain analytical insight for this behavior, we recall that the velocity of the shock wave front $dx_{s}/dt$ is very generally given by
the Rankine\textendash{}Hugoniot conditions \cite{debnath2011nonlinear,landau1987fluid}, which express it 
as the ratio of the spatial discontinuity of the (vacancy) drift current, and the spatial discontinuity of the density 
profile across the shock viz. $dx_{s}/dt=\Delta j/\Delta u\left|_{x_{s}}\right.$.
Within the VEOVM model \cite{Rozenberg2010}, we  obtain the following {\em nonlinear rate equation} 
 (see Appendix ~\ref{sec:DynofSW} for details), which describes the dynamics of the shock wave front:
\begin{eqnarray}
\frac{dx_{s}}{dt} & = & \frac{2Du_{-}\sinh\left(IA_{SB}u_{-}\right)-2Du_{+}\sinh\left(IA_{SB}u_{+}\right)}{\Delta u}, 
\label{eq:shock wave velocity}
\end{eqnarray}
where $D$ is a prefactor related to the activation energy for vacancy migration (Arrhenius factor) (see Eq.\ref{burgers} 
and Ref.\cite{Rozenberg2010}), 
and $u_{-/+}$ are the density of vacancies at the two sides of the shock wave front (see Fig.\ref{fig:shock wave dynamics}). 
The density $u_{-}$ depends on the shock wave front position via: $u_{-}= Q/x_{s} + u_{+} $, where $Q$ is the 
total number of vacancies carried by the shock wave, which remains a constant parameter
through the propagation phase ($t <\tau_1$) and $u_{+}$ is a constant 
background density which can be written as $u_{+}=Q_{B}/x_{int}$, $Q_{B}$ standing for the total number of background vacancies.  

Our description of the propagation phase is fully consistent with our numerical
simulations. As shown in the inset of Fig.\ref{fig:shock wave dynamics} (top panel), the resistance remains 
essentially constant until the wave front reaches the SB-bulk interface after a (current dependent) time $\tau_1$, 
and then begins to drop. Moreover, we also achieved a good fit to the shock front velocity
by using Eq.\ref{eq:shock wave velocity}, 
as is shown in Fig.\ref{fig:shock wave dynamics} (see Appendix ~\ref{sec:fitVEOVM} for details). 

\begin{figure}[h!]
\centering{}\includegraphics[width=0.6\textwidth]{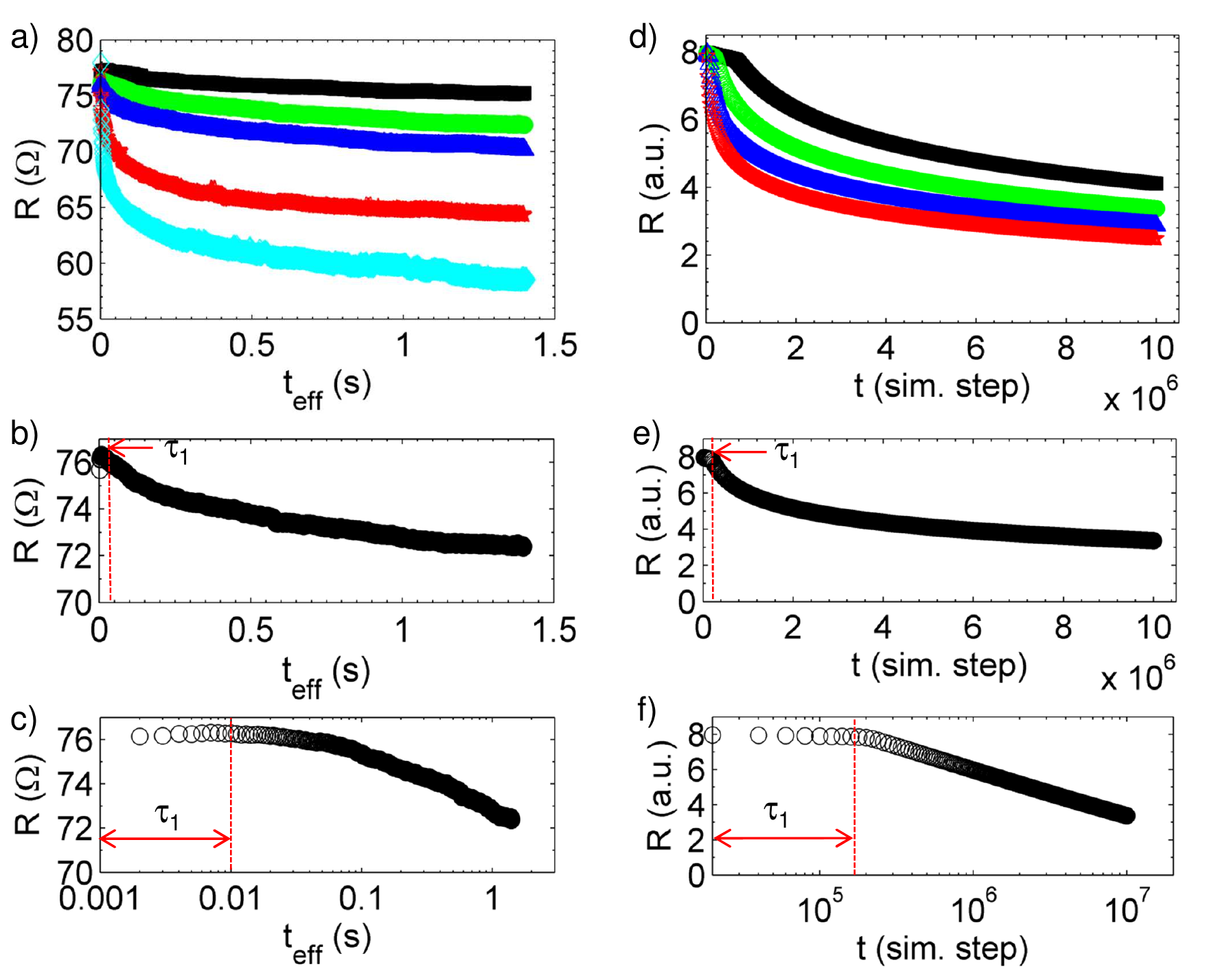}
\caption{ Time dependence of the resistive change $R(t)$ for various external current intensities.
Top left: experimental data measured on an Ag/LPCMO/Ag memristor with $I$ = 37.5 mA (black), 40 mA (green), 50 mA (blue), 
80 mA (red) and 100 mA (cyan).  $t_{eff}$ is the effective time duration of the applied currents (see Appendix ~\ref{sec:expdet} for details). 
The initial state was reset by applying an intense negative polarity current of 350 mA. Note that the initial value of the
resistance, 78 $\Omega$ is recovered within $\pm 1 \Omega$.
Top right: model simulations with applied current: $I[a.u.]$ = 58.5 (black), 71.5 (green), 84.5 (blue) and 97.5 (red).
Middle panels: experimental data for $I$ = 40 mA (left) and simulations for $I[a.u.]$ = 71.5 (right).
$\tau_1$ is defined as the time interval from the beginning of the applied pulse until the resistance starts to drop.
Bottom panels: Idem as middle panels in a semilog scale.}
\label{fig.R(t)}
\end{figure}

\section{Resistance switching: the ``leakage phase''}
After the shock front reaches the boundary point $x_{int}$ the resistance begin to drop. To understand this 
behaviour, we note that the total resistance of the Schottky barrier is given by the {\em  total number 
of vacancies} within the barrier region viz. from (\ref{resistance}) $\ensuremath{R_{SB}=\int_{SB}dxA_{SB}u(x)}$. As a result, the 
resistance drop per unit time is approximately given by the ionic vacancy-current passing through the SB-bulk 
interface at $x_{int}$,
\begin{eqnarray}
dR(t)/dt = -A_{SB}j(x=x_{int})
\end{eqnarray}
since $R_{SB} \gg R_B$ as $A_{SB} \gg A_B$.
Notice that during the propagation phase the ionic current through the interface $x_{int}$ is negligibly small. 
This is because the initial vacancy concentration there, and hence the local field, are also negligibly small. 
However, when the shock wave front eventually reaches 
the end of the SB region, after travelling for a time $\tau_{1}$, we do expect a sudden resistance 
drop as a large number of ionic vacancies begin to leak out into the bulk region.

We shall now focus on the detailed description of  the resistive drop. In Fig.\ref{fig.R(t)} we show 
the systematic dependence of $R(t)$ as a function of the applied external (electronic) current.
Along with the simulations of the VEOVM, we also present our experimental results measured on 
a manganite-based (La$_{0.325}$Pr$_{0.300}$Ca$_{0.375}$MnO$_3$) memristive device \cite{Quintero2007,Jin,Scherff,Park,Herpers} . The experimental setup and device are described in detail in Appendix~\ref{sec:expdet} and in \cite{FGM,PhysRevX.6.011028}.
The set of curves were obtained for applied current intensities just above the threshold for the onset of the resistance switch. 
The goal here was not to demonstrate the fast switching speed of the device, 
but rather on the contrary, achieve
relatively slow switching speeds in order to access the different time scales. 
In addition, this also minimizes thermal heating effects \cite{FGM}. 
We observe that in both, simulation 
and experiments, the resistance-change rapidly becomes larger and faster with 
the increase of the applied electric stress
intensity.
We also observe an overall good qualitative agreement between experiments and model simulations. This 
is also highlighted by the semi-log plots, which clearly display the two-stage process 
involved in the resistive switch, before and after the impact time $\tau_1$.

Remarkably, within shock wave scenario, we may also obtain explicit expressions that quantify 
the resistance change during the leakage phase. Our analysis may be simplified by first noting, 
from general considerations of shock waves, that their shape at long times becomes "flat", i.e. 
the gradient of the local density rapidly decreases ($\partial_{x}u\rightarrow0$) at all points 
that were overtaken by the shock wave front\cite{debnath2011nonlinear,courant1962methods}.  
Indeed, our data is fully consistent with this 
observation, as the  vacancy density profile within the SB remains approximately "flat" (i.e. spatially constant $u(t,x) = u_{SB}(t)$) 
at all times after the shock front reaches the interface (see Fig.\ref{fig:shock wave dynamics}). 
Then, within the VEOVM the SB resistance is simply proportional to the total vacancy concentration within the barrier and we
have,  $R(t) \approx R_{SB}(t) = A_{SB}x_{int}u_{SB}(t)$. Since the electronic current $I$ is held fixed, 
the vacancy (i.e. ionic) current through the interface depends only on the vacancy concentration $u_{SB}$ 
(cf Eq.\ref{model}). Thus, within the VEOVM we obtain a {\em nonlinear rate equation}, 
describing the resistance drop during the "leakage phase":
\begin{eqnarray}
\frac{dR}{dt}  = -\frac{2DR}{x_{int}}\sinh\left(\frac{IR}{x_{int}}\right).\label{eq:R(t)}
\end{eqnarray}
Similarly as we showed before for the propagation phase, this equation may be validated by
a quantitative fit to the simulation results (see Appendix ~\ref{sec:fitVEOVM}). 
Note that due to the strong nonlinear form of this rate equation, the $R(t)$ response is significantly different 
from the simple exponential decay expected in the familiar linear case (e.g. in standard RC circuits). 
Therefore, within the short time scale associated with the initial fast drop of resistance
and where the RS is significant ($IR/x_{int}\gg1$), 
the present type of nonlinear system is dominated by the activated process and
the approximation $\sinh\left(IR/x_{int}\right)\approx\frac{1}{2}\exp\left(IR/x_{int}\right)$ is valid.
This enables the approximate solution of the Eq.\ref{eq:R(t)}. 
\begin{eqnarray}
R  =  R_{HI}-\frac{x_{int}}{I}\ln\left(1+\frac{t^{*}}{\tau_2\left(I\right)}\right),\label{eq:approximate solution}
\end{eqnarray}
where the time $t^{*}$ is measured from the ``impact'' time $\tau_{1}$ and
$\tau_2\left(I\right)= \frac{x_{int}^2}{DIR_{HI}} \exp\left(-IR_{HI}/x_{int}\right)$ (see Appendix ~\ref{sec:scaling}) 
is the current-dependent characteristic time for the resistance drop.

\section{Resistivity scaling}
An interesting consequence of Eq.\ref{eq:approximate solution} is that it suggests
the scaling behaviour of the curves $R(t^{*})$.  
In fact, one may define the normalized resistance drop 
$\delta R\left(t^{*}\right)=[R-R\left(\tau_{2}\right)]/[\left(R_{HI}-R\left(\tau_{2}\right)\right)]$ 
and see from Eq.\ref{eq:approximate solution} that obeys it the scaling form:
\begin{eqnarray}
\delta R\left(t^{*}\right)=1-\ln\left(1+t^{*}/\tau_{2}\right)/\ln\left(2\right), \label{eq:scaling}
\end{eqnarray}

\begin{figure}[h!]
\centering{}\includegraphics[width=0.6\textwidth]{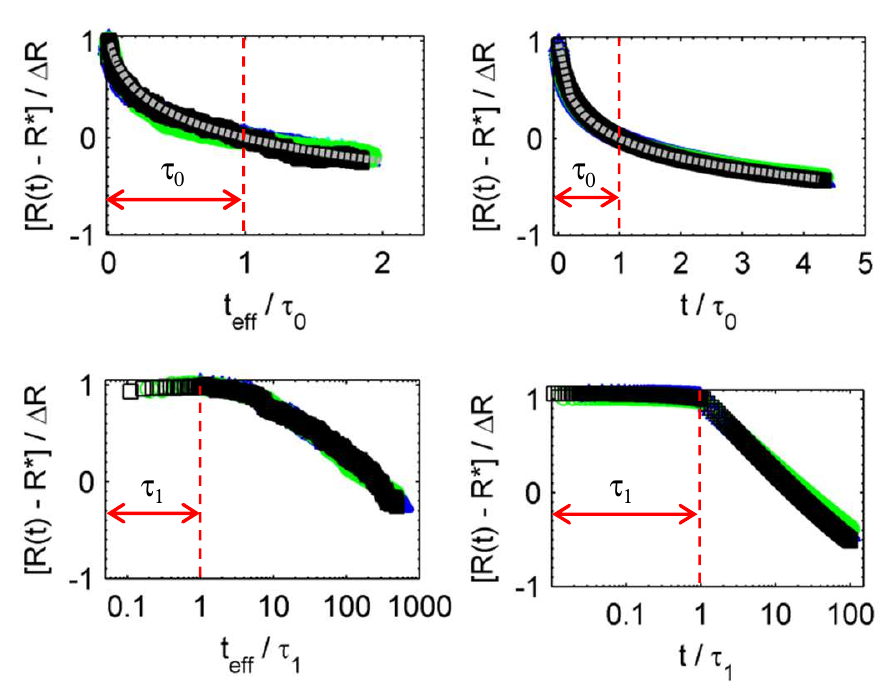}
\caption{ Scaled curves of the $R(t)$ data sets of Fig.\ref{fig.R(t)}.
The left panels show the collapsed experimental data and the right ones the numerical simulations. 
The time $\tau_0$ is an auxiliary scaling variable, which
is proportional to the characteristic time $\tau_2$ (see Appendix ~\ref{sec:scaling} for details on the
scaling procedure).
The scaled data were fitted (white dotted line) with a generalized version of Eq.\ref{eq:scaling}
(see Appendix ~\ref{sec:scaling}).
The lower panels show the same data sets scaled with the
shock wave impact time $\tau_1$ (the experimental curves show only three
data sets for the lower current values. At higher currents our electronics could not resolve $\tau_1$). 
To achieve the scaling of the lower panels, we assumed for each plot the normalization
value of $\Delta R$ determined from the previous scaling (top panels).
}
\label{scaling}
\end{figure}

In Fig.\ref{scaling} we demonstrate that this striking feature is indeed present 
in both, our experiments and simulations data.
In the upper panels of the figure we show the excellent scaling that is achieved, where all the 
experimental and the simulation curves $R(t)$ from Fig.\ref{fig.R(t)} were respectively collapsed 
onto a single one.
Moreover, the collapsed data can also be fitted with a slightly more general form of 
Eq.\ref{eq:scaling}, that we discuss in Appendix ~\ref{sec:scaling}.
Remarkably, in the lower panels of Fig.\ref{scaling} we show that a collapse of the data $R(t)$ can also be obtained
using the impact time $\tau_1$ as the scaling variable. This is significant, because it shows that a single scaling behavior
may include the two phases of the resistive switching process, namely before and after $\tau_1$.

We should mention that the scaling scenario was derived with the assumption of an ohmic behaviour 
in the $I$-$V$ characteristics. While this may not be the case in general \cite{FGM}, within the present set
of experiments, which are performed near the current threshold of RS, our results indicate that this is a reasonable 
assumption or at least a valid approximation.

\section{Conclusions}

To conclude, from quite general considerations of migration of ionic defects under strong electric fields
in solids, we have argued that the dynamics of the spatial profile of defect concentration should be governed
by a Burgers'-type nonlinear equation and develop shock waves.
We demonstrated that this scenario is indeed realized within a concrete realization, namely
a ionic migration model that was previously applied to describe resistive switching phenomena in manganite 
based memristive devices. In those systems, a key role is played by the migration of oxygen vacancies,
which are the ionic defects relevant to the electronic transport properties.
We thus predicted a two-stage process for the resistive switch phenomenon. An initial one, where the
oxygen-vacancy concentration profile develops a shock wave that propagates throughout a highly 
resistive (Schottky barrier) region near the electrode. During this phase the resistance essentially
does not change. 
This is followed by a second phase, where the shock wave emerges from the high resistive region and 
the ionic defects leak into the conductive bulk.
Our scenario was further validated by novel experimental data on a manganite based memristor device.
A remarkable results of our study is that both, the numerical simulations and the experimental curves,
obeyed a scaling behaviour, providing decisive support to our theory.
The present work provides novel insights on the physical mechanism behind the commutation speed on 
novel non-volatile electronic memories,  unveiling an unexpected connection between a phenomenon
of technological relevance and a classic theme of nonlinear dynamical systems. Furthermore, 
and from a practical point of view, our study unveils some of the key physical parameters that control the 
resistive switching properties of devices. Of foremost importance are the characteristic time scales, $\tau_1$ 
and $\tau_2$. Interestingly, we find that both depend on on the same combination of 
parameters, namely,  $x_{int}^2/(DIR_{HI})$ and $x_{int}/(IR_{HI})$, which appears 
in the exponential factor (see Eq.~\ref{eq:approximate solution} , Eq.~\ref{eq:tau2} 
and ~\ref{eq_tau1_scal} in Appendix ~\ref{sec:scaling}). While we may intuitively understand that a shorter interface or a stronger current 
pulse would increase the commutation speed, this expression also provides concrete guidance for the 
engineering effort of devices, which is less evident. For instance, it shows that shortening the length 
of the interface by a factor of three should at least gain a full order magnitude in speed. It also points 
towards adopting materials where the diffusion constant of oxygen vacancies is large. While this is a 
priori fixed by the specific chemistry of the material, one may also envision that the diffusion might 
be boosted by preparing samples with smaller grain sizes and with extended columnar defects that 
could be induced by heavy ion irradiation. 

A final important point to consider regards the general validity of the present results for
other type of resistive switching systems, such as for instance those based on 
HfO$_x$,TaO$_x$ and TiO$_{2-x}$, which are currently receiving a great deal of attention. 
Those compounds, in their stoichiometric form, are, unlike PCMO, very good insulators. Thus, in order
to make RS devices one possibility is to perform an initial "electroforming" procedure, which creates
metallic filaments that may subsequently be "burned" and regenerated, irrespective to the
applied polarity of the electric pulses. That mechanism is not relevant for the present study.
There is, however, another possibility to make RS devices with these compounds, which
is to heavily dope very thin films ( $\sim$ tens of nm) with oxygen vacancies \cite{ghenzi2015}. 
In this case, the systems become more conductive than their respective stoichiometric compound and display
a qualitatively different type of resistive switching under electric pulsing, which is bipolar and
relevant here. 
In fact, in these systems the mechanism relays on oxygen ionic migration under the strong electric fields,
thus it is interesting to note that the shock wave formation may be at work, shaping the evolution of the 
spatial profile of oxygen density.
Indeed, a local increase of oxygen concentration brings these systems (locally) closer to their respective 
stoichiometric composition, thus it locally increases the resistivity. Hence we realize that
for such systems one may fulfill the conditions for onset of shock wave propagation. In fact, an analogous
role to that of oxygen-vacancy density in PCMO (that renders the system more resistive), may be played
in the simpler binary compounds by the oxygen density  (see Appendix ~\ref{sec:binaryox} for details and the simulation results of \cite{Strukov}).
It would  be very interesting indeed to explore the fast commuting behavior in thin films of TaO$_x$, HfO$_x$ 
and related compounds, to search for evidence of shock waves propagation. 
These are exciting directions that are now open for future investigations.

\chapter{Conclusions and Outlook}

\textit{\emph{In chapters $2$ and $3$, }}we have found analytically
an inextricable link between the healing of gap and charge disturbances
in strongly correlated superconductors, suggesting that this phenomenon
is generic to any system close to Mott localization.\textit{\emph{
Furthermore, we have shown how the weak-coupling AG theory of $T_{c}$
suppression and normal state resistivity by dilute non-magnetic impurities
is modified in a strongly correlated metal.}} Given its simplicity,
we suggest that this phenomenon is generic to other systems close
to Mott localization. An important experimental test of these findings
would be provided by STM studies\textit{\emph{ on correlated superconductors
with impurities. Perhaps it is also interesting to check how the normal
state conductivity and the superconducting transition temperature
$T_{c}$ varies with the impurity concentration at different doping
levels.}}

Consider systems with pairing symmetry other than the $d$-wave. We
may check their robustness against non-magnetic disorders via similar
techniques. Another interesting direction is to study the correlated
systems where impurities are not dilute.

In chapter $4$, we show the important role of defects in the correlated
system for practical applications. Due to generality of the theory,
we may study the resistive switching phenomena based on binary oxides,
which are currently receiving a great deal of attention.  When heavily
doped with oxygen vacancies, the systems become more conductive than
their respective stoichiometric compound and display the bipolar type
of resistive switching under electric pulsing.  

In fact, in these systems the mechanism relays on oxygen ionic migration
under the strong electric fields, thus it is interesting to note that
the shock wave formation may be at work, shaping the evolution of
the  spatial profile of oxygen density. It would  be very interesting
indeed to explore the fast commuting behavior in thin films of $\textrm{TaO\ensuremath{_{x}}}$
, $\textrm{HfO\ensuremath{_{x}}}$  and related compounds, to search
for evidence of shock waves propagation.  These are exciting directions
that are now open for future investigations.

As our results are obtained through the classical phenomenological
model, it is crucial to establish underlying microscopic guidelines.
Furthermore, we may predict whether the candidate material is good
for fabricating fast switching devices under certain situations. For
example, as some of the materials are correlated bad metal, we may
apply similar slave boson methodology discussed previously and try
to capture the relevant physics.

\renewcommand*{\appendixtocname}{Appendices}
\appendix

\chapter{The Linear Approximation}

\label{sec:linearappox}

\section{$U(1)$ slave boson approach}

Although many results have been obtained within the weak coupling
framework, only few controlled methods for strongly correlated system
are established. Among these theoretical tools, we pick the slave-boson
method as our main tool, which can be treated in mean-field level
\cite{nagaosa1999quantum,nagaosa1999quantums} with Gaussian fluctuations.
There are several representations of the slave boson theory \cite{baskaran1987resonating,PhysRevLett.95.057001,kotliar1986new}
and the key idea is to introduce auxiliary bosons. Certain constraints
are further imposed in order to force the matrix elements of the Hamiltonian
equal in the original and the enlarged Hilbert space. The physics
contained in this method is equivalent to that included in the simple
Gutzwiller approximation. The main advantage of the slave boson method
is the inequality constraint, namely, the electron occupancy is less
than one (away from half filling) can be treated easily in the path
integral formalism with one Lagrange multiplier $\lambda_{i}$ on
each site. The functional integral over $\lambda_{i}$ would reproduce
the delta function of the constraint on each site. However, we ignore
the imaginary time dependence of $\lambda_{i}$ and the constraint
is obeyed in average sense. At mean-field level, we absorb its saddle
point value to the chemical potential and neglect the thermal fluctuations
of the $\lambda$ field which is a gauge field \cite{wen2004quantum,ioffe1989gapless}.

Different representations are suitable for different problems. Kotliar's
representation \cite{kotliar1986new} makes the Hubbard model simple
to analyze, while in the large $U$ limit, Coleman's representation
\cite{coleman1984new} is much more convenient since it has already
projected out the double occupancy. In the following discussion, we
would focus on this so called $U(1)$ slave boson theory and introduce
one bosonic field which is related directly to the hole doping density.
While the $U(1)$ theory has a clearer relation between auxiliary
field and physical quantity, the system we are interested in has the
$SU(2)$ symmetry, especially in studying the $d$ wave superconducting
state. Although the phase diagram obtained by $U(1)$ or $SU(2)$
slave boson theory does not fully agree with the experimental results,
they do capture important features of the strongly correlated Fermi
liquid state.

In the large $U$ limit of the standard Hubbard model, we can obtain
the effective $t-J$ Hamiltonian which can is used to describe doped
antiferromagnets \cite{RevModPhys.63.1} and the doped Mott insulator.
In this model, we have the following Hamiltonian with $n_{i}\equiv\sum_{\sigma}c_{i\sigma}^{\dagger}c_{i\sigma}$:

\[
H=P_{s}TP_{s}+J\sum_{ij}\mathbf{S}_{i}\cdot\mathbf{S}_{j}+\sum_{i}(\varepsilon_{i}-\mu_{0})n_{i},
\]
where $T$ denotes the kinetic energy $-t\sum_{ij\sigma}c_{i\sigma}^{\dagger}c_{j\sigma}+h.c.$
with $h.c.$ as the abbreviation for ``hermitian conjugate'', and
$P_{s}$ is the projection operator for projecting out the double
occupant states. We then obtain the explicit form \cite{altland2010condensed}:

\[
P_{s}TP_{s}=-t\sum_{ij\sigma}(1-n_{i\bar{\sigma}})c_{i\sigma}^{\dagger}c_{j\sigma}(1-n_{j\bar{\sigma}})+h.c..
\]
Below, we discuss how various quantities transform within the slave
boson approach. According to \cite{lee2006doping}, we have

\begin{equation}
c_{i\sigma}^{\dagger}=f_{i\sigma}^{\dagger}b_{i}+\epsilon_{\sigma\sigma^{,}}f_{i\sigma^{,}}d_{i}^{\dagger},\label{eq:operator 1}
\end{equation}
and the identity:

\begin{equation}
\sum_{\sigma}f_{i\sigma}^{\dagger}f_{i\sigma}+b_{i}^{\dagger}b_{i}+d_{i}^{\dagger}d_{i}=\mathbf{1}.\label{eq:operator 2}
\end{equation}
In the above equations, $f_{i\sigma}$ are the auxiliary fermions
and $b,d$ are the auxiliary bosons. Combine Eq.(\ref{eq:operator 1})
and Eq.(\ref{eq:operator 2}), we can easily recover the anti-commutation
algebra among $c$ operators. After projecting out the double occupant
states, we obtain:

\begin{equation}
c_{i\sigma}^{\dagger}=f_{i\sigma}^{\dagger}b_{i}\label{eq:operator 3}
\end{equation}

\begin{equation}
\sum_{\sigma}f_{i\sigma}^{\dagger}f_{i\sigma}+b_{i}^{\dagger}b_{i}=\mathbf{1}\label{eq:operator 4}
\end{equation}
However, Eq.(\ref{eq:operator 3}) is not an operator identity since
we combine it with the Eq.(\ref{eq:operator 4}), we can no longer
recover the anti-commutation relations among $c$ operators. Although
it is not an operator identity, we can safely use it in our Hamiltonian
as the matrix elements are all faithful. Alternatively, we can utilize
Eq.(\ref{eq:operator 1}) and then perform the projection, both routes
will provide us the same results. Comparing the original and the enlarged
Hilbert space, we have relations $c_{i\sigma}^{\dagger}c_{j\sigma}=f_{i\sigma}^{\dagger}f_{j\sigma}b_{i}^{\dagger}b_{j}$
and $c_{i\sigma}^{\dagger}c_{i\sigma}=f_{i\sigma}^{\dagger}f_{i\sigma}$.

In the mean-field level, we set $\lambda_{i}(\tau)=\lambda$ with
auxiliary fields $\chi$ and $\Delta$ as order parameters in order
to decouple the four fermion terms, i.e. $J\sum_{ij}\mathbf{S}_{i}\cdot\mathbf{S}_{j}$.
As we require the saddle point approximation with Gaussian fluctuations
around it to reproduce the mean-field theory, cautions should be paid
to choose the parameters in the Lagrangian. In the following discussion,
we set $\widetilde{J}=\frac{3}{8}J$ according to the fact we are
using the multi-component Hubbard Stratonovich transformation \cite{lee2006doping}.

In the homogenous state, we have $\left\langle b_{i}^{\dagger}b_{j}\right\rangle =1-n=x$
as a parameter keeping track of the hole density. The mean-field conditions
as $\chi_{ij}=\sum_{\sigma}\left\langle f_{i\sigma}^{\dagger}f_{j\sigma}\right\rangle $
and $\Delta_{ij}=\left\langle f_{i\uparrow}f_{j\downarrow}-f_{i\downarrow}f_{j\uparrow}\right\rangle $
can be derived from $\frac{\delta Z}{\delta\chi_{ij}^{*}}=\frac{\delta Z}{\delta\Delta_{ij}^{*}}=0$.
The partition function can be expressed in the path integral formalism
as follows,

\[
Z=\int Df^{\dagger}DfDb^{*}DbD\lambda D\chi D\Delta\exp(-S),
\]
with action $S=\int_{0}^{\beta}\mathcal{L}(\tau)d\tau$ and the Lagrangian
in imaginary time formalism obtained from the Hamiltonian:

\begin{eqnarray*}
\mathcal{L}(\tau) & = & \widetilde{J}\sum_{ij}(\left|\chi_{ij}\right|^{2}+\left|\Delta_{ij}\right|^{2})+\sum_{i\sigma}f_{i\sigma}^{\dagger}\partial_{\tau}f_{i\sigma}-\widetilde{J}[\sum_{ij}\chi_{ij}^{*}(\sum_{\sigma}f_{i\sigma}^{\dagger}f_{j\sigma})+h.c.]\\
 & + & \widetilde{J}[\sum_{ij}\Delta_{ij}(f_{i\uparrow}^{\dagger}f_{j\downarrow}^{\dagger}-f_{i\downarrow}^{\dagger}f_{j\uparrow}^{\dagger})+h.c.]-[\sum_{ij\sigma}t_{ij}b_{i}b_{j}^{*}f_{i\sigma}^{\dagger}f_{j\sigma}+h.c.]+\sum_{i}b_{i}^{*}\partial_{\tau}b_{i}\\
 & + & \sum_{i}\lambda_{i}(\sum_{\sigma}f_{i\sigma}^{\dagger}f_{i\sigma}+b_{i}^{*}b_{i}-1)+\sum_{i\sigma}(\varepsilon_{i}-\mu_{f})f_{i\sigma}^{\dagger}f_{i\sigma},
\end{eqnarray*}
where $\mu_{f}=\mu-d\widetilde{J}$ with $d$ as the dimension of
the lattice. Below, we use the conventional parametrization of order
parameters described in many contexts \cite{mahan2000many,ubbens1992flux}.
Based on the Matsubara frequency representation we have: $f(\tau)=\frac{1}{\sqrt{\beta}}\sum_{n}f(\omega_{n})e^{-i\omega_{n}\tau}$.
In radial gauge, $b_{i}^{*}=b_{i}=r_{i}$ which indicates that we
only consider the amplitude of the bosonic field. Below, we first
obtain the effective action by integrating out the fermion field,
then derive the free energy and mean-field equations as follows \cite{nagaosa1999quantum,nagaosa1999quantums}:

\begin{eqnarray*}
S & = & \beta\widetilde{J}\sum_{ij}(\left|\chi_{ij}\right|^{2}+\left|\Delta_{ij}\right|^{2})+\beta\sum_{i}\lambda_{i}(r_{i}^{2}-1)\\
 &  & +\sum_{n\sigma}\sum_{ij}f_{i\sigma}^{\dagger}(\omega_{n})(-i\omega_{n}+\varepsilon_{i}-\mu_{f}+\lambda_{i})\delta_{ij}f_{j\sigma}(\omega_{n})\\
 &  & -\sum_{n}\sum_{ij\sigma}f_{i\sigma}^{\dagger}(\omega_{n})(t_{ij}+t_{ji})r_{i}r_{j}f_{j\sigma}(\omega_{n})\\
 &  & +\widetilde{J}\sum_{n}\sum_{ij}\left[\begin{array}{cc}
f_{i\uparrow}^{\dagger}(\omega_{n}) & f_{i\downarrow}(-\omega_{n})\end{array}\right]\left[\begin{array}{cc}
-\chi_{ij}^{*} & \Delta_{ij}\\
\Delta_{ij}^{*} & \chi_{ij}
\end{array}\right]\left[\begin{array}{c}
f_{j\uparrow}(\omega_{n})\\
f_{j\downarrow}^{\dagger}(-\omega_{n})
\end{array}\right]\\
 &  & +\widetilde{J}\sum_{n}\sum_{ij}\left[\begin{array}{cc}
f_{i\downarrow}^{\dagger}(-\omega_{n}) & -f_{i\uparrow}(\omega_{n})\end{array}\right]\left[\begin{array}{cc}
-\chi_{ij}^{*} & \Delta_{ij}\\
\Delta_{ij}^{*} & \chi_{ij}
\end{array}\right]\left[\begin{array}{c}
f_{j\downarrow}(-\omega_{n})\\
-f_{j\uparrow}^{\dagger}(\omega_{n})
\end{array}\right],
\end{eqnarray*}

\begin{eqnarray*}
S & = & \beta\widetilde{J}\sum_{ij}(\left|\chi_{ij}\right|^{2}+\left|\Delta_{ij}\right|^{2})+\beta\sum_{i}\lambda_{i}(r_{i}^{2}-1)\\
 &  & +\sum_{n\sigma}\sum_{ij}f_{i\sigma}^{\dagger}(\omega_{n})(-i\omega_{n}+\varepsilon_{i}-\mu_{f}+\lambda_{i})\delta_{ij}f_{j\sigma}(\omega_{n})\\
 &  & -\sum_{n}\sum_{ij\sigma}f_{i\sigma}^{\dagger}(\omega_{n})(t_{ij}+t_{ji})r_{i}r_{j}f_{j\sigma}(\omega_{n})\\
 &  & +\widetilde{J}\sum_{n}\sum_{ij}\left[\begin{array}{cc}
f_{i\uparrow}^{\dagger}(\omega_{n}) & f_{i\downarrow}(-\omega_{n})\end{array}\right]\left[\begin{array}{cc}
-\chi_{ij}^{*}-\chi_{ji} & 2\Delta_{ij}\\
2\Delta_{ij}^{*} & \chi_{ij}+\chi_{ji}^{*}
\end{array}\right]\left[\begin{array}{c}
f_{j\uparrow}(\omega_{n})\\
f_{j\downarrow}^{\dagger}(-\omega_{n})
\end{array}\right]\\
 & = & \beta\widetilde{J}\sum_{ij}(\left|\chi_{ij}\right|^{2}+\left|\Delta_{ij}\right|^{2})+\beta\sum_{i}\lambda_{i}(r_{i}^{2}-1)\\
 &  & +\sum_{n\sigma}\sum_{ij}f_{i\sigma}^{\dagger}(\omega_{n})(-i\omega_{n}+\varepsilon_{i}-\mu_{f}+\lambda_{i})\delta_{ij}f_{j\sigma}(\omega_{n})\\
 &  & +\sum_{n}\sum_{ij}\left[\begin{array}{cc}
f_{i\uparrow}^{\dagger}(\omega_{n}) & f_{i\downarrow}(-\omega_{n})\end{array}\right]\left[\begin{array}{cc}
-2\widetilde{J}\chi_{ij}^{*}-(t_{ij}+t_{ji})r_{i}r_{j} & 2\widetilde{J}\Delta_{ij}\\
2\widetilde{J}\Delta_{ij}^{*} & 2\widetilde{J}\chi_{ij}+(t_{ij}+t_{ji})r_{i}r_{j}
\end{array}\right]\left[\begin{array}{c}
f_{j\uparrow}(\omega_{n})\\
f_{j\downarrow}^{\dagger}(-\omega_{n})
\end{array}\right],
\end{eqnarray*}

\begin{eqnarray*}
S & = & \beta\widetilde{J}\sum_{ij}(\left|\chi_{ij}\right|^{2}+\left|\Delta_{ij}\right|^{2})+\beta\sum_{i}\lambda_{i}(r_{i}^{2}-1)\\
 &  & -\sum_{n}\sum_{ij}\left[\begin{array}{cc}
f_{i\uparrow}^{\dagger}(\omega_{n}) & f_{i\downarrow}(-\omega_{n})\end{array}\right]\left[\begin{array}{cc}
[G_{ij}^{-1}]_{11} & [G_{ij}^{-1}]_{12}\\{}
[G_{ij}^{-1}]_{21} & [G_{ij}^{-1}]_{22}
\end{array}\right]\left[\begin{array}{c}
f_{j\uparrow}(\omega_{n})\\
f_{j\downarrow}^{\dagger}(-\omega_{n})
\end{array}\right].
\end{eqnarray*}
In order to derive the mean-field equations, we only consider the
imaginary time dependence of the fermion field which has been integrated
over. Ignoring the imaginary time dependence of the slave boson, Lagrange
multiplier and order parameters, we obtain the free energy as follows:

\[
F=\widetilde{J}\sum_{ij}(\left|\chi_{ij}\right|^{2}+\left|\Delta_{ij}\right|^{2})+\sum_{i}\lambda_{i}(r_{i}^{2}-1)-kT\sum_{n}\textrm{tr}\ln-G^{-1},
\]
which then leads to the mean-field equations presented in the main
text.

\section{Linearized equations}

Our linear approximation approach consists of expanding the mean-field
equations (\ref{eq:chieq}-\ref{eq:sum rule}) of the main text to
first order in the site energies $\varepsilon_{i}$. Denoting linear
deviations in the various fields by $\delta$ we get

\begin{eqnarray}
\delta\chi_{ij} & = & 2kT\sum_{nl}\left(-g_{il}g_{lj}+G_{1il}G_{1lj}\right)\left(\delta\lambda_{l}+\varepsilon_{l}\right)+2kTr\sum_{nlm}\left(-g_{il}g_{mj}+G_{1il}G_{1mj}\right)\left(\delta r_{l}h_{lm}+\delta r_{m}h_{lm}\right)\nonumber \\
 &  & -2kT\sum_{nlm}\left(-g_{il}g_{mj}+G_{1il}G_{1mj}\right)\left(\widetilde{J}\delta\chi_{lm}\right)+2kT\sum_{nlm}\left(g_{il}G_{1mj}+g_{mj}G_{1il}\right)\left(\widetilde{J}\delta\Delta_{lm}\right),\label{eq:deltachi}
\end{eqnarray}

\begin{eqnarray}
\delta\Delta_{ij} & = & -2kT\sum_{nl}\left(g_{il}G_{1lj}+g_{lj}G_{1il}\right)\left(\delta\lambda_{l}+\varepsilon_{l}\right)-2kTr\sum_{nlm}\left(g_{il}G_{1mj}+g_{mj}G_{1il}\right)\left(\delta r_{l}h_{lm}+\delta r_{m}h_{lm}\right)\nonumber \\
 &  & +2kT\sum_{nlm}\left(g_{il}G_{1mj}+g_{mj}G_{1il}\right)\left(\widetilde{J}\delta\chi_{lm}\right)-2kT\sum_{nlm}\left(G_{1il}G_{2mj}+g_{il}g_{mj}\right)\left(\widetilde{J}\delta\Delta_{lm}\right),\label{eq:deltadelta}
\end{eqnarray}

\begin{eqnarray}
-r\delta r_{i} & = & kT\sum_{nl}\left(-g_{il}g_{li}+G_{1il}G_{1li}\right)\left(\delta\lambda_{l}+\varepsilon_{l}\right)+kTr\sum_{nlm}\left(-g_{il}g_{mi}+G_{1il}G_{1mi}\right)\left(\delta r_{l}h_{lm}+\delta r_{m}h_{lm}\right)\nonumber \\
 &  & -kT\sum_{nlm}\left(-g_{il}g_{mi}+G_{1il}G_{1mi}\right)\left(\widetilde{J}\delta\chi_{lm}\right)+kT\sum_{nlm}\left(g_{il}G_{1mi}+g_{mi}G_{1il}\right)\left(\widetilde{J}\delta\Delta_{lm}\right),\label{eq:deltar}
\end{eqnarray}

\begin{eqnarray}
\lambda\delta r_{i}+r\delta\lambda_{i}+\sum_{l}h_{il}\chi_{il}\delta r_{l}+r\sum_{l}h_{il}\delta\chi_{il} & = & 0,\label{eq:deltalambda}
\end{eqnarray}
where $G_{1ij}\equiv[G_{ij}]_{11}$, $G_{2ij}\equiv[G_{ij}]_{22}$,
$g_{ij}\equiv[G_{ij}]_{12}=[G_{ij}]_{21}$ are the Green's functions
of the clean system, $n$ is the fermionic Matsubara frequency index
and $\widetilde{J}=\frac{3}{8}J$. The latter choice is made, in the
presence of correlations, so that the multi-channel Hubbard-Stratonovich
transformation we used reproduces, at the saddle-point level, the
mean-field results \cite{lee2006doping}\footnote{The usual choice $\widetilde{J}=\frac{1}{4}J$ does not change the
analytical results and would give rise to hardly noticeable changes
in the numerical plots.}. In general, the clean Green's functions in $\mathbf{k}$-space are
given by 
\begin{equation}
G_{1}(\omega_{n},\mathbf{k})=\frac{i\omega_{n}+e(\mathbf{k})}{(i\omega_{n})^{2}-e^{2}(\mathbf{k})-\widetilde{J}^{2}\Delta^{2}(\mathbf{k})},\label{eq:g1}
\end{equation}

\begin{equation}
G_{2}(\omega_{n},\mathbf{k})=\frac{i\omega_{n}-e(\mathbf{k})}{(i\omega_{n})^{2}-e^{2}(\mathbf{k})-\widetilde{J}^{2}\Delta^{2}(\mathbf{k})},\label{eq:g2}
\end{equation}

\begin{equation}
g(\omega_{n},\mathbf{k})=\frac{\widetilde{J}\Delta(\mathbf{k})}{(i\omega_{n})^{2}-e^{2}(\mathbf{k})-\widetilde{J}^{2}\Delta^{2}(\mathbf{k})},\label{eq:gsmall}
\end{equation}
where the renormalized dispersion is

\begin{eqnarray}
e(\mathbf{k}) & = & -2\left(xt+\chi\widetilde{J}\right)\left[\cos\left(k_{x}a\right)+\cos\left(k_{y}a\right)\right]-4xt^{\prime}\cos\left(k_{x}a\right)\cos\left(k_{y}a\right)-\mu,\label{eq:disp}
\end{eqnarray}
we have absorbed the clean $\lambda$ in the chemical potential, and
\begin{equation}
\Delta\left(\mathbf{k}\right)=2\Delta_{0}\left[\cos\left(k_{x}a\right)-\cos\left(k_{y}a\right)\right].\label{eq:gap}
\end{equation}
Notice that the dimensionful gap function is $\Delta_{phys}\left(\mathbf{k}\right)=\widetilde{J}\Delta\left(\mathbf{k}\right)$.
As we focus on the asymptotic long-range behavior of the different
fields, their variations are dominated by the corresponding clean-limit
symmetry channel. We therefore define local order parameters as $\delta\chi_{i}\equiv\frac{1}{2d}\sum_{j}\delta\chi_{ij}\Gamma(s)_{ij}$,
$\delta\Delta_{i}\equiv\frac{1}{2d}\sum_{j}\delta\Delta_{ij}\Gamma(d_{x^{2}-y^{2}})_{ij}$.
Thus, \emph{defining vectors and matrices in the lattice site basis
with bold-face letters}, Eqs.~(\ref{eq:deltachi}-\ref{eq:deltalambda})
can be recast as
\begin{equation}
\left(\boldsymbol{A}+r^{2}\boldsymbol{B}\right)\delta\boldsymbol{\Phi}=r^{2}\boldsymbol{C},\label{eq:matrixequations}
\end{equation}
where 
\begin{equation}
\boldsymbol{A}=\left(\begin{array}{cccc}
\boldsymbol{M}_{11} & \boldsymbol{M}_{12} & \boldsymbol{M}_{13} & \boldsymbol{M}_{14}\\
\boldsymbol{M}_{21} & \boldsymbol{M}_{22} & \boldsymbol{M}_{23} & \boldsymbol{M}_{24}\\
\boldsymbol{M}_{31} & \boldsymbol{M}_{32} & \boldsymbol{M}_{33} & \boldsymbol{M}_{34}\\
0 & 0 & \lambda\mathbf{1}-\frac{\lambda}{2d}\boldsymbol{\Gamma}(s) & 0
\end{array}\right),\boldsymbol{B}=\left(\begin{array}{cccc}
0 & 0 & 0 & 0\\
0 & 0 & 0 & 0\\
0 & 0 & 0 & 0\\
-2dt\mathbf{1} & 0 & 0 & \mathbf{1}
\end{array}\right),\delta\boldsymbol{\Phi}=\left(\begin{array}{c}
\delta\boldsymbol{\chi}\\
\delta\boldsymbol{\Delta}\\
r\delta\boldsymbol{r}\\
\delta\boldsymbol{\overline{\lambda}}
\end{array}\right),\boldsymbol{C}=\left(\begin{array}{c}
0\\
0\\
0\\
\boldsymbol{\varepsilon}
\end{array}\right).\label{eq:matrices}
\end{equation}
Here, the elements of (the vector) $\boldsymbol{\varepsilon}$ are
the disorder potential values $\varepsilon_{i}$, $\mathbf{1}$ is
the identity matrix, $\delta\overline{\lambda}_{i}=\delta\lambda_{i}+\varepsilon_{i}$,
and

\begin{equation}
M_{11ij}=-\delta_{ij}-\frac{\widetilde{J}kT}{d}\sum_{nml}\Gamma(s)_{il}\left(-g_{ij}g_{ml}+G_{1ij}G_{1ml}\right)\Gamma(s)_{jm}\label{eq:m11}
\end{equation}

\begin{eqnarray}
M_{12ij} & = & \frac{\widetilde{J}kT}{d}\sum_{nml}\Gamma(s)_{il}\left(g_{ij}G_{1ml}+g_{ml}G_{1ij}\right)\Gamma(d_{x^{2}-y^{2}})_{jm}
\end{eqnarray}

\begin{eqnarray}
M_{13ij} & = & \frac{kT}{d}\sum_{nml}\Gamma(s)_{il}\left(-g_{ij}g_{ml}+G_{1ij}G_{1ml}-g_{im}g_{jl}+G_{1im}G_{1jl}\right)h_{jm}
\end{eqnarray}

\begin{equation}
M_{14ij}=\frac{kT}{d}\sum_{nl}\Gamma(s)_{il}\left(-g_{ij}g_{jl}+G_{1ij}G_{1jl}\right)
\end{equation}

\begin{eqnarray}
M_{21ij} & = & -\frac{\widetilde{J}kT}{d}\sum_{nml}\Gamma(d_{x^{2}-y^{2}})_{il}\left(g_{ij}G_{1ml}+g_{ml}G_{1ij}\right)\Gamma(s)_{jm}
\end{eqnarray}

\begin{eqnarray}
M_{22ij} & =\delta_{ij}+ & \frac{\widetilde{J}kT}{d}\sum_{nml}\Gamma(d_{x^{2}-y^{2}})_{il}\left(G_{1ij}G_{2ml}+g_{ij}g_{ml}\right)\Gamma(d_{x^{2}-y^{2}})_{jm}
\end{eqnarray}

\begin{eqnarray}
M_{23ij} & = & \frac{kT}{d}\sum_{nml}\Gamma(d_{x^{2}-y^{2}})_{il}\left(g_{ij}G_{1ml}+g_{ml}G_{1ij}+g_{im}G_{1jl}+g_{jl}G_{1im}\right)h_{jm}
\end{eqnarray}

\begin{equation}
M_{24ij}=\frac{kT}{d}\sum_{nl}\Gamma(d_{x^{2}-y^{2}})_{il}\left(g_{ij}G_{1jl}+g_{jl}G_{1ij}\right)
\end{equation}

\begin{eqnarray}
M_{31ij} & = & -\widetilde{J}kT\sum_{nm}\left(-g_{ij}g_{mi}+G_{1ij}G_{1mi}\right)\Gamma(s)_{jm}
\end{eqnarray}

\begin{equation}
M_{32ij}=\widetilde{J}kT\sum_{nm}\left(g_{ij}G_{1mi}+g_{mi}G_{1ij}\right)\Gamma(d_{x^{2}-y^{2}})_{jm}
\end{equation}

\begin{eqnarray}
M_{33ij} & = & \delta_{ij}+kT\sum_{nm}h_{jm}\left(-g_{ij}g_{mi}+G_{1ij}G_{1mi}-g_{im}g_{ji}+G_{1im}G_{1ji}\right)
\end{eqnarray}

\begin{equation}
M_{34ij}=kT\sum_{n}\left(-g_{ij}g_{ji}+G_{1ij}G_{1ji}\right)\label{eq:m34}
\end{equation}

In writing Eqs.~(\ref{eq:matrixequations}), we have made explicit
the $r$ dependence of Eqs.~(\ref{eq:deltachi}-\ref{eq:deltalambda}).
We note, however, that there is also an implicit dependence on $r$
through the dispersion (\ref{eq:disp}) (where $x=r^{2}$), which
enters the various Green's functions in Eqs.~(\ref{eq:g1}-\ref{eq:gsmall}).

Since the matrix elements in Eqs.~(\ref{eq:m11}-\ref{eq:m34}) are
all calculated in the \emph{translation-invariant clean system}, Eqs.~(\ref{eq:matrixequations})
can be easily solved in $\mathbf{k}$-space by matrix inversion. Normal
state results are obtained by removing the second row and column and
setting $\Delta\left(\mathbf{k}\right)$ to zero. Non-correlated results
correspond to the absence of slave bosons and constraints, so we just
remove the third and fourth rows and columns and set $x=1$ and $\lambda_{i}=0$.
In every case, the clean limit is first solved self-consistently for
$\chi$, $\Delta$, $\lambda$ and $\mu$, and then the fluctuations
in the presence of impurities are obtained.

In discussing the solution to Eqs.~(\ref{eq:matrixequations}), we
rely on the fact that all quantities in Eqs.~(\ref{eq:m11}-\ref{eq:m34})
are non-singular and finite as $x\to0$. Thus, we can write their
formal solution as
\begin{equation}
\delta\boldsymbol{\Phi}=r^{2}\left(\boldsymbol{A}+r^{2}\boldsymbol{B}\right)^{-1}\boldsymbol{C}=r^{2}\boldsymbol{A}^{-1}\boldsymbol{C}+{\cal O}\left(r^{4}\right).\label{eq:solution}
\end{equation}
\emph{It follows that $\delta\chi_{i}$, $\delta\Delta_{i}$, $r\delta r_{i}$,
and $\delta\overline{\lambda}_{i}=\delta\lambda_{i}+\varepsilon_{i}$
are all of order $r^{2}=x$}

\chapter{The Irrelevance of Spinon Fluctuations and the ``Minimal Model''}

\label{sec:irrelevancespinon}

We can shed light on the strong healing effect by studying a simplified
case obtained by ``turning off'' the $\delta\chi_{i}$ fluctuations.
In this case, we need to solve the smaller set of equations
\begin{equation}
\left(\begin{array}{ccc}
\boldsymbol{M}_{22} & \boldsymbol{M}_{23} & \boldsymbol{M}_{24}\\
\boldsymbol{M}_{32} & \boldsymbol{M}_{33} & \boldsymbol{M}_{34}\\
0 & \lambda\mathbf{1}-\frac{\lambda}{2d}\boldsymbol{\Gamma}(s) & r^{2}\mathbf{1}
\end{array}\right)\left(\begin{array}{c}
\delta\boldsymbol{\Delta}\\
r\delta\boldsymbol{r}\\
\delta\overline{\boldsymbol{\lambda}}
\end{array}\right)=\left(\begin{array}{c}
0\\
0\\
r^{2}\boldsymbol{\varepsilon}
\end{array}\right).\label{eq:SCnochi}
\end{equation}
The healing factor obtained in this simplified model is almost identical
to the full solution, as shown by the red and green curves of the
left panel of Fig.~2 of the main text. This shows that the spinon
field fluctuations are utterly irrelevant for the strong healing.

A further fruitful simplification is obtained by setting $M_{32}$
to zero in Eqs.~(\ref{eq:SCnochi}). This defines what we called
the ``minimal model'' (MM). In this case, the ``strong-correlation
sub-block'' of $\delta r_{i}$ and $\delta\overline{\lambda}_{i}$
fluctuations decouples and suffers no feed-back from the gap fluctuations.
In fact, the MM corresponds to breaking up the solution to the problem
into two parts: (i) the spatially fluctuating strong correlation fields
$r_{i}$ and $\lambda_{i}$ are first calculated for \emph{fixed,
uniform} $\Delta$ and $\chi$, and then (ii) the effects of their
spatial readjustments are fed back into the gap equation to find $\delta\Delta_{i}$. 

\begin{figure}
\begin{centering}
\includegraphics[scale=1.3]{figures/0\lyxdot 2_powerspectrum}
\par\end{centering}

\centering{}\includegraphics[scale=1.3]{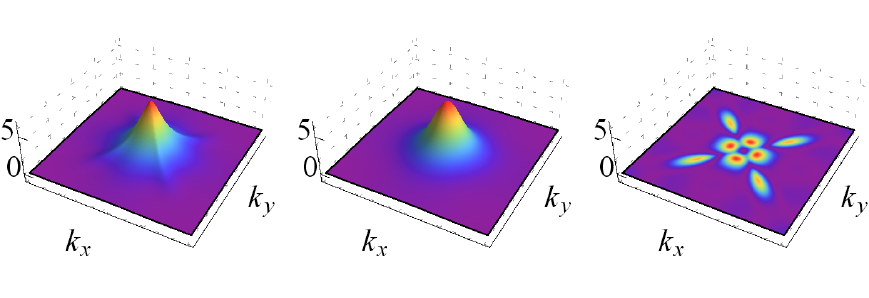}\caption{\label{fig:gapflucMM}Power spectra of gap fluctuations $S(\mathbf{k})$,
$S(\mathbf{k})_{loc}$ and $S(\mathbf{k})_{nonloc}$ (first to third
columns) for $x=0.2$ in the presence of correlations. The top figures
were obtained from the full solution of the linearized Eqs.~(\ref{eq:matrixequations}),
whereas the bottom ones correspond to the minimal model (Eqs.~(\ref{eq:gapflucminimal})).
Note that the healing factors are $h=0.74\%$ (top) and $h=0.69\%$
(bottom).}
\end{figure}

Strikingly, the healing factor in this case is \emph{numerically indistinguishable}
from the one obtained from Eqs.~(\ref{eq:SCnochi}) (green curve
of the left panel of Fig.~2 of the main text). Furthermore, the full,
local and non-local PS of gap fluctuations are also captured quite
accurately by the MM, as seen in Fig.~\ref{fig:gapflucMM}. We conclude
that the MM, which incorporates only the effects of strong correlations,
is able to describe with very high accuracy the healing process in
the $d$-wave SC state.

The MM also permits us to obtain simple and physically transparent
expressions. In particular, it follows immediately that 
\begin{equation}
r\delta r\left(\mathbf{k}\right)=\frac{r^{2}}{\lambda a\left(\mathbf{k}\right)-r^{2}M_{33}\left(\mathbf{k}\right)/M_{34}\left(\mathbf{k}\right)}\varepsilon\left(\mathbf{k}\right),\label{eq:rdeltar}
\end{equation}
where $a\left(\mathbf{k}\right)=1-\Gamma_{s}\left(\mathbf{k}\right)/4$,
and we used the Fourier transform of $\boldsymbol{\Gamma}(s)$, $\Gamma_{s}\left(\mathbf{k}\right)=2\left[\cos\left(k_{x}a\right)+\cos\left(k_{y}a\right)\right]$.
Moreover, 
\begin{eqnarray}
\delta\Delta\left(\mathbf{k}\right) & = & \frac{r^{2}\left[M_{24}\left(\mathbf{k}\right)M_{33}\left(\mathbf{k}\right)-M_{23}\left(\mathbf{k}\right)M_{34}\left(\mathbf{k}\right)\right]}{M_{22}\left(\mathbf{k}\right)\left[\lambda a\left(\mathbf{k}\right)M_{34}\left(\mathbf{k}\right)-r^{2}M_{33}\left(\mathbf{k}\right)\right]}\varepsilon\left(\mathbf{k}\right),\label{eq:gapflucminimal}\\
 & = & \frac{\left[M_{24}\left(\mathbf{k}\right)\frac{M_{33}\left(\mathbf{k}\right)}{M_{34}\left(\mathbf{k}\right)}-M_{23}\left(\mathbf{k}\right)\right]}{M_{22}\left(\mathbf{k}\right)}r\delta r\left(\mathbf{k}\right).\label{eq:gapflucminimal2}\\
 & = & \chi_{pc}^{MM}\left(\mathbf{k}\right)\delta n\left(\mathbf{k}\right),\label{eq:gapflucminimal3}
\end{eqnarray}
where we used $n_{i}=1-r_{i}^{2}\Rightarrow\delta n_{i}=-2r\delta r_{i}$,
and 
\begin{equation}
\chi_{pc}^{MM}\left(\mathbf{k}\right)=-\frac{\left[M_{24}\left(\mathbf{k}\right)\frac{M_{33}\left(\mathbf{k}\right)}{M_{34}\left(\mathbf{k}\right)}-M_{23}\left(\mathbf{k}\right)\right]}{2M_{22}\left(\mathbf{k}\right)}.\label{eq:chipcMM}
\end{equation}
The local part of the response, which we have shown to be the dominant
one, can be studied by looking at the long wavelength limit. As $k\to0$,
$a\left(\mathbf{k}\right)\sim k^{2}/4$ and 
\begin{equation}
\delta\Delta_{loc}\left(\mathbf{k}\right)\approx-\chi_{pc}^{MM}\left(\mathbf{k}=0\right)\frac{8r^{2}/\lambda}{k^{2}+\xi_{S}^{-2}}\varepsilon\left(\mathbf{k}\right),\label{eq:gapfluclorentzian}
\end{equation}
where
\begin{equation}
\frac{1}{\xi_{S}}=\sqrt{-\frac{4r^{2}}{\lambda}\frac{M_{33}\left(\mathbf{k}=0\right)}{M_{34}\left(\mathbf{k}=0\right)}}.\label{eq:csiSMM}
\end{equation}
Eqs.~(\ref{eq:chipcMM}) and (\ref{eq:csiSMM}) give us the expressions
for the pair-charge correlation function and the healing length within
the MM.

\chapter{The Normal State and the ``Minimal Model''}

\label{sec:normalstate}

\begin{figure*}[t]
\begin{centering}
\includegraphics[scale=0.7]{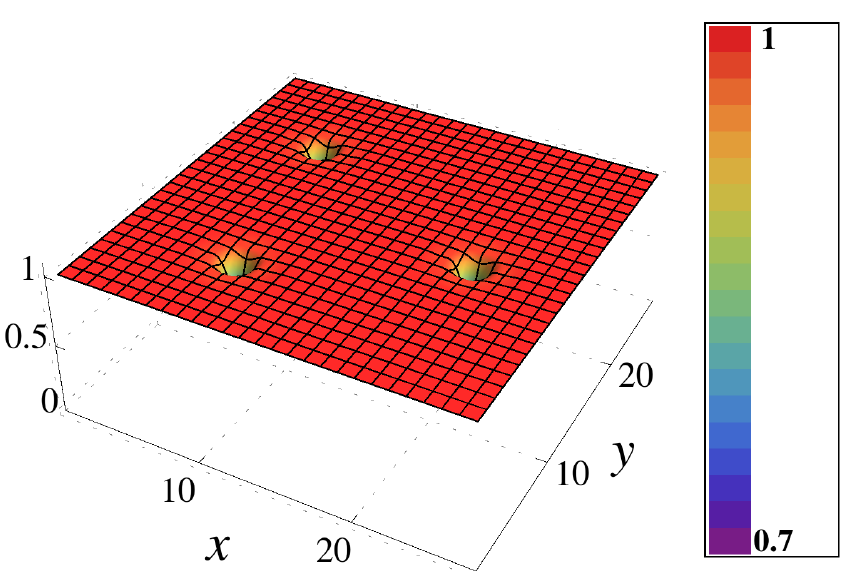}\includegraphics[scale=0.9]{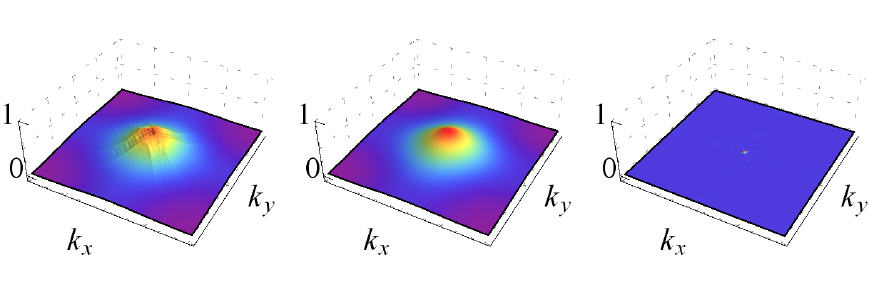}
\par\end{centering}

\centering{}\includegraphics[scale=0.7]{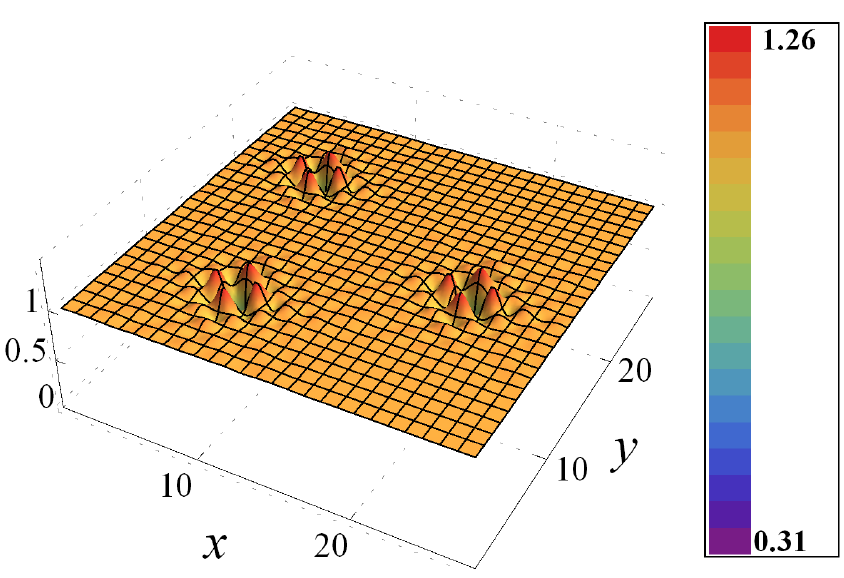}\includegraphics[scale=0.9]{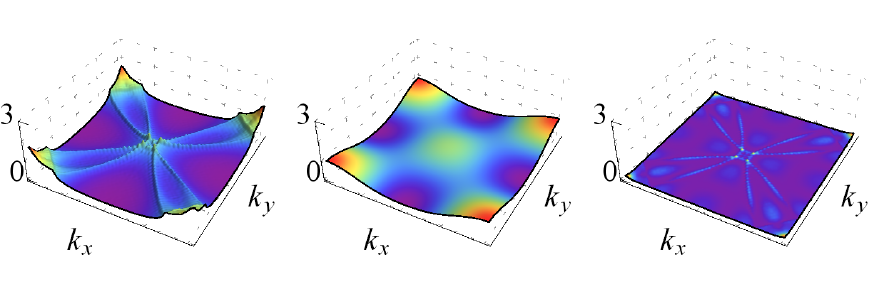}\caption{\label{fig:densityfluct-1}Spatial variations of normalized local
density $\frac{\delta n_{i}}{n_{0}}$ in the normal state for three
impurities (first column) and the corresponding power spectra $N(\mathbf{k})$,
$N(\mathbf{k})_{loc}$ and $N(\mathbf{k})_{nonloc}$ (second to fourth
columns), in the presence (top) and in the absence (bottom) of strong
correlations for $x=0.2$. The strong suppression of density oscillations
by correlations is accompanied by the dominance of the spherically
symmetric local power spectrum {[}$N_{loc}\left(\mathbf{k}\right)${]}
over the anisotropic non-local one {[}$N_{nonloc}\left(\mathbf{k}\right)${]}.}
\end{figure*}

It is instructive to analyze also the behavior of the charge fluctuations
in the normal state. This can be achieved by suppressing the second
row and column of Eqs.~(\ref{eq:matrixequations}) and setting $\Delta\left(\mathbf{k}\right)$,
and thus $g\left(i\omega_{n},\mathbf{k}\right)$, to zero. Even after
these simplifications, the full solution is long and cumbersome. However,
accurate insight can be gained from a MM of the normal state, in which
we also set the $\delta\chi_{i}$ to zero by hand. As before, the
strong-correlation sub-block decouples and Eq.~(\ref{eq:rdeltar})
is still valid (albeit with matrix elements calculated in the normal
state). The local part of the charge response is given by an expression
similar to Eq.~(\ref{eq:gapfluclorentzian})
\begin{equation}
\delta n_{loc}\left(\mathbf{k}\right)\approx-\frac{8r^{2}/\lambda}{k^{2}+\xi_{N}^{-2}}\varepsilon\left(\mathbf{k}\right),\label{eq:chargefluclorentzian}
\end{equation}
where $\xi_{N}$ is given by Eq.~(\ref{eq:csiSMM}), again with matrix
elements calculated in the normal state. The behavior of $\xi_{N}$
as a function of doping is shown by the green curve of the right panel
of Fig.~2 of the main text.

In addition, just like in the Coulomb gas, the density fluctuations
also show Friedel-like oscillations coming from the singularity in
the response function at $2k_{F}$. Thus, expanding Eq.~(\ref{eq:rdeltar})
in $r^{2}$, 
\begin{equation}
\delta n_{nonloc}\left(\left|\mathbf{k}\right|\approx2k_{F}\right)\approx-\frac{2r^{2}}{\lambda a\left(\left|\mathbf{k}\right|\approx2k_{F}\right)}\left[1+\frac{r^{2}M_{33}\left(\left|\mathbf{k}\right|\approx2k_{F}\right)/M_{34}\left(\left|\mathbf{k}\right|\approx2k_{F}\right)}{\lambda a\left(\left|\mathbf{k}\right|\approx2k_{F}\right)}\right]\varepsilon\left(\mathbf{k}\right).\label{eq:chargeflucnonlocal}
\end{equation}
Since 
\begin{eqnarray}
M_{34}\left(\mathbf{k}\right) & = & \mathbf{\Pi}\left(\mathbf{k}\right),\\
M_{33}\left(\mathbf{k}\right) & = & 1+\mathbf{\Pi}^{b}\left(\mathbf{k}\right),\\
\mathbf{\Pi}\left(\mathbf{k}\right) & = & \frac{1}{V}\sum_{\mathbf{q}}\frac{f\left[\widetilde{h}\left(\mathbf{q}+\mathbf{k}\right)\right]-f\left[\widetilde{h}\left(\mathbf{q}\right)\right]}{\widetilde{h}\left(\mathbf{q}+\mathbf{k}\right)-\widetilde{h}\left(\mathbf{q}\right)},\\
\mathbf{\Pi}^{b}\left(\mathbf{k}\right) & = & \frac{1}{V}\sum_{\mathbf{q}}\frac{f\left[\widetilde{h}\left(\mathbf{q}+\mathbf{k}\right)\right]-f\left[\widetilde{h}\left(\mathbf{q}\right)\right]}{\widetilde{h}\left(\mathbf{q}+\mathbf{k}\right)-\widetilde{h}\left(\mathbf{q}\right)}\left[h\left(\mathbf{q}+\mathbf{k}\right)+h\left(\mathbf{q}\right)\right],\\
h\left(\mathbf{k}\right) & = & -t\Gamma_{s}\left(\mathbf{k}\right)-4t^{\prime}\cos\left(k_{x}a\right)\cos\left(k_{y}a\right),
\end{eqnarray}
the leading divergent behavior is 
\begin{equation}
\frac{M_{33}\left(\left|\mathbf{k}\right|\approx2k_{F}\right)}{M_{34}\left(\left|\mathbf{k}\right|\approx2k_{F}\right)}\approx\frac{1}{\mathbf{\Pi}\left(\left|\mathbf{k}\right|\approx2k_{F}\right)}.
\end{equation}
The two contributions from Eqs.~(\ref{eq:chargefluclorentzian})
and (\ref{eq:chargeflucnonlocal}) together give, in real space, 
\begin{eqnarray}
\frac{\delta n_{i}}{n_{0}} & = & x\sum_{j}\left(c_{1}\frac{e^{-r_{ij}/\xi}}{\xi^{(d-3)/2}(r_{ij})^{(d-1)/2}}+c_{2}x\left[\boldsymbol{\Pi^{-1}}\right]_{ij}\right)\varepsilon_{j},\label{eq:chargepert}
\end{eqnarray}
where $r_{ij}$ is the distance between sites $i$ and $j$, and $c_{1}$
and $c_{2}$ are constants that depend on $t,t^{\prime},J$ and $x$.

We stress that in the full solution of the linearized equations in
which $\delta\chi_{i}\neq0$, the structure of Eq.~(\ref{eq:rdeltar})
is still preserved, with the factor $M_{33}/M_{34}$ being replaced
by a long combination of several $M_{ij}$ elements, which, however,
has a \emph{finite negative }$\mathbf{k}\to0$ \emph{limit and a singularity
at} $2k_{F}$. Therefore, the results of Eqs.~(\ref{eq:chargefluclorentzian}),
(\ref{eq:chargeflucnonlocal}) and (\ref{eq:chargepert}) remain valid
in the general case. The spatial charge fluctuations for three impurities
and the PS in the normal state in the full solution are shown in Fig.~\ref{fig:densityfluct-1}
both in the absence and in the presence of strong correlations. Note
how the non-local part is down by an additional factor of $x$ as
compared to the local part {[}see Eqs.~(\ref{eq:chargeflucnonlocal})
and (\ref{eq:chargepert}){]}.

\chapter{Shock Wave Properties}
\section{Method of characteristics} 
\label{sec:charact}

In this section, we demonstrate that shock waves are  typically formed in presence of 
sufficient nonlinearity, and analyze the problem via the \textit{method of characteristics}.
Considering the vacancy migration in highly resistive barriers, where we can neglect the 
normal diffusion term  $D\partial_{xx}u$, we obtain  a first order, 
%
%
partial differential equation of the general form 
\begin{equation}
\partial_{t}u+c(u,t)\partial_{x}u=0\label{Burger's type eq}  .
\end{equation}
 Here, the $u$ dependence in $c(u,t)$ gives rise to the nonlinearity effect for the onset of  shock wave formation. In order to determine the density profile evolution, we need to track the motion of points corresponding to specific values of $u$ (see Fig.\ref{fig:shock-sketch}, left panel).  We can solve the trajectories (the so-called "characteristics") followed by such point in the $t-x$ plane by solving the following 
 equation  \cite{debnath2011nonlinear,courant1962methods} \footnote{It's straightforward to
see that $u$ is constant along the characteristics, since $du/dt=dx/dt\partial_{x}u+\partial_{t}u=0$.}:
\begin{equation}
\frac{dx}{dt}=c(u,t)\label{eq:characteristics}   .
\end{equation}
As different characteristics might intersect with
each other in the $t-x$ plane (see the inset of Fig.\ref{fig:shock-sketch}), the emergence of intersecting 
trajectories indicates that the solution $u(t,x)$ becomes multivalued, since the points 
can be traced back along each of the characteristics to different initial values of $u(t=0,x)$. 
The intersection which happens chronologically first determines the formation of the shock wave, 
as shown in Fig.\ref{fig:shock-sketch} . As an illustration, we provide an example 
with $c(u,t)=u^{2}$ and an initial Gaussian distribution $u(t=0,x)=0.5\exp\left[-\left(x-5\right)^{2}/4\right]$.
The time dependent wave profile is solved numerically with fixed boundary
conditions, and characteristics are computed according to Eq.\ref{eq:characteristics}.

We should stress, however, that this qualitative behavior is a robust feature 
of any such nonlinear diffusion equation, provided that the prefactor 
$c(u,t)$ is a monotonically {\em increasing} function of $u$ (i.e. $\partial_{u}c(u,t)>0$). 
To see this, note that Eq. (\ref{Burger's type eq}) takes the form of a simple wave equation, 
thus its solution is a traveling wave, with speed locally proportional to $c(u,t)$. 
Therefore, at any given time, each point on the wave front
moves with a different speed proportional to $c(u,t)$ and the ``crest''
having the largest value of $u$ moves {\em fastest}, which leads to a ``kink''
type shock. Alternatively, it is easy to check that if $\partial_{u}c(u,t)\leq0$,
the shock wave will not form. In the case
where $c(u,t)$ is any function independent of $u$, the characteristics are curves
of  the same shape but parallel  to each other, and hence they
would not have any intersections; the shock wave would again not form. 
Finally, it can be shown  that if the diffusion 
term $D\partial_{xx}u$, is non-zero but is parametrically small (as compared to the drift term), 
the diffusion will prevent  the wave from topping over (to undergo "self-breaking"), and the shock wave front will remain sharp as it propagates \cite{debnath2011nonlinear,courant1962methods}. 

\section{Dynamics of shock waves }
\label{sec:DynofSW}

In this section, we provide a simple derivation of the Rankine\textendash{}Hugoniot
condition \cite{debnath2011nonlinear,landau1987fluid} which determines
the equation of motion of the shock wave front. An important
feature of the shock wave are the spatial discontinuities of $u$ and 
$j$. Considering a shock wave front propagating on the interval
$\left[0,d\right]$ in one dimension, we define $u_{+}\equiv u(t,x_{s}(t)+\epsilon)$,
$u_{-}\equiv u(t,x_{s}(t)-\epsilon)$, $j_{+}\equiv j(t,x_{s}(t)+\epsilon)$
and $j_{-}\equiv j(t,x_{s}(t)-\epsilon)$, where $x_{s}(t)$
is the coordinate of the shock wave front and $\epsilon\rightarrow0^{+}$
is a infinitesimal positive quantity. According to the continuity
equation $\partial_{t}u+\partial_{x}j=0$, we can write:

\begin{equation}
\frac{d}{dt}\left(\int_{0}^{x_{s}(t)}udx+\int_{x_{s}(t)}^{d}udx\right)=j(0)-j(d),
\end{equation}

\begin{eqnarray}
\frac{dx_{s}}{dt}u_{-}-\frac{dx_{s}}{dt}u_{+}+\int_{0}^{x_{s}(t)}\partial_{t}udx+\int_{x_{s}(t)}^{d}\partial_{t}udx & = & j(0)-j(d).
\end{eqnarray}
Using again the continuity equation, we can immediately determine the shock wave front velocity via the following equation:

\begin{eqnarray}
v_{s}=\frac{dx_{s}}{dt}  =\frac{j_{+}-j_{-}}{u_{+}-u_{-}}=  \frac{\Delta j}{\Delta u}\left|_{x_{s}}\right., 
\end{eqnarray}
which is often called Rankine\textendash{}Hugoniot condition.

\chapter{Analysis of Numerical and Experiment Results}
\section{VEOVM model }
\label{sec:VEOVMmod}

As it was mentioned in the main text, to simulate the vacancies dynamics we adopted \emph{the voltage enhanced oxygen vacancy migration model} (VEOVM), which is a well validated model for the RS effect \cite{Rozenberg2010,ghenzi2010hysteresis}. 
Taking a capacitor-like device, this model considers a 1 dimensional conductive channel connecting the two contacts, along which oxygen vacancies can migrate through (see Fig. \ref{fig:Fig-Sim}). This channel is divided in small cells corresponding to physical nano-domains, and at the same time the whole device is divided in two regions: the active interfacial region close to the metal contact (i.e. high resistive Schottky Barrier) and the high conductive bulk region. A diagram of the model with a single active contact, as used in the simulations, is presented at the top panel on Fig.\ref{fig:Fig-Sim}. The resistance of each cell in the channel $r(x)$ is proportional to the local density of vacancies: $r(x)=u(x)A_{\alpha}$, where $u(x)$ is the density of vacancies on the cell located at $x$ and $A_{\alpha}$ is a proportionality constant whose magnitude depends on the region of the device: $\alpha=SB,B$, where $SB$ stands for Schottky barrier and $B$ for bulk, 
and where $A_{SB}\gg A_{B}$.
Migration is assumed to occur only between neighbours cells, and in each step of the simulation the probability for vacancy migration from a cell at $x$ to its neighbour at $x+\Delta x$ is computed according to:

\begin{equation}
P(x, x+\Delta x)=u(x)(1-u(x+\Delta x))\exp(\Delta V(x)-V_0),
\label{Prob}
\end{equation}

where $\Delta V(x)$ is the drop of voltage (per unit length) at $x$ and $V_0$ is an activation constant for vacancy migration. In each step of the simulation the migration from cell to cell is computed and the new resistance of each cell is calculated. The total resistance of the device is calculated simply as the addition of all the cells in the channel. For the values used in the simulations, the middle term $(1-u(x+\Delta x))$ can and will be neglected in the theoretical analysis explained in the following sections.
As mentioned in the main text, this model corresponds to the case of granular materials with activated transport process. To put it in the context of the $generalized-Burgers' equation$ description made in main text, it is easy to see that the drift current originated from this model satisfies the general form $j_{drift}\sim \sinh\left(E\right)$. The net current of vacancies generated by the action of an external electric field $E(x)=\partial V(x)/\partial x$, between a cell at $x$ and its neighbour cell at $x+\Delta x$  can be written simply as Fick's-law for the migration probability: $j_{drift}= \partial P/\partial x \approx [P(x,x+\Delta x) -P(x+\Delta x,x)]/\Delta x$ , where in the last term the discrete character of the model has been introduced. In particular for our model we take $\Delta x=1$. Using the definition of $P$ from the model (neglecting its middle term), we get from here that $j_{drift}=P(x, x+\Delta x) - P( x+\Delta x,x) = 2Du\sinh\left(\Delta V(x)\right) $, where $D$ stands for the Arrhenius factor $\exp(-V_0)$. If the external electric stress is a controlled current $I$, then $\Delta V(x)=Ir(x)=IA_{\alpha}u(x)$. 

\section{Simulations details}
\label{sec:simdet}

\begin{figure}[h]
  \centering{}\includegraphics[width=1.0\textwidth]{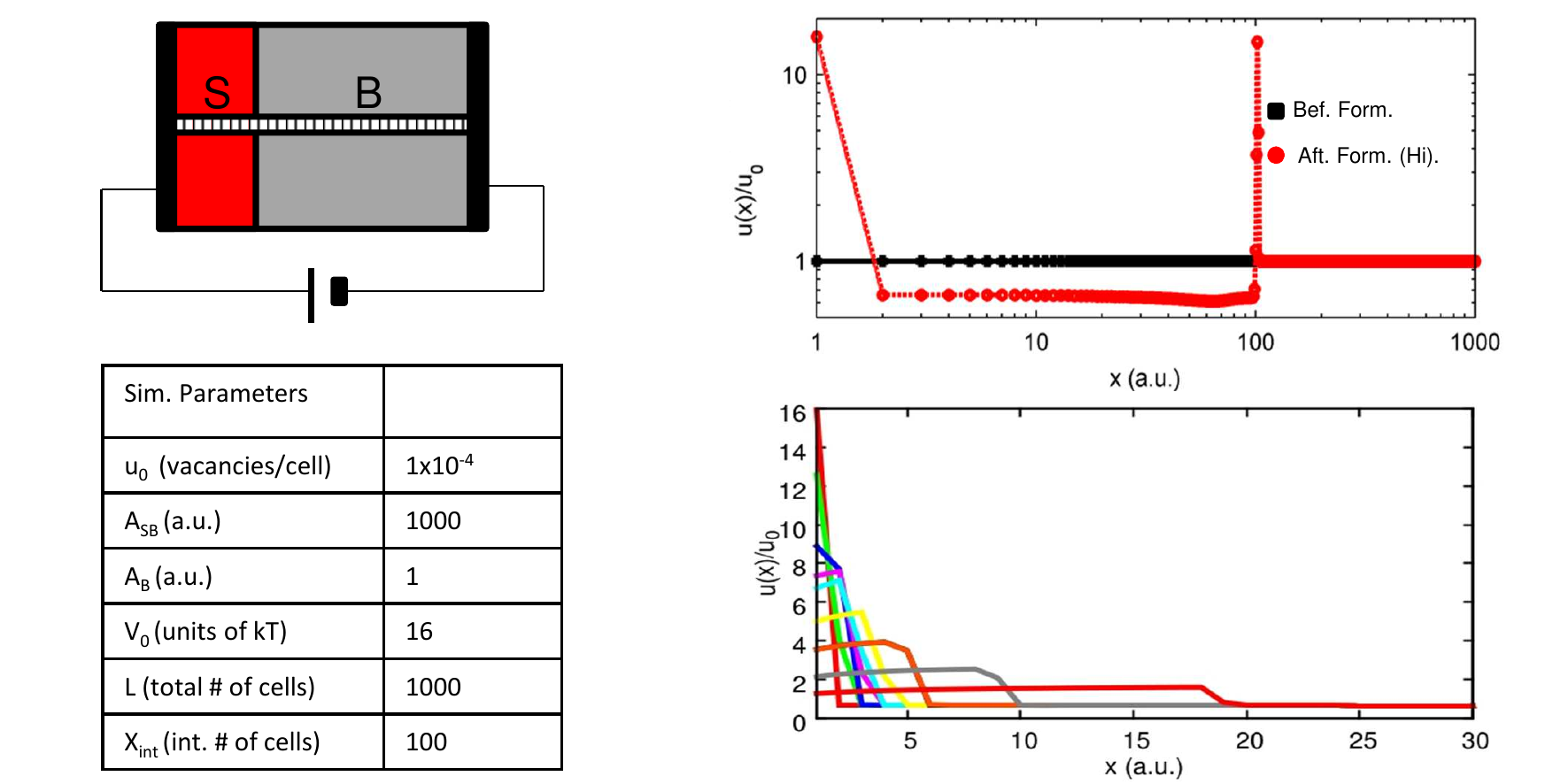}
  \caption{ Top panels: top-left: Schematic diagram of the VEOVM model with a single active contact. The two regions SB and B correspond to the high resistance interface (Schottky Barrier) and the more conductive central bulk, respectively. The small cells within the channel indicate the domains. Top-right: vacancy distribution along the conductive channel before (black squares) and after (red circles) the forming process. The distribution shown after the forming process correspond to a Hi resistance state. Starting from a uniform distribution, during the forming process a characteristic distribution is obtained where vacancies initially at the interface migrates to the low-field bulk region generating a pile-up of vacancies close to the limit between the two regions. In the Hi state, part of these vacancies migrates back to the interfacial region and are accumulated in the vicinity of to the metal contact.Bottom-left (table): parameters used in the simulations. Bottom-right: Snapshots of the shock wave formation in the early stage of simulations (between the first and second 
snapshot of Fig.~\ref{fig:shock wave dynamics}).
   }  
  \label{fig:Fig-Sim}
\end{figure}

In order to analyze the existence of a shock wave scenario, the Hi to Lo process was simulated for different external currents. Previously, an initial forming process was performed at which current of different polarities was applied and well defined Hi and Lo states were obtained. In the top-right panel of Fig.\ref{fig:Fig-Sim} the vacancy profile is shown before and after the forming process. The device starts with a uniform distribution along the channel, and by the end of the forming process the profile shows a characteristic distribution where vacancies migrated out from  the interfacial region and accumulated in the neighboring   low-field bulk region \cite{Rozenberg2010}. The figure shows the formed Hi resistance state where it can be seen a second accumulation of vacancies next to the left metal contact in the interfacial region. This second accumulation of vacancies constitutes the initial distribution for the Hi to Lo process and it will evolve to form the shock wave during switching as shown in the main text.
To improve the stability of the simulations the currents were applied using a rise time (from 0 to their actual values) always negligible compared with the characteristic times of the process $\tau_{1}$ and $\tau_{2}$ (with rise times in the order of a few thousands of steps).
The simulated device contained a total of 1000 cells with a 100 cells long interfacial region (left interface). For the initial distribution a uniform density of $u_0=1\times10^{-4}$ per cell was used. The value of the activation constant $V_0$ was set to 16,  which corresponds to the activation
energy of $0.4 eV$ reported in similar manganite 
systems \cite{nian2007evidence} ( i.e. , $0.4 eV / k_{B}T = 0.4 /0.026$ at room temperature). 
For the resistivity constants we used $A_{SB}=1000$ and $A_B=1$ (for a table with the parameters see Fig.\ref{fig:Fig-Sim}). 
For the analysis developed in this paper we solely consider the 
shock wave after it is formed.  In the bottom panel of Fig.\ref{fig:Fig-Sim} 
we show a standard formation process of the shock wave: the initial 
narrow accumulation against the metal contact rapidly evolves into 
the shock wave profile.

\section{Fitting for VEOVM model }
\label{sec:fitVEOVM}

In this section, we explain the details of the comparison between simulation
and theory and the procedures used for fitting. Consider the drift current within
left interfacial region in VEOVM model:

\begin{eqnarray}
j(u,t,x) = P(x, x+\Delta x) - P( x+\Delta x,x) = 2Du\sinh\left(IA_{SB}u\right)
\label{eq_j}
\end{eqnarray}

where the term $1-u(x+\Delta x)$ in the probability  has been neglected as 
explained before \footnote{The typical value of $u(x)$ in simulation varies between $10^{-4}$ and $10^{-3}$.}.
From Rankine\textendash{}Hugoniot conditions an expression for the shock-wave front velocity $\frac{dx_{s}}{dt}$ can be obtained, as provided in Eq.~(\ref{eq:shock wave velocity}) in main text. Performing a redefinition of parameters (for practical reasons only) we can rewrite such equation as:

\begin{eqnarray}
\frac{dx}{dt} =  \frac{2D}{x_{int}}\left[\left(1+\frac{\alpha}{\beta}x\right)\sinh\left(\frac{I}{\beta}+\frac{I}{\alpha x}\right)-\frac{\alpha}{\beta}x\sinh\left(\frac{I}{\beta}\right)\right],
\label{eq:shock wave velocity1}
\end{eqnarray}

where we used that $u_{-}=u_{+}+\Delta u$ , $Q=\Delta u x_{s}$ is the total number of vacancies carried
by shock wave, $Q_{B}\equiv u_{+}x_{int}$ is total number of background vacancies in the left region (being $x_{int}$ the length of the interfacial region),  $x\equiv x_{s}/x_{int}$ is the normalized coordinate, and where we defined the new parameters $\alpha \equiv x_{int}/A_{SB}Q$ and $\beta \equiv x_{int}/A_{SB}Q_{B}$.
From the previous definitions we can write the high resistance value as $R_{HI}=A_{SB}\left(Q_{B}+Q\right)$
where $R_{HI}$ is a constant determined by the vacancy concentration and independent of $I$
 \footnote{There is in fact a small migration of background vacancies into the bulk during the period of shock wave propagation which causes a small variation of the resistance during this phase.}.

We can solve then this equation numerically with the material dependent parameters $x_{int}$, $\alpha$,$\beta$ and $D$. 
Considering the initial rise time for the current and the non-flatness of the shock wave, an accurate test of the $x_s(t)$ prediction can be performed for $x>x_{0}\approx0.4$ using the integral-form equation: 

\begin{eqnarray}
t-t_{0} & = & \intop_{x_{0}}^{x}\frac{\left(x_{int}/D\right)dy}{2\left(1+\frac{\alpha}{\beta}y\right)\sinh\left(\frac{I}{\beta}+\frac{I}{\alpha y}\right)-\frac{2\alpha}{\beta}y\sinh\left(\frac{I}{\beta}\right)}\label{eq:t-x simulation}
\end{eqnarray}

which is used for the fit of the simulation data for $t <\tau_1$ in the lower left panel of Fig.\ref{fig:shock wave dynamics} in the main text.

\begin{figure}[h]
\centering{}\includegraphics[width=0.7\columnwidth]{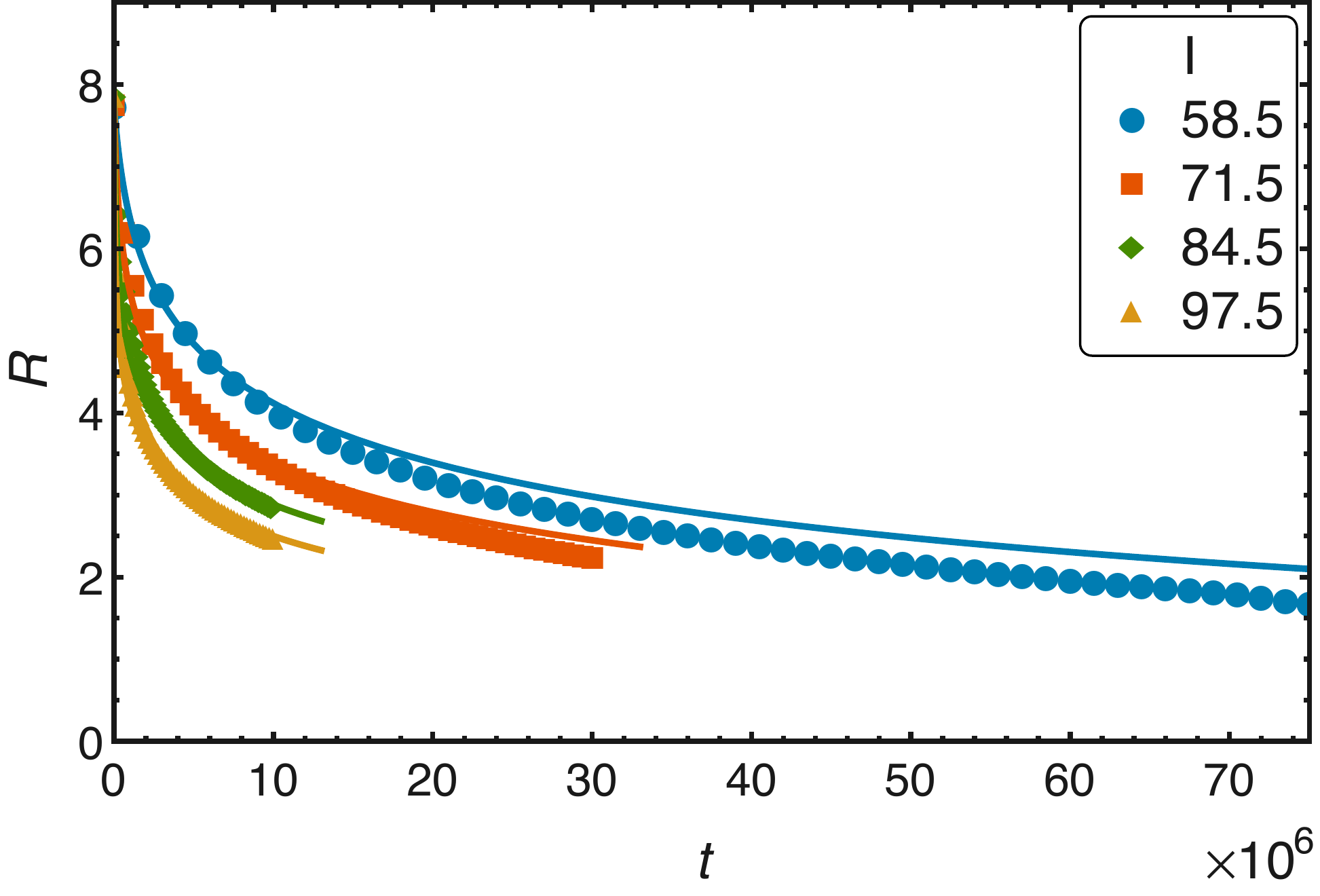}

\caption{Evolution of the resistance during the switching phase for different applied currents 
according to simulations (dots) and theory (lines) from Eq.~(\ref{eq:R(t)}). 
The currents shown are $I=58.5a.u., 71.5a.u.,84.5a.u.$ and $97.5a.u.$}
\label{fig:R-fit}
 \end{figure}

On the other hand, we have also the rate equation for the switching (i.e. "leakage") phase 
(cf. Eq.\ref{eq:R(t)} from main text). The numerical solution of this equation is used to fit the simulation data in the resistance switching phase
for $\tau_1 < t $, shown in Fig.\ref{fig:R-fit}.

The values of the parameters that enter the equations Eq.\ref{eq:t-x simulation} and Eq.\ref{eq:R(t)} were extracted directly 
from the simulation results. We find $Q_{B}=63.4u_0$, $Q=15.3u_0$, $A_{SB}=1000$, $R_{HI}=7.9a.u.$ 
and $D=1.12\times 10^{-7}$. The very good fits of the simulation data shown in Fig.~\ref{fig:shock wave dynamics}  and Fig.\ref{fig:R-fit} were achieved 
by slightly 
relaxing  the value of the single parameter $x_{int}$=100 to the values 94.3 (Fig.~\ref{fig:shock wave dynamics} ) in the propagation phase
and 82.2 (Fig.\ref{fig:R-fit}) in the leakage phase.

The small discrepancy between the relaxed parameters and its actual value mainly comes from the non-flatness of the density 
profile both during the propagation of the shock wave and during the leaking phase, as well as from the fact that a small part 
of background vacancies leaks into the bulk during the propagating phase.

 \section{Experimental details}
 \label{sec:expdet}
 
 \begin{figure}[h!]
   \centering{}\includegraphics[width=0.7\textwidth]{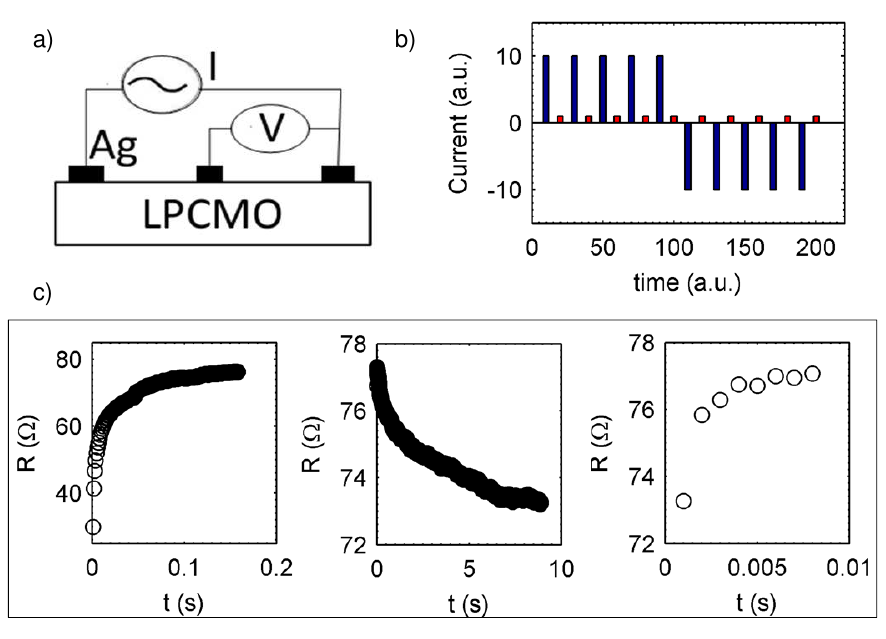}
   \caption{ (a) Diagram of the experimental set up. (b) Sketch of the pulsed current protocol used in the experiments. A high current Write pulse (blue) is followed by a low current Read pulse (red). The first pulse generates the RS while the second pulse measures the non-volatile resistance. (c) R vs t curves from a complete RS process. The system is first taken to a Hi resistance state under a current of -350mA (left panel). Then the Hi to Lo RS is measured under an applied current of 37.5mA (middle panel). Finally the system is taken back, ie ``re-initialized'', to a Hi resistance state with the application of a -350mA current.
    }
   
   \label{fig:Fig-Exp}
 \end{figure}
 
 For the experimental validation of our theory, the RS phenomenon was studied in a bulk La$_{0.325}$Pr$_{0.300}$Ca$_{0.375}$MnO$_3$ (LPCMO) polycrystalline sample with hand-painted Ag contacts of approximately 1mm diameter. The device is a parallelepiped of $1\times1\times10 \text{mm}^3$, with the contacts painted over the 10mm surface. The distance between contacts is also in the order of 1mm. Measurements were performed using a three wire configuration in order to develop a single contact analysis.
 
 To induce the RS effect an external constant current was used as the electric stimulus. To generate this current and to acquire the data a Keithley 2612 Sourcemeter was employed. The measurements were done using a three wire configuration in order to measure a single interface resistance (Fig.\ref{fig:Fig-Exp}.a). The Hi to Lo RS was studied for currents of different magnitude, as described in the main text.  More information about the measurement technique and simultaneous contact analysis can be seen in ref. \cite{Quintero2007}.
 
 The application of current is done following a pulsed protocol: a high current pulse (Write) is followed by a low current pulse (Read). A schematic diagram of the pulsed protocol is shown in (Fig.\ref{fig:Fig-Exp}.b). Each pulse lasts 1ms. Between two pulses there is an interval of about 0.5s, meant to reduce  possible heating effects. The time axis exhibited in the $R$ vs $t$ curves (both in Fig.\ref{fig.R(t)}  and in Fig.\ref{fig:Fig-Exp}.c, below) is the effective time elapsed during the actual application of the Write pulses, i.e. disregarding both the 0.5 s timeout and the Reading elapsed time. 
 The Write pulses possess enough strength to change the resistive state of the system, while the Read pulse (low current) measures the remnant (stable, non-volatile) resistance of the device without affecting it. In the experiments shown in main text, the Write pulses are in the order of the mA while the Read pulses are in the order of the $\mu$A.
 
 The complete measurement process is shown in Fig.\ref{fig:Fig-Exp}.c. To obtain the initial Hi state, pulses of -350mA were applied ( Fig.\ref{fig:Fig-Exp}.c, left panel). Next, the desired  accumulation experiment is performed, by applying positive pulses of constant amplitude which decrease the resistance (Fig.\ref{fig:Fig-Exp}.c, middle panel). The initial Hi -resistance state is recovered by applying -350 mA pulses (Fig.\ref{fig:Fig-Exp}.c, right panel). In every case the initial Hi-resistance state obtained was of the same magnitude within a range of about 1$\%$.

\section{Scaling}
\label{sec:scaling}

In this section, we show that the scaling behaviour
is a direct consequence of the strong non-linearity of the drift current
$j(u,t,x)$, which depends on local electric field exponentially. 
In the time range where significant resistive switching occurs, we have ($IR/x_{int}\gg1$)
and we may approximate $\sinh\left(IR/x_{int}\right)\approx\frac{1}{2}\exp\left(IR/x_{int}\right)$.
From Eq.(\ref{eq:R(t)}), using a normalized resistance $\widetilde{R}=R/R_{HI}$,
we have:

\begin{eqnarray}
\frac{d\widetilde{R}}{dt} & = & -\frac{D}{x_{int}}\exp\left[\left(\frac{IR_{HI}\widetilde{R}}{x_{int}}\right)\left(1+\lambda\right)\right],
\end{eqnarray}
where $\lambda=\frac{x_{int}}{R_{HI}}\ln\widetilde{R}/\left(I\widetilde{R}\right)$
is a small parameter as long as $\widetilde{R}$ close is to $1$ (i.e. at the beginning of the resistive change). 
For simplicity we consider the leading order as we set $\lambda=0$:

\begin{eqnarray}
\widetilde{R} & = & 1-\frac{x_{int}}{IR_{HI}}\ln\left(1+\frac{t^{*}}{\tau_2}\right),\label{eq:approximate solution,Appendix}
\end{eqnarray}

where $t^{*}= t-\tau_1$ is the time measured from the impact time as explained in main text, and  $\tau_2$ is a characteristic time for the resistance switch and is dominated by an exponential dependence
of the applied current as follows: 

\begin{equation}
\tau_2=\frac{x_{int}^2}{DIR_{HI}}\exp\left(-\frac{IR_{HI}}{x_{int}}\right).
\label{eq:tau2}
\end{equation}

\begin{figure}[h]
\centering{}\includegraphics[width=5in,height=5in,keepaspectratio]{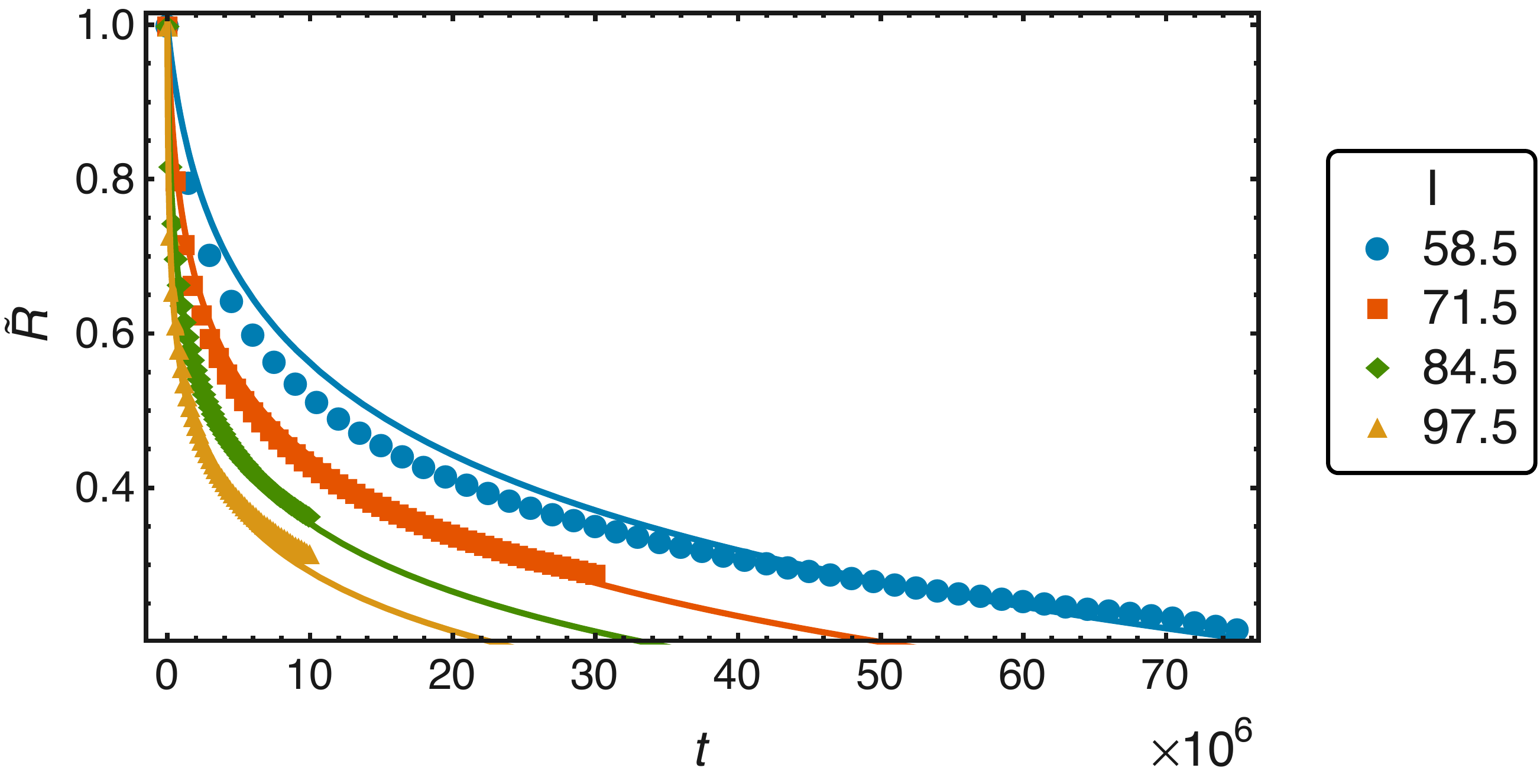}\caption{Fitting results for the evolution of the normalized resistance according to the Eq.~(\ref{eq:approximate solution,Appendix}).}
\label{FittRbar}

\end{figure}

In Fig.\ref{FittRbar} we use this approximate expression to fit the $R(t)$ simulation data. Comparison with the previous fit done with
Eq.\ref{eq:R(t)} and shown in Fig.\ref{fig:R-fit} allows us to check that this approximate solution is relatively accurate within the time domain we are interested in for fast switching devices.

Now, using Eq.~(\ref{eq:approximate solution,Appendix}), we can show that the $I$-dependent family curves $R(t)$
should obey scaling. 
We first consider the normalized resistive change $\delta R(t)$ defined as
\begin{eqnarray}
\delta R(t) & = & \frac{R(t) -R(\tau_{0})} {R_{HI}-R\left(\tau_{0}\right)}
\label{eq:scaling1}
\end{eqnarray}
where  $\tau_0$ is some yet unspecified time (and we drop the $*$ from $t^*$).   
Then, replacing with Eq.\ref{eq:approximate solution,Appendix} we have,
\begin{eqnarray}
\delta R\left(t\right) & = & \frac{R_{HI} - \frac{x_{int}}{I} \ln(1+ \frac{t}{\tau_2}) - R_{HI} + \frac{x_{int}}{I} \ln(1+ \frac{\tau_0}{\tau_2})}
{R_{HI} - R_{HI} + \frac{x_{int}}{I} \ln(1+ \frac{\tau_0}{\tau_2})}
\end{eqnarray}
And rescaling the time by $\tau_0$,
\begin{eqnarray}
\delta R\left(t/\tau_0\right) & = & 1 - \frac{\ln(1+ \frac{\tau_0}{\tau_2}\frac{t}{\tau_0})} {\ln(1+ \frac{\tau_0}{\tau_2})}
\label{eq:scaling2}
\end{eqnarray}

Notice that with the natural choice of simply setting $\tau_0$ as $\tau_2$, 
we obtain the scaling form that we presented in the main text.

However, in the experiment we do not know, a priori, how to determine the characteristic time 
scale $\tau_2$, which is a strong function of the current $I$ and other material parameters. 
So we adopt the following strategy. We use $\tau_0$
as a free scaling parameter, one for each $I$-dependent curve $R(t)$, that we rescale according to 
Eq.\ref{eq:scaling2}.  We vary the set of values $\tau_0[I]$ until we obtain
a collapse of all the experimental and the simulation curves. 
The results of the successful collapse are shown in Fig.\ref{scaling} of the main text. In our experience, the collapse is unique. We
found a sole way to properly collapse the whole set of
curves, which gave us further confidence on the adopted
procedure. 
A crucial point now is that the collapsed set of curves could be fitted with the expression
\begin{equation}
F(t)=1-\frac{\ln\left(1+ct\right)}{\ln\left(1+c\right)}.
\label{eq:scaling3}
\end{equation}
with $c$ a {\it current independent} constant. We find the values
$c=15.2$ for the experimental data and $c=29.4$ for the simulation ones. 
Hence, in regard of Eqs.\ref{eq:scaling2} and \ref{eq:scaling3} we observe that the constant $c$ is
nothing but the ratio between the analytically established characteristic time $\tau_2$ and the
empirically determined $\tau_0$, which are simply proportional to one another.

\begin{figure}[h]
\centering{}\includegraphics[width=6in,height=6in,keepaspectratio]{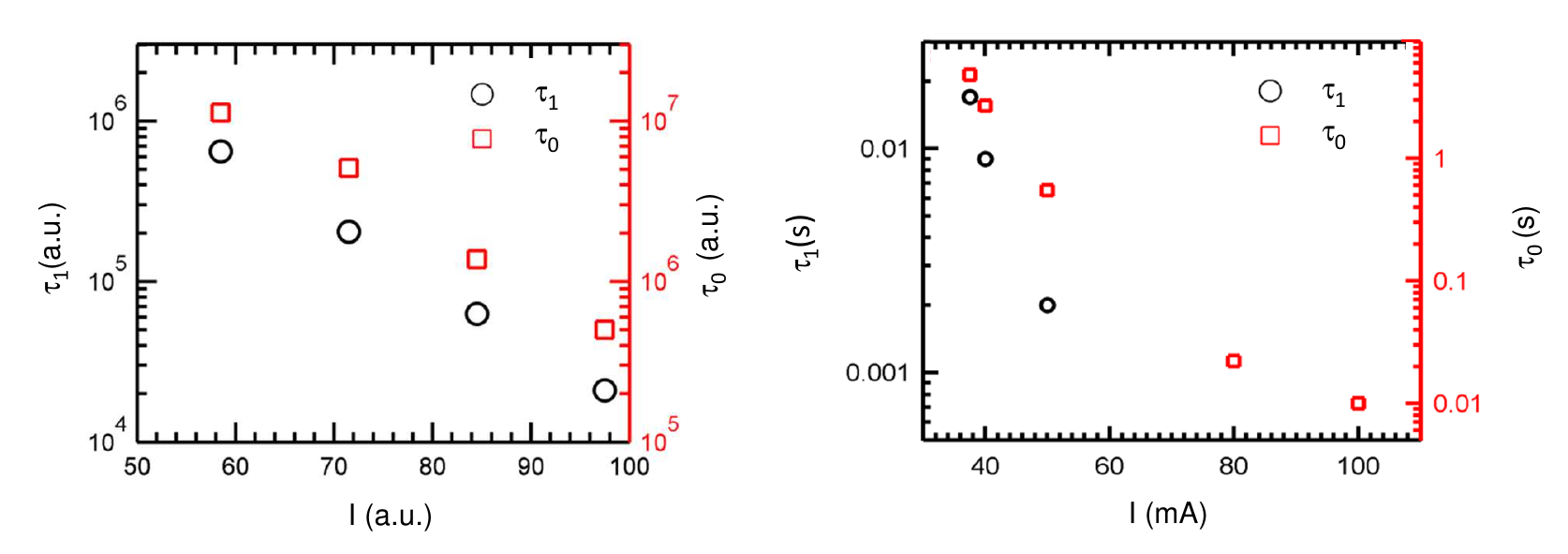}
\caption{Evolution of the characteristic times for different applied currents for both simulations (left panel) and 
experiments (right panel). It can be seen that both times follow an 
approximate exponential dependence with $I$, and that there exist a relative 
proportionality between them as predicted by our analysis.}
\label{Taustogether}

\end{figure}

In Fig.\ref{Taustogether} we plot the dependence of the scaling time as a function of the applied
current $\tau_0(I)$.
In the case of the simulation results, the data follow the same exponentially decreasing 
behaviour as deduced for $\tau_2(I)$ (see Eq.\ref{eq:tau2}). 
Moreover, we find the ratio $\tau_0(I)/\tau_2(I)$ in good agreement with the constant $c=29.4$, which
validates our practical scaling procedure for the determination of the characteristic switching time in
the experimental case.

Under the same set of approximations that we have assumed in this section, plus the additional one
of neglecting the effect of the background distribution of vacancies, one may  also derive an
explicit expression for the characteristic time $\tau_1$.
We start from Eq.~(\ref{eq:shock wave velocity1}), and similarly as before, adopt the 
approximation $\sinh\left(IR/x_{int}\right)\approx\frac{1}{2}\exp\left(IR/x_{int}\right)$. Then, making 
the substitution $y^{\prime} = 1/y$ in Eq.~(\ref{eq:t-x simulation}), and assuming that the integral is dominated 
by the exponential factor (i.e. neglecting the change in lower order factors), the integral has an analytical solution, 
and we get for the case with no background vacancies (i.e. $1/\beta =0$)
 \begin{equation}
 \tau_1 \approx C(t_0,x_0) + \frac{x_{int}^2}{DIR_{HI}} \exp\left(-\frac{IR_{HI}}{x_{int}}\right).
 \label{eq_tau1_scal}
 \end{equation}
 
 where the first term is the integration constant determined by the initial conditions $t_0$  and $x_0$. If this constant can be neglected (for example for fast forming shock waves) then we have the surprising result that $\tau_1 = \tau_2$. 

In the case of the model simulations, the results displayed in Fig.\ref{Taustogether} shows that, in fact, both characteristic times
have the same $I$-dependent exponentially decaying behavior. Moreover, the ratio of the two characteristic times
is approximately 25, which is close to the constant $c=29.4$ quoted before, therefore our simulations
also validate the equality between the two characteristic times $\tau_1$ and $\tau_2$ predicted by the shock wave scenario.

As may be expected, the experimental situation is only qualitatively consistent with the previous discussion. As seen
in Fig.\ref{Taustogether}, for the experimentally determined characteristic times, we observe that they have
approximately a similar dependence with the applied current, however, it is less clear if the equality between them also holds.

We show and compare the evolution of both characteristic times under different applied current in Fig.\ref{Taustogether} for both simulations and experiments, observing a good agreement with the analysis presented here.

\section{Quantitative Comparison: Experiments \& Simulations}
\label{sec:quantcomp}

\begin{figure}[h]
  \centering{}\includegraphics[width=1.0\textwidth]{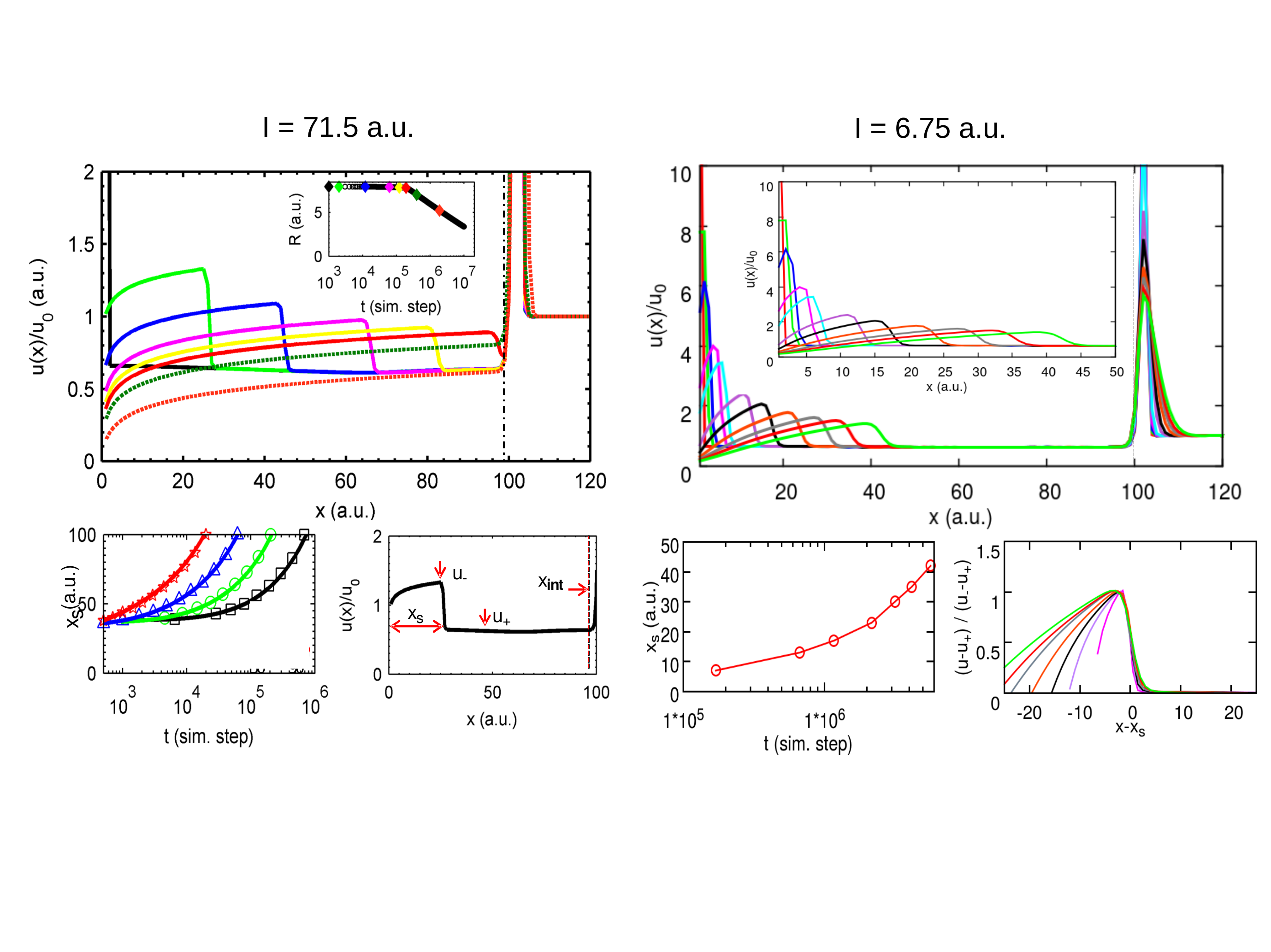}
  \caption{Shock wave dynamics for two different applied currents, which differ by nearly an order of magnitude. The lower current case (right panel) corresponds to applied voltages comparable to the experimental ones, and the higher voltage cases (left panel, same as FIG. 2 ) show  similar shock wave dynamics. The two lower-right panels demonstrate that (i) that the propagation of the shock wave 
front remains qualitatively the same as for higher applied currents, and (ii) the collapse of 
the snap-shots of successive profiles indicates the formation of the shock-wave. }
  \label{fig:Low_vs_High_I}
\end{figure}

The magnitudes of the main parameters used in the simulations can be related to actual experimental magnitudes. In our model all voltages are normalized by $k_{B}T/e$ and  
energies by $k_{B}T$. As experiments were performed at room temperature, we have 
$k_{B}T/e\approx 26$ mV and $k_{B}T\approx 0.026$ eV. In the simulations the activation energy constant was given a value of $V_0=16~k_{B}T \approx 0.4eV$, which is consistent with experimental values reported for similar PCMO devices \cite{nian2007evidence}. A correspondence can also be established for the voltages adopted in experiments and simulations. In the former the applied voltages used for the resistive-commutation were in the order of 1V. In units of $k_{B}T/e$ this leads to the dimensionless voltage values in the order of a few tens (1V/26mV = 38.5). However, in order to perform the extensive computational 
work required by our study within reasonable computational times, 
the values adopted in the simulations were higher by about one order
of magnitude. For evident practical reasons (equipment and device limitations) such an increase of 
applied voltage cannot be performed in the experiments. 
Nevertheless, it should be clear from our theoretical discussion
of the mathematical nature of the equations, that the same qualitative results, namely formation of
shock waves,  also remain valid at lower applied voltages. 
Despite the significantly higher computational cost (days versus hours), we present here a set of runs 
for voltages that are close to the experimental values. In figure \ref{fig:Low_vs_High_I}, we show the results for a simulation with an applied dimensionless current $I=6.75$, which corresponds to
a physical units $V=1.4$V, since  the initial (dimensionless) $R=8$ and then $V=IRk_{B}T/e=6.75\times8\times26$mV $ \approx 1.4$V. For convenience, in the figure 
we also reproduce the
corresponding results from the main text for a simulated voltage of $V=14.87 V$ ($I=71.5a.u.$).

\chapter{Shock-wave Formation in Binary Oxides}
\label{sec:binaryox}

In this section we describe under what conditions one may expect 
the formation of shock waves (SW) to occur in binary insulating transition metal oxides
such as $\textrm{HfO}_{2}$, $\textrm{TiO}_{2}$, $\textrm{TaO}$, which are
of current interest.
These insulating systems are qualitatively different from the doped manganite compound
that we consider in the main text. Nevertheless, it is interesting to realize that they may
also sustain the formation of a SW front according to our general formalism.
A key point to realize is, however, that the SW may occur for the propagation of 
the {\it oxygen} density profile and not the oxygen vacancy as in the manganite case.
This is because in the latter oxygen vacancies increase the local resistivity by means
of creating defects in the oxygen-metal-oxygen bonds, hence in the conduction bands,
while, in contrast in the former case of binary systems the oxygen vacancies dope
electron carriers to an otherwise good insulator.
Hence, the crucial features to realize that SW may occur in binary systems are the
following: (i) the electroforming step renders the highly insulating system
poorly conductive by creating a path with a massive production of oxygen vacancies (OV).
In an extreme case, this path may become a purely metallic filament, which leads
to the so called {\it non-polar} resistive switching mode. However, if this is not
so drastic, one gets to the {\it bi-polar} switching mode with an intermediate density of OV.
(ii) Within such a conductive path, oxygen ions move through the OV sites by means
of the applied electric field. As local oxygen density increases, the system locally
approaches the stoichiometric formula, hence, becoming locally more insulator
with a local increase of resistivity. (iii) The local increase of resistivity leads to
higher {\it local} voltage drops, hence higher local electric fields, which further promote
a higher motion of oxygen ions, leading to the formation of the SW.
Interestingly, this phenomenon may have occurred in the simulation study of
Strukov et al. (cf. Figs.~2(f) and 2(h) of  \cite{Strukov}).

We shall now describe how the expressions of our SW scenario derived for
conductive manganites may be recast for insulating binary compounds. Lets consider $u_{O}+u_{OV}=n$ where $n$ is the integer for the oxygen
atoms in the unit formula, and $u_{O}$ and $u_{OV}$ are the actual number
of oxygen atoms and oxygen vacancies per unit formula in the sample
respectively, hence they are proportional to the respective local densities.
In the main text we have derive the expressions in terms of $u$, the 
oxygen vacancies that corresponds to $u_{OV}$. As we said before
we shall now focus on the oxygen density $u_{O}$. Similarly as in the main text (MT), we have a continuity equation for
the oxygen $\partial_{t}u_O+\partial_{x}j_O(t,x)=0$.
From the components of drift and diffusion currents we get
\begin{equation}
\partial_{t}u_O+f\left(u_O\right)\partial_{x}u_O=D\partial_{xx}u_O,\label{burgers2}
\end{equation}
where $f\left(u_O\right)\equiv\partial_{u_O}j_{O,drift}\left(u_O,I\right)$, 
$j_{O,diffusion}=-D\partial_{xx} u_O$ and
$I(t)$ is the magnitude of the electronic current.

Notice that this equation is the generalized Burgers' equation for the case of
oxygen ion movement. The key point here is that, as pointed out in the MT,
the function $f\left(u_O\right)$ should be any monotonically increasing
function of $u_O$. Again, following the MT, $j_{O,drift} = u_O g(E)$, where
$E$ is the magnitude of the local electric field and $g$ clearly is an
increasing function of it. The specific form of $g(E)$ 
is material dependent but the important point already mentioned is that
a local increase of $u_O$ leads to a local increase of $\rho(u_O)$, and hence
of $E$.
Therefore, from analogous arguments as discussed in the MT a SW front
may also occur in the binary insulating compounds. As we said before,
they may occur for the case of systems in bi-polar mode RS.
An interesting issue would be to investigate the temperature or current
intensity dependence of the SW formation. A priori, temperature may play
two competing roles. On one hand, higher currents or fields would lead to
higher temperature, hence to higher oxygen ion mobility. This may favour
the formation of a SW. However, temperature also decreases the local resistivity
of an insulator, hence may prevent the local increase of the field intensity.

\begin{figure}[h]
\centering{}\includegraphics[width=4.5in,height=4.5in,keepaspectratio]{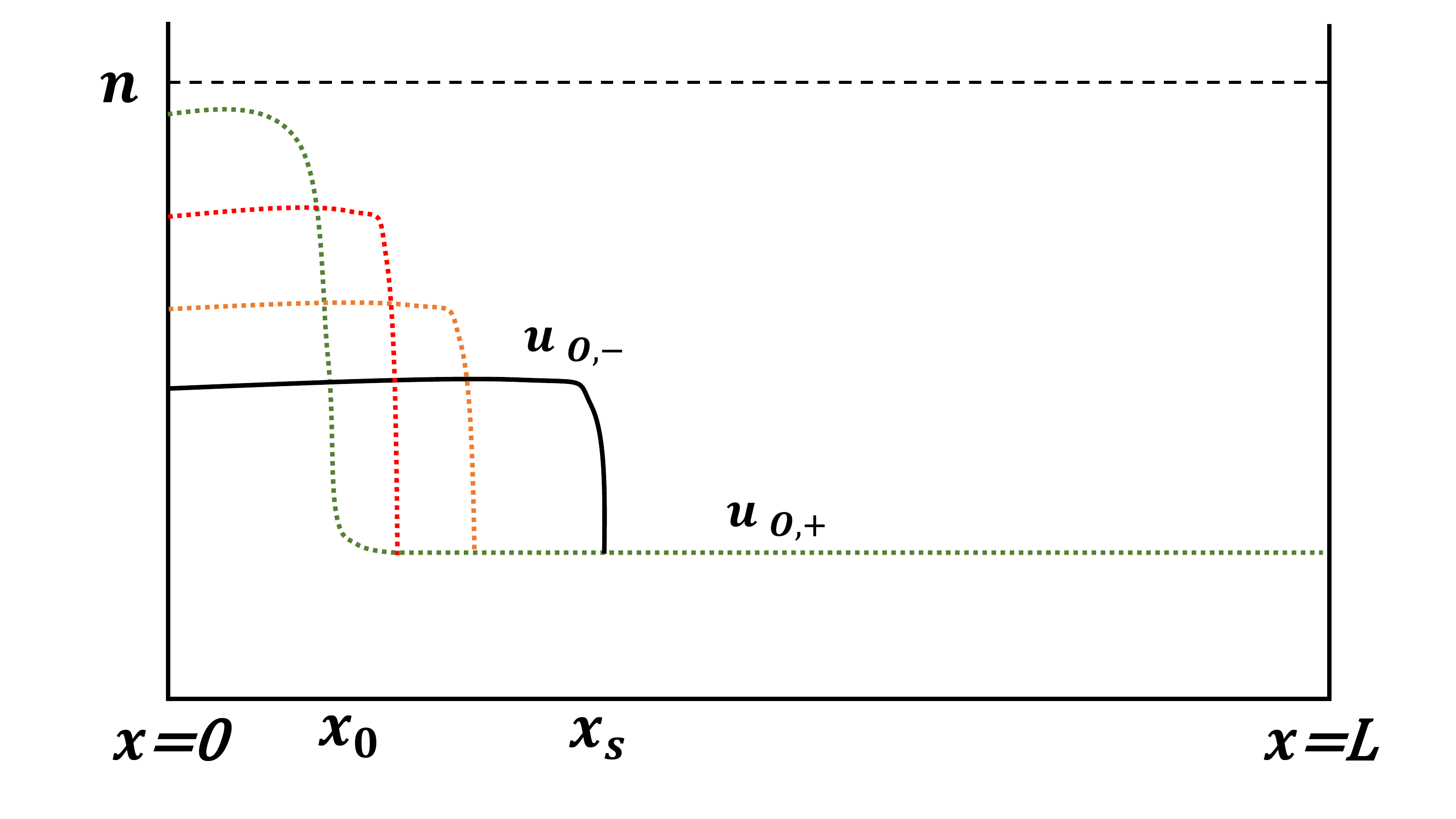}

\caption{The schematic plot for the oxygen shock wave front propagation. The
green curve is the initial distribution where oxygen are piled up
at $x<x_{0}$.\label{fig:The-schematic-plot}}
\end{figure}

Having discussed and established under which general conditions one may expect
the formation of a SW, we now present the specific formulas for the behavior of
the resistance $R(t)$ during the evolution of the SW front.  
Figure \ref{fig:The-schematic-plot} is a schematic representation of the
time evolution of the SW and serves for the definition of the various
relevant variables.

We assume that electroforming has created a  
path rich in oxygen vacancies, which shall dominate the conduction
through the system. Thus there is a constant background of density of
oxygen $u_O^0 = u_{O,+}$ on top of which the SW evolves.
The initial state is a high resistive state with a pile-up of oxygen atom
(at the l.h.s.) that brings the local density close to the stoichiometric value, and
hence a highly insulating region. As a strong (positive) voltage is applied at
the electrode at $L$, the oxygen ions will migrate rapidly forming a 
plateau of density $u_{O,-}$ with a sharp front of magnitude $u_{O,-} - u_{O,+}$
at the position $x_s$. The front moves with velocity $v_s$.

The total (two point) resistance of the system (we neglect the effect of the
electrode contacts) is given by,

\begin{equation}
R\left(t\right)=\int_{x=0}^{x=L}\rho\left(u_{O}\left(x\right)\right)dx,
\end{equation}
using the continuity equation for oxygen, we can describe the decrease
rate of the resistance  from the high resistive state $R_{HI}$ to the low resistive
state $R_{LO}$ as,
\begin{eqnarray}
\frac{dR\left(t\right)}{dt} & = & -\int_{0}^{x_{s}}\left(\frac{d\rho}{du_{O}}\right)\partial_{x}j_{O}\left(u_{O},t,x\right)dx-\int_{x_{s}}^{L}\left(\frac{d\rho}{du_{O}}\right)\partial_{x}j_{O}\left(u_{O},t,x\right)dx\nonumber \\
 &  & +\rho\left(u_{O}\left(x_{s}-\varepsilon\right)\right)v_{s}-\rho\left(u_{O}\left(x_{s}+\varepsilon\right)\right)v_{s},
\end{eqnarray}
where the quantity $\left(\frac{d\rho}{du_{O}}\right)$ is a function
of $u_{O}\left(x,t\right)$.
The shock wave velocity $v_{s}$ can be obtained via the Rankine\textendash Hugoniot condition
as described in the Appendix~\ref{sec:DynofSW}. 
Since we assume a large initial pile-up distribution of oxygen density
near $x=0$, this creates a large local resistivity and thus
large drift force, hence we expect the shock wave will form quickly. 
Therefore, the distribution of $u_{O}\left(x,t\right)$ for $x<x_{s}$ 
soon becomes approximately flat and will be denoted as $u_{O}\left(t,x_{s}-\varepsilon\right)
\approx u_{O,-}(t)$.

Using the boundary condition  $j_{O}\left(u_{O},t,x=0\right)=j_{O}\left(u_{O},t,x=L\right)=0$,
and defining $j_{O}\left(u_{O},t,x_{s}\pm\varepsilon\right)\equiv j_{O,\pm}$,
$\rho_{\pm}\equiv\rho\left(u_{O}\left(x_{s}\pm\varepsilon\right)\right)$,
$\rho^{\prime}\left(u_{O}\right)_{\pm}\equiv\frac{d\rho}{du_{O}}\left|_{x_{s}\pm\varepsilon}\right.$,
$u_{O}\left(x_{s}\pm\varepsilon\right)\equiv u_{O,\pm}$, we have:
\begin{eqnarray}
\frac{dR\left(t\right)}{dt} & = & -\left[\rho^{\prime}\left(u_{O}\right)_{-}j_{O,-}-\rho^{\prime}\left(u_{O}\right)_{+}j_{O,+}\right]+\left[\rho_{-}-\rho_{+}\right]v_{s},\label{eq:main}
\end{eqnarray}
Recalling that we assumed the background oxygen density approximately constant,
we define $Q_{O}$ as the total number of active oxygen ions,
which is a constant during lapse of motion of the shock wave:
\begin{equation}
Q_{O}=\left[u_{O,-}\left(t\right)-u_{O,+}\right]x_{s}\left(t\right).\label{eq:QO}
\end{equation}
In terms of the parameters that we defined we have,
\begin{equation}
R\left(t\right)=\rho\left(u_{O,-}\left(t\right)\right)x_{s}\left(t\right)+\rho\left(u_{O,+}\right)\left(L-x_{s}\left(t\right)\right).\label{eq:R}
\end{equation}
Then taking the time derivative and comparing to Eq.\ref{eq:main}, we obtain
\begin{eqnarray}
\frac{du_{O,-}}{dt} & = & \frac{j_{O,+}\rho^{\prime}\left(u_{O}\right)_{+}/\rho^{\prime}\left(u_{O}\right)_{-}-j_{O,-}}{x_{s}\left(t\right)},
\end{eqnarray}
where $u_{O,+}$ is time independent and known and similarly for
$\rho^{\prime}\left(u_{O}\right)_{+}$, $j_{O,+}$. 
To make further progress we now need the specific form of $\rho\left(u_{O}\right)$ and $j_{O}\left(u_O,t,x\right)$,
which depend on the specific transport properties of the physical system.  Eliminating $x_s$ we
could then obtain a differential equation for $u_{O,-}\left(t\right)$:
\begin{equation}
\frac{du_{O,-}}{dt}=\frac{\left(j_{O,+}\rho^{\prime}\left(u_{O}\right)_{+}/\rho^{\prime}\left(u_{O}\right)_{-}-j_{O,-}\right)\left(u_{O,-}-u_{O,+}\right)}{Q_{O}},\label{eq:u}
\end{equation}
and then, finally, can obtain the expression for $R\left(t\right)$ through equation Eq.\ref{eq:R}.

For a concrete illustration we may now assume a specific model for the conduction, 
where the local resistivity is a increasing function $u_{O}$, as expected for the case of
the binary insulators: 
\begin{eqnarray}
j_{O} & = & 2Du_{O}\sinh\left[I\rho\left(u_{O}\right)\right]\nonumber \\
\rho\left(u_{O}\right) & = & \frac{\rho_{0}}{1+A_{0}\left(n-u_{O}\right)}\label{eq:model}
\end{eqnarray}
This is motivated by the expected approximate linearity of the {\it conductivity} with
the concentration of dopants, namely, oxygen vacancies, therefore we have,
$\sigma = 1/ \rho \sim u_{OV} = (n - u_{O}) $. The parameter $A_0$ is a suitable constant and $\rho_0$ is the intrinsic
resistivity of the stoichiometric system (i.e., for $u_{OV}$=0).

From the schematic Fig.\ref{fig:The-schematic-plot}, we have $Q_{O}=\left[u_{O,-}\left(t=0\right)-u_{O,+}\right]x_{0}$
and we can solve Eq.\ref{eq:u} using the explicit model defined in Eq.\ref{eq:model}. 
We may then solve (numerically) for the evolution of $u{}_{O,-}\left(t\right)$, and finally obtain the resistance as,
\begin{eqnarray}
R\left(t=0\right) & = & R_{HI}=\frac{\rho_{0}x_{0}}{1+A_{0}\left(n-u_{O,-}\left(t=0\right)\right)}+\frac{\rho_{0}\left(L-x_{0}\right)}{1+A_{0}\left(n-u{}_{O,+}\right)}\nonumber \\
R\left(t\right) & = & \frac{\rho_{0}x_{s}}{1+A_{0}\left(n-u{}_{O,-}\left(t\right)\right)}+\frac{\rho_{0}\left(L-x_{s}\right)}{1+A_{0}\left(n-u{}_{O,+}\right)}\nonumber \\
R\left(t=t_{final}\right) & = & R_{LO}=\frac{\rho_{0}L}{1+A_{0}\left(n-u{}_{O,-}\left(t_{final}\right)\right)}.
\end{eqnarray}
As a final remark, one may notice that from Eq.\ref{eq:QO} the resistance can be reparametrized
as a function of the shock wave front position $x_{s}$,
\begin{equation}
R\left[x_{s}\left(t\right)\right]=\frac{\rho_{0}x_{s}}{1+A_{0}\left(n-\frac{Q_{O}}{x_{s}}-u_{O,+}\right)}+\frac{\rho_{0}\left(L-x_{s}\right)}{1+A_{0}\left(n-u{}_{O,+}\right)},
\end{equation}
and $R_{HI}=R\left[x_{s}\left(t\right)=x_{0}\right]$, $R_{LO}=R\left[x_{s}\left(t\right)=L\right]$.
It is straightforward to see that $R\left[x_{s}\left(t\right)\right]$ is a monotonically decreasing function of $x_{s}\left(t\right)$ and
thus $R_{LO}<R_{HI}$.


\renewcommand*{\bibname}{References}


 \bibliographystyle{unsrtnat}
 \bibliography{shao_refs}

\end{document}